# Noise will be noise: Or phase optimized recursive filters for interference suppression, signal differentiation and state estimation (extended version)

Hugh L. Kennedy

*Abstract*— **The increased temporal and spectral resolution of oversampled systems allows many sensor-signal analysis tasks to be performed (e.g. detection, classification and tracking) using a filterbank of low-pass digital differentiators. Such filters are readily designed via flatness constraints on the derivatives of the complex frequency response at dc, pi and at the centre frequencies of narrowband interferers, i.e. using maximally-flat (MaxFlat) designs. Infinite-impulse-response (IIR) filters are ideal in embedded online systems with high data-rates because computational complexity is independent of their (fading) 'memory'. A novel procedure for the design of MaxFlat IIR filterbanks with improved passband phase linearity is presented in this paper, as a possible alternative to Kalman and Wiener filters in a class of derivative-state estimation problems with uncertain signal models. Butterworth poles are used for configurable bandwidth and guaranteed stability. Flatness constraints of arbitrary order are derived for temporal derivatives of arbitrary order and a prescribed group delay. As longer lags (in samples) are readily accommodated in oversampled systems, an expression for the optimal group delay that minimizes the white-noise gain (i.e. the error variance of the derivative estimate at steady state) is derived. Filter zeros are optimally placed for the required passband phase response and the cancellation of narrowband interferers in the stopband, by solving a linear system of equations. Low complexity filterbank realizations are discussed then their behaviour is analysed in a Teager-Kaiser operator to detect pulsed signals and in a state observer to track manoeuvring targets in simulated scenarios.**

*Index Terms*— **Acoustic signals, biomedical devices, cybernetics, digital electronics, digital signal processing, humans and machines, linear state-space systems, signals and systems, process models**

## Notation

$i = \sqrt{-1}$: Complex unit.
$s = \sigma + \Omega i$: Complex $s$-plane coordinate, reached via the Laplace transform.
$\sigma$: Real part of $s$ (reciprocal seconds).
$\Omega$: Imaginary part of $s$, angular frequency (radians per second).
$\tau = 1/\sigma$: Coherence duration (seconds).
$\lambda = 2\pi/\Omega$: Wave period (seconds).
$z$: Complex $z$-plane coordinate, reached via the $\mathcal{Z}$ transform.
$\omega = \Omega/F_s$: Normalized angular frequency (radians per sample).
$f = \omega/2\pi$: Normalized frequency (cycles per sample).
$F$: Frequency (cycles per second or Hz).
$F_s$: Sampling frequency i.e. sampling rate (cycles per second or Hz).
$T_s = 1/F_s$: Sampling period (seconds).
$t$: Time (seconds).
$n$: Time index, into a sampled sequence (samples, $0 \le n < N$, $t = nT_s$).
$m$: Delay index, into a sample history (samples, $0 \le m < M$, $t = nT_s - mT_s$).
$q$: Group delay parameter (samples, $-\infty < q < +\infty$, $t = nT_s - qT_s$).
$p$: Pole position in the complex $z$-plane.
$k$: Basis function, state vector, or operator, index ($0 \le k < K$).
$\psi$: Basis-function.
LSS: linear state-space
LTI: Linear time-invariant
$\boldsymbol{A}$, $\boldsymbol{B}$ & $\boldsymbol{C}$: Continuous-time LSS matrices of an LTI system.
$\mathcal{H}(s)$: Continuous-time transfer-function of an LTI system.
$\mathcal{H}(\Omega)$: Continuous-time frequency-response of an LTI system.
$h(t)$: Continuous-time impulse-response of an LTI system.
$\boldsymbol{G}$, $\boldsymbol{H}$ & $\boldsymbol{C}$: Discrete-time LSS matrices of an LTI system.
$\mathcal{H}(z)$: Discrete-time transfer-function of an LTI system.
$\mathcal{H}(\omega)$: Discrete-time frequency-response of an LTI system.
$h[m]$: Discrete-time impulse-response of an LTI system.
$(\blacksquare)$: Denotes a function of continuous argument.
$[\blacksquare]$: Denotes a sampled function of integer argument.
$|\blacksquare|$: Magnitude of a complex variable or the determinant of a matrix.
$\angle\blacksquare$: Angle of a complex variable, e.g. $\omega = \angle z$.
$\mathrm{Re}(\blacksquare)$: Real part of a complex variable, e.g. $\sigma = \mathrm{Re}(s)$
$\mathrm{Im}(\blacksquare)$: Imaginary part of a complex variable, e.g. $\Omega = \mathrm{Im}(s)$.
$\lfloor\blacksquare\rfloor$: Rounds down to the nearest integer.
$E\langle\blacksquare\rangle$: Expectation operator.
$\blacksquare^{\mathrm{T}}$: Transpose of a real matrix or vector.
$\blacksquare^{\dagger}$: Hermitian transpose of a complex matrix or vector.
$\overline{\blacksquare}$: Complex conjugation operator.
$\blacksquare^{(k)}$: The $k$th derivative, e.g. with respect to (w.r.t) time or frequency.
$\dot{\blacksquare}$: First derivative w.r.t. time.
$\blacksquare_c$: Denotes a critical value, constant, or parameter, of a process or filter.
$\blacksquare!$: Factorial operator.

## 1 Introduction

The extraction of signal from noise is a fundamental and enduring problem in signal processing. A revolution, in the field (and in control & communication) occurred towards the end of the pre-digital era when Norbert Wiener showed the post-war world how to design continuous-time systems, i.e. electronic devices made from analogue circuits, that incorporate continuous-time linear models of natural processes, using the Laplace transform (to reach the complex $s$-domain). Over the last 50 years, discrete-time formulations (on the complex $z$-plane) for sampled time-series and digital circuits have risen to prominence and during this new

H L. Kennedy is with DST Group, Edinburgh, SA 5111 Australia (e-mail: hugh.kennedy@dst.defence.gov.au) and the University of South Australia, Mawson Lakes, SA 5095, Australia (e-mail: hugh.kennedy@unisa.edu.au).

era, computer implementations of the Kalman filter and Wiener filter became the *de-facto* standards for optimal state estimation and filtering problems where continuous-time models for signal and interfering processes are known [1],[2],[3].

The recursive (discrete-time) Kalman filter is a minimum mean-square error (MMSE) solution that leverages prior knowledge of second-order moments of noise inputs and initial state estimates, for a recursive realization with a variable gain [1],[2]. When statistical stasis prevails (e.g. when sampling for a long time at a fast and constant rate) the steady-state gain of the filter is computed by solving the Riccati equations. Solution for simple integrating processes (e.g. constant velocity or constant acceleration) is straightforward [2],[4]; however, reasonable solutions may be difficult to reach for high-order (state-transition and state-observation) process models.

The (discrete-time) Wiener filter does not utilize covariance matrices therefore it is an appropriate MMSE solution when reliable statistical priors are unavailable. As its poles are fixed (in the $z$-domain) for a constant-gain, Riccati solutions are unnecessary, thus high-order models may be employed for signal and interference processes. However, unlike the Kalman filter, it is unsuitable for process models that are not wide-sense stationary such as integrating processes (that have poles at $s = 0$), or other marginally stable or unstable processes with poles on the imaginary axis (in the $s$-domain, where $s = i\Omega$) because the power spectral density is undefined for such linear systems [1].

An alternative approach for high-order models that are only partially known, and not necessarily wide-sense stationary, is presented here. It is suggested that maximally flat (MaxFlat) design procedures are suitable in many problems where Kalman or Wiener filters would usually be applied. MaxFlat filters are designed by constraining the complex frequency response at dc (i.e. where $|z| = 1$ in the complex plane and $\angle z = \omega = 0$), for minimal signal distortion; and optionally elsewhere (i.e. $\omega_c < \omega \leq \pi$, where $\omega_c$ is the filter bandwidth) for maximal suppression of band-limited interference (i.e. coloured noise). This simple design approach may be used to craft frequency responses that are the same as or better than those reached via *ab-initio* procedures for partially known process models.

For digital filter design, MaxFlat procedures are attractive because, in the finite impulse-response (FIR) case, they yield closed-form solutions or a simple system of linear equations that are readily solved for the filter coefficients without using bespoke optimizers [5],[6],[7],[8],[9],[10],[11],[12],[13],[14]. MaxFlat procedures for infinite impulse-response (IIR) filters have received less attention than their FIR counterparts [14],[15],[16],[17],[18],[19],[20]; however, IIR filters are an attractive design alternative in oversampled systems with feeble computers (e.g. in embedded biomedical devices, autonomous drone swarms, or low-orbit satellite constellations).

Existing design procedures for IIR MaxFlat filters have some inconvenient shortcomings. For the procedures described in [14],[17],[18],[19], stable solutions are only obtained for some group delays, which must be pre-specified as a design parameter. Furthermore, in [17],[18],[19], the bandwidth is set implicitly by balancing the flatness order at $\omega = 0$ and $\omega = \pi$ or at another frequencies in the passband. Recursive Laguerre filters also lack a bandwidth parameter and flatness is only achieved in the near dc region because all poles are real [15],[34],[35],[36]. The bandwidth of the discrete-time Butterworth filters is adjustable; however, the magnitude response (and phase response) is not otherwise configurable, e.g. notches cannot readily be placed to cancel interferers and the passband group-delay cannot be reduced. The method used to design MaxFlat filters in [20] does allow the bandwidth to be explicitly set via a frequency parameter; however, it does not lead to closed-form solutions thus an iterative optimization is required to find the filter coefficients; furthermore, the group delay is not determined automatically during the design procedure thus it must be set to satisfy system latency requirements or manually adjusted by trial and error.

In the MaxFlat design procedure presented here, Butterworth poles (for guaranteed causal stability, regardless of the pass-band group delay) are used to set the filter bandwidth ($\Omega_c$ or $\omega_c$) and the filter zeros are set (for a specified group delay, $q$) to satisfy derivative constraints at arbitrary frequencies, to compute the derivatives of low-frequency signals, and to cancel interference. An expression for the optimal group delay is also derived, that minimizes the white-noise gain of the filter or the variance of the estimation error at steady state.

An introductory overview of optimal state estimators for linear processes and systems is provided in Section 2 and expressions for their frequency responses are derived then possible FIR and IIR filtering alternatives from the digital signal processing literature (i.e. MaxFlat filters) are discussed in Section 3. In Section 4, a novel procedure for the derivation of IIR MaxFlat filter coefficients is described (see Section 4.1) then filter realization alternatives are considered (see Section 4.2). After a brief discussion of the philosophy of parsimonious process modelling and observer design in Section 5, the proposed MaxFlat filter is used in a Teager-Kaiser (TK) operator to detect pulsed signals (instead of a Wiener filter) in Section 6 and in a state observer to track manoeuvring targets (instead of a Kalman filter) in Section 7. These hypothetical scenarios are used to illustrate tuning considerations and possible applications of the proposed filters. Computer code for the design procedure is provided in the Appendices.

## 2 Linear state-space models and the Kalman filter

For linear processes with dominant poles that are close to the origin and a bandwidth that is much lower than the sampling rate – e.g. observations of ballistic, orbital or celestial, bodies – cascading temporal integrators are generally an adequate signal model. Some alternative process models for wider-band or narrow-band signals – e.g. generated by highly manoeuvrable targets or oscillatory phenomena – are discussed in [21],[22]&[25]. A simple first-order lag is undoubtedly the most used coloured-noise model in the literature [2]; however, it is mainly a pedagogical tool as the separation of a low-frequency signal from low-frequency noise is a fundamentally intractable problem. The first-order Nyquist resonator (with $\Omega/F_s = \omega = \pi$) is proposed in [22] and [23] as a more practical and useful first-order noise model. Second-order interference models with complex poles are used to model higher-order spherical-harmonic gravity accelerations on the orbit of a satellite in [24]; a third-order model with real poles is used to model atmospheric jitter in [25]. In all cases, linear state-space (LSS) models are used to represent the signal and noise processes.

A LSS system definition is used to define how the output $y$ and internal states $\boldsymbol{w}$ (a $K \times 1$ vector) of a linear time-invariant (LTI) system responds to a given input $x$:

$\dot{\boldsymbol{w}}(t) = \boldsymbol{A}_{K\times K}\boldsymbol{w}(t) + \boldsymbol{B}_{K\times 1}x(t)$ (1a)
$y(t) = \boldsymbol{C}_{1\times K}\boldsymbol{w}(t)$ (1b)
in the time domain or
$s\boldsymbol{W}(s) = \boldsymbol{A}_{K\times K}\boldsymbol{W}(s) + \boldsymbol{B}_{K\times 1}X(s)$ (1c)
$Y(s) = \boldsymbol{C}_{1\times K}\boldsymbol{W}(s)$ (1d)
in the $s$-domain (from a zero initial-state).

In the above continuous-time LSS definition (1a) are the state-propagation equations and (1b) are the measurement (or observation) equations. This system may be used to model the dynamics of both signal and noise processes in natural phenomena and analogue electronic circuits (realized using a network of ideal integrators, $s^{-1}$). For this continuous-time system the ($s$-plane) poles of the process are equal to the eigenvalues of $\boldsymbol{A}$ and the $(K \times K)$ Laplace transform of the fundamental matrix $\boldsymbol{\mathcal{H}}(s)$ is

$\boldsymbol{\mathcal{H}}(s) = (s\boldsymbol{I}_{K\times K} - \boldsymbol{A})^{-1}$ . (2a)

It is used to derive the continuous-time transfer-function of the single-input/single-output system, the frequency response, and the power spectrum density, as follows:

$\mathcal{H}(s) = \boldsymbol{C}\boldsymbol{\mathcal{H}}(s)\boldsymbol{B}$ (2b)
$\mathcal{H}(\Omega) = \mathcal{H}(s)|_{s=i\Omega}$ and (2c)
$P(\Omega) = \bar{\mathcal{H}}(\Omega)\mathcal{H}(\Omega)$ . (2d)

The fundamental matrix also determines the (Dirac-delta) impulse response and (unit) step-response from a zero initial-state:

$h(t) = \boldsymbol{C}\boldsymbol{G}(t)\boldsymbol{B}$ and (3a)
$g(t) = \boldsymbol{C}\boldsymbol{H}(t)$ (3b)

and for a non-zero initial state and a non-zero (constant) input:

$\boldsymbol{w}(t) = \boldsymbol{G}(t)\boldsymbol{w}(0) + \boldsymbol{H}(t)x(0)$ (3c)
$y(t) = \boldsymbol{C}\boldsymbol{w}(t)$ (3d)
where the $\boldsymbol{G}(t)$ is the fundamental matrix, i.e.
$\boldsymbol{G}(t) = \mathcal{L}^{-1}\{\boldsymbol{\mathcal{H}}(s)\}$ and (3e)
$\boldsymbol{H}(t) = \mathcal{L}^{-1}\{\boldsymbol{\mathcal{H}}(s)\boldsymbol{B}s^{-1}\}$. (3f)

When system step inputs and system output samples are synchronous and uniform (with a period of $T_s$)

$\boldsymbol{G} = \boldsymbol{G}(t)|_{t=T_s}$ (4a)
which is the state transition matrix and
$\boldsymbol{H} = \boldsymbol{H}(t)|_{t=T_s}$ . (4b)

For an input sequence $x(t)$, which is a contiguous train of rectangular pulses of duration $T_s$, the output sequence $y[n]$ is then simply found using the following discrete-time LSS recursion [2],[26]:

$\boldsymbol{w}[n] = \boldsymbol{G}_{K\times K}\boldsymbol{w}[n-1] + \boldsymbol{H}_{K\times 1}x[n]$ (4c)
$y[n] = \boldsymbol{C}_{1\times K}\boldsymbol{w}[n]$ . (4d)

The $\mathcal{Z}$ transform of this system (for a zero initial-state) is

$$\boldsymbol{W}(z) = z^{-1}\boldsymbol{G}_{K\times K}\boldsymbol{W}(z) + \boldsymbol{H}_{K\times 1}X(z) \tag{4e}$$
$$Y(z) = \boldsymbol{C}_{1\times K}\boldsymbol{W}(z)\ . \tag{4f}$$

This discrete-time LSS system may also be used to model the dynamics of natural processes (at times $t = nT_s$) or digital electronic circuits (realized using a network of unit delays $z^{-1}$). The ($z$-plane) poles of this discrete-time LSS system are equal to the eigenvalues of $\boldsymbol{G}$ and the discrete-time transfer-function of the input to the internal states $\boldsymbol{\mathcal{H}}(z) = \boldsymbol{W}(z)/X(z)$ is found by re-arranging (4e) to yield

$$\boldsymbol{\mathcal{H}}(z) = z\{z\boldsymbol{I}_{K\times K} - \boldsymbol{G}\}^{-1}\boldsymbol{H}\ . \tag{4g}$$

For the input to output, the discrete-time transfer-function $\mathcal{H}(z) = Y(z)/X(z)$, the frequency response, and impulse response, of this system are:

$$\mathcal{H}(z) = \boldsymbol{C}\boldsymbol{\mathcal{H}}(z) \tag{5a}$$
$$\mathcal{H}(\omega) = \mathcal{H}(z)|_{z=e^{i\omega}} \text{ and} \tag{5b}$$
$$h[m] = \mathcal{Z}^{-1}\{\mathcal{H}(z)\}\ . \tag{5c}$$

For composite systems, e.g. signal plus interference (i.e. coloured noise), the scalar inputs and the state vectors of the signal (sig) and interference (int) processes of order $K_{\text{sig}}$ and $K_{\text{int}}$ are stacked to form an augmented system of order

$$K = K_{\text{sig}} + K_{\text{int}} \text{ with} \tag{6a}$$
$$\boldsymbol{x} = \begin{bmatrix} x_{\text{sig}} \\ x_{\text{int}} \end{bmatrix} \text{ and} \tag{6b}$$
$$\boldsymbol{w} = \begin{bmatrix} \boldsymbol{w}_{\text{sig}} \\ \boldsymbol{w}_{\text{int}} \end{bmatrix}. \tag{6c}$$

The Kalman filter considers a slightly modified system with one output and two inputs: process noise $\boldsymbol{x}_Q(t) = [x_{\text{sig}} \quad x_{\text{int}}]^{\text{T}}$ and additive measurement noise $x_R(t)$ i.e.

$$\dot{\boldsymbol{w}}(t) = \boldsymbol{A}_{K\times K}\boldsymbol{w}(t) + \boldsymbol{B}_{K\times 1}\boldsymbol{x}_Q(t) \tag{7a}$$
$$y(t) = \boldsymbol{C}_{1\times K}\boldsymbol{w}(t) + x_R(t)\ . \tag{7b}$$

When all inputs are Gaussian-distributed white-noise sequences with variance $\sigma_{\text{sig}}^2$, $\sigma_{\text{int}}^2$ & $\sigma_R^2$, the Kalman filter is the optimal MMSE estimator of the state vector. For known and constant input noise parameters and an initial state with a reasonable covariance matrix the variable gain of the Kalman filter expedites the convergence of state estimates and reduces the duration of start-up transients. At steady-state, the Kalman filter also has the LSS form of (4c) & (4d), with $\hat{\boldsymbol{w}}_{\text{KF}}$, $\boldsymbol{G}_{\text{KF}}$, $\boldsymbol{H}_{\text{KF}}$ and $\boldsymbol{C}_{\text{KF}}$, where $\boldsymbol{G}_{\text{KF}}$ and $\boldsymbol{H}_{\text{KF}}$ incorporate the steady-state Kalman gain vector $\boldsymbol{K}_{\text{KF}}$, as determined by solving the Riccati equations; $\boldsymbol{C}_{\text{KF}}$ ignores the interference states and applies the desired time shift of $qT_s$ seconds into the future or past. These system matrices are defined as follows:

$$\boldsymbol{H}_{\text{KF}} = \boldsymbol{K}_{\text{KF}} \tag{8a}$$
$$\boldsymbol{G}_{\text{KF}} = \boldsymbol{G} - \boldsymbol{K}_{\text{KF}}\boldsymbol{C}\boldsymbol{G} \tag{8b}$$
$$\boldsymbol{C}_{\text{KF}} = [\boldsymbol{C}_{\text{sig}} \quad \boldsymbol{0}_{1\times K_{\text{int}}}]\boldsymbol{G}(t)|_{t=-qT_s}\ . \tag{8c}$$

This state estimator is configured using the delay parameter $q$ (in units of samples). The Kalman 'filter' is reached using $q = 0$, the fixed-lag Kalman 'smoother' uses $q > 0$, whereas the (phase-lead) 'predictor' uses $q < 0$ [1],[2]. Filter bandwidth (for signal transmission) is determined by $\sigma_{\text{sig}}^2$ and the number of signal-process poles at $s = 0$; whereas the severity of filter notches (for interference suppression) is determined by $\sigma_{\text{int}}^2$. It is suggested in this paper that reasonable state estimators with these properties may also be designed directly in the frequency domain, without solving the Riccati equations, without the assumption of Gaussian noise, and without prior knowledge of noise variance. This parsimonious design procedure may be appropriate when process models are only partially known, or when high-order process models with complex poles impede the solution of the Riccati equations. Candidate design procedures are considered in Section 3, then extended and adapted for the derivative state-estimation problem in Section 4.

## 3 FIR and IIR MaxFlat filters

### *3.1 FIR versus IIR filters*

FIR and IIR filters are complementary because the computational complexity of an FIR filter is determined by the duration of its impulse response, not by the complexity of the underlying process model whereas the computational complexity of an IIR filter is determined by the order of the process model, not by the duration of its impulse response. Thus, assuming there are sufficient degrees of freedom afforded by its internal states (i.e. the order of the FIR or IIR filter): zeros may be placed on (or near) the unit circle to suppress multiple narrowband interferers using a low-order FIR filter [23]; whereas, poles may be placed near the unit circle to enhance multiple narrowband signals using a low-order IIR filter [22]. The impulse response of an FIR filter may be shaped arbitrarily, for perfect symmetry or anti-symmetry, to yield a frequency response with perfect phase linearity; whereas the impulse response of a (causal) IIR filter is restricted to forms that may be generated recursively, i.e. a monomial multiplied by a damped sinusoid, thus temporal symmetry and perfect phase linearity are impossible in causal realizations.

### *3.2 From FIR to IIR filters*

The coefficients of FIR filters are usually derived using an iterative procedure that minimizes maximum errors (i.e. MiniMax) when a sharp (step-like) transition between pass and stop bands is required [27]. The achieved frequency response $\mathcal{H}(\omega)$, is matched to a desired frequency response $D(\omega)$, in a way that minimizes the maximum error $\Delta(\omega)$, where $\Delta(\omega) = |D(\omega) - \mathcal{H}(\omega)|$. Non-iterative least-squares procedures, involving integrals of weighted squared errors [28], optionally with equality constraints on derivatives of the frequency response are a reasonable alternative when wider transition bands are acceptable or desirable [12],[13].

IIR filters are usually designed by minimizing a quadratic squared-error cost, subject to inequality constraints on pole radii for causal stability and optionally on worst-case phase or magnitude errors [29],[30],[31],[32]. Unfortunately, the cost function is a non-linear function of the pole positions, thus bespoke solvers are required, or approximate linearized cost functions are employed [20],[33]. Procedures that use derivative constraints to specify the frequency response of an IIR filter (i.e. MaxFlat designs) lead to a linear system of equations, that also obviates the need for iterative solvers. For all IIR procedures, the passband group-delay is set by modulating the desired frequency response by a complex sinusoid $e^{-iq\omega}$, where $q$ is the delay (in samples). It may be included as a parameter to be optimized by the solver (subject to optional inequality constraints) along with the pole and zero positions [29],[30], manually optimized by trial and error, or simply set to an arbitrary value to satisfy other system latency requirements [31],[33].

The response of an FIR filter (with all poles at the origin) may be expressed as a linear combination of basis functions. Those functions are integer delays in the time domain, for $0 \le m < M$, where $m$ is the delay index and $MT_s$ is the duration of the impulse response (in seconds); and complex sinusoids in the frequency domain $\psi_k(\omega) = e^{-ik\omega}$, where $k = m$ and $i$ is the imaginary unit.

Laguerre filters have repeated real poles in the complex $z$-plane at $p$ (for $0 \le p < 1$) and are conveniently realized via a network of so-called 'leaky' (first-order) integrators [34],[35],[36]. When the outputs of this network are weighted appropriately and summed, this simple structure may be used to recursively implement a projection onto the family of discrete associated Laguerre polynomials that are ortho-normal with respect to a $m^\kappa p^m$ weight [15]. They are perfect for processing low-frequency signals (relative to the sampling rate) that are well represented in the time domain by local low-order Taylor series expansions or polynomials. The response of these IIR filters is a linear combination of: monomials multiplied by a (real) exponential in the time domain; and $\psi_k(\omega) = e^{i\omega}/(e^{i\omega} - p)^k$ terms in the frequency domain. Laguerre filters have both practical and pedagogical value, as they provide an intuitive link between IIR and FIR filters. As $p \to 1$ the duration of each exponentially decaying term dilates (in the time domain) and as $p \to 0$ it contracts; then when $p = 0$ is reached, each term becomes a unit impulse (i.e. infinitesimally narrow) delayed by $k$ samples, for an FIR filter.

The MaxFlat IIR filters developed in this paper are an extension of these repeated-pole expansions. Instead of repeated poles at the origin (as used in FIR filters) or on the positive real axis (as used in Laguerre IIR filters), a phalanx of complex Butterworth poles, with $\psi_k(\omega) = e^{i\omega}/(e^{i\omega} - p_k)$, are used for recursive IIR filters with configurable bandwidth. Placing poles for the desired bandwidth then optimally assigning zeros for the required passband phase response and the suppression of narrowband interference in the stopband, linearizes the IIR design procedure so that simpler linear (FIR-like) design procedures may be used [15],[16].

### *3.3 MaxFlat filters*

Maximally flat (MaxFlat) procedures are a reasonable alternative when wider transition bands are acceptable or desirable. The 'MaxFlat' term is used here to describe filters where all (or nearly all) degrees of freedom are used to satisfy derivative constraints. Constraints at $\omega = 0$ (i.e. dc) are used to ensure that polynomial signals from an integrating process are passed without attenuation for the estimation derivatives in the time domain [9]. Smoothed (i.e. low-pass filtered) estimates are obtained using long tapered windows [10], white-noise gain minimization [11],[12], or coloured-noise gain minimization [13],[15],[16],[23]. Constraints at $\omega = \pi$ (i.e. pi) or at other frequencies in the stopband may also be specified to ensure that narrowband interferers at/near those frequencies are nullified/attenuated [5],[6],[7],[8],[14],[16],[17],[18],[20],[23]; and for these filters, the width of the passband, stopband, or notch, increases with the number of constraints applied at a given frequency (i.e. $\omega = 0$, $\omega = \pi$, or $\omega = \omega_{\text{int}}$, respectively). Narrow transition-bands are desirable in radio-frequency (RF) applications where steady-state performance for stationary periodic signals is the priority. Broad transition bands (in the frequency domain) are preferable in image/video processing [16], target tracking [22], feedback control, and biomedical applications, where the transient response (in the temporal or spatial domain) is also important, i.e. well damped with prolonged ringing due to the Gibbs phenomenon suppressed [13]. Some wavelet families may be interpreted as being FIR filterbanks with different bandwidths and flatness constraints at $\omega = 0$ and $\omega = \pi$ for so-called 'vanishing moments'.

As MaxFlat filters with sufficient derivative constraints at dc are guaranteed to have the properties required for the unbiased estimation of temporal derivatives at steady state, they are ideal for the types of signal analysis tasks considered here. For FIR filters and *non-causal* IIR filters [16], the dc 'flatness' constraints required for unbiased differentiators are trivial and independent of the group delay. More general high-order flatness constraints for *causal* IIR filters with arbitrary group delay and phase linearity in the low-frequency passband are derived and presented here.

### *3.4 Group delay*

When sequentially analysing uniformly sampled signals in online systems, an upper bound is usually placed on processing latency to guarantee robust stability (essential, in closed-loop feedback systems) and to minimize response times (desirable, in open-loop supervisory/surveillance systems). However, when sampling periods ($T_s = 1/F_s$ in seconds) are orders of magnitude less than the dominant time-constant ($\tau_c$ in seconds) of process dynamics (e.g. of the actuator, plant, sensor, target or interferer), group-delay requirements are less restrictive. As the relative bandwidth ($\omega_c = \Omega_c/F_s$ in radians per sample) contracts in the frequency domain, the relative memory ($\tau_c/T_s$ in samples) expands in the time domain. This temporal relaxation allows other response requirements to receive greater attention using the filter group-delay ($q$ in samples) as a free parameter.

FIR filters are usually designed to have a group delay equal to half the impulse-response duration, i.e. $q = (M - 1)/2$, for a symmetric or anti-symmetric impulse response, thus a linear-phase filter with a constant group-delay over the entire frequency domain [6],[7],[9],[11],[13],[28]. Procedures for the design of nonlinear-phase FIR filters with reduced group-delay, i.e. $q < (M - 1)/2$, have been presented [5],[8],[20]; however in such cases, linear-phase FIR filters of reduced order (with $M = \lfloor 2q + 1 \rfloor$) or IIR filters should also be considered. Non-causal linear-phase IIR filters are design using $q = 0$.

In oversampled systems, selection of the desired group-delay over the passband of a causal IIR filter is less straightforward than the FIR case. If there is no reason to favour a low delay over a high delay, then how should the optimal delay of an IIR filter be determined? An appropriate group delay supports magnitude flatness and phase linearity over the passband. It is not possible to realize a digital (FIR and IIR) filter with a satisfactory frequency response when the applied group delay is unreasonably large, small, or negative (for a phase lead in a predictive filter). As a general 'rule-of-thumb' $K/2 < q < K$ is sometimes recommended, where $K$ is the order of the IIR filter [32],[33]. As an alternative to iterative optimization or searching on a discrete grid, a simple (non-iterative) procedure for the determination of the optimal passband group delay for recursive digital smoothers and differentiators of arbitrary order is derived in the section that follows.

## 4 MaxFlat filterbanks with Butterworth poles and an optimal group-delay

### *4.1 Filter Design*

The Laplace transform ($t \rightarrow s$) of an ideal $k_t$th-order differentiator (w.r.t time) is $s^{k_t}$ (for $k_t \geq 0$). Its continuous-time frequency-response is found by evaluating this continuous-time transfer-function along the imaginary axis of the complex $s$-plane by substituting $s = \sigma + i\Omega$ with $\sigma = 0$ yielding

$$D_{k_t}(\Omega) = D_{k_t}(s)\big|_{s=i\Omega} = s^{k_t}|_{s=i\Omega} = (i\Omega)^{k_t} \text{ where} \qquad (9)$$

$D_{k_t}(\Omega)$ is the (desired) continuous-time frequency-response of a $k_t$th-order differentiator
$\Omega$ is the angular frequency ($-\infty \leq \Omega \leq \infty$) in units of radians per second.

The corresponding discrete-time frequency-response is then found by substituting $\Omega = \omega F_s = \omega/T_s$, yielding

$$D_{k_t}(\omega) = D_{k_t}(\Omega)\big|_{\Omega=\omega/T_s} = (i\Omega)^{k_t}|_{\Omega=\omega/T_s} = \left(\frac{i\omega}{T_s}\right)^{k_t} \qquad (10)$$

where
$D_{k_t}(\omega)$ is the (desired) discrete-time frequency-response of a $k_t$th-order differentiator
$\omega$ is the angular frequency ($-\pi \leq \omega \leq \pi$) in units of radians per sample and
$T_s$ is the sampling period in units of seconds per sample.

The discrete-time transfer-function of an all-pass delay of $q$-samples is $z^{-q}$. Its discrete-time frequency-response is found by substituting $z = e^{i\omega}$, yielding

$D(\omega;q)=D_q(z)\big|_{z=e^{i\omega}}=z^{-q}\big|_{z=e^{i\omega}}=e^{-iq\omega}$ where (11)
$q$ is in units of samples.

The discrete-time frequency response of an '*ideal*' all-pass differentiator with a (fractional) delay of $q$ samples is therefore found by multiplying (10) and (11), yielding

$$D_{k_t}(\omega;q)=D(\omega;q)D_{k_t}(\omega)=\;e^{-iq\omega}\left(\frac{i\omega}{T_s}\right)^{k_t}. \quad (12)$$

The $k_\omega$th derivative (w.r.t $\omega$) of this discrete-time frequency-response, evaluated at $\omega=0$ is reached via $k_\omega$ applications of the product rule, yielding

$$\mathcal{D}^{\mathrm{dc}}_{k_\omega,k_t}(q)=\left\{\frac{d^{k_\omega}}{d\omega^{k_\omega}}D_{k_t}(\omega;q)\right\}\Big|_{\omega=0}=\begin{cases}0 & \text{for } k_\omega<k_t\\ i^{k_\omega}(-q)^{k_\omega-k_t}\left(\frac{1}{T_s}\right)^{k_t}\frac{k_\omega!}{(k_\omega-k_t)!} & \text{for } k_\omega\geq k_t\end{cases}. \quad (13)$$

These complex derivatives of the frequency response at dc specify the phase and magnitude requirements of a (causal or non-causal, FIR or IIR) low-pass differentiator. They allow recursive digital differentiators with an arbitrary group delay to be designed in the frequency domain via a Taylor series expansion around dc. The first $K_\omega^{\mathrm{dc}}$ derivatives (w.r.t frequency and evaluated at dc) of the *realized* filter (with $0\leq k_\omega<K_\omega^{\mathrm{dc}}$) are matched to the corresponding derivatives of the *ideal* differentiator ($k_t>0$) or smoother ($k_t=0$). For a filterbank of $K_t$ differentiators (with $0\leq k_t<K_t$) the phase linearity, bandwidth, and the white-noise gain, increase with the number of dc flatness constraints ($K_\omega^{\mathrm{dc}}\geq K_t$).

However, attempting to replicate the response of the ideal differentiator in (12) is not recommended because it does not have the desired properties of a practical discrete-time estimator. On the one hand, non-negligible bandwidth is necessary for non-infinite memory, i.e. to concentrate the response in time; and on the other hand, non-negligible memory is also necessary for (coloured and white) noise attenuation, i.e. to concentrate the response in frequency. Thus, the frequency response of the filter away from the near-dc region should be shaped to meet other design objectives, for instance: the bias versus variance trade-off in target trackers or the temporal/spatial scale of analysis in signal/image processors.

Additional frequency constraints play an important role in this regard, for instance at the Nyquist frequency ($\omega=\pi$) or elsewhere (outside the signal band, at $\omega=\omega_{\mathrm{nb}}$). In addition to setting the bandwidth of the wideband (wb) low-frequency signal process, they may be used to supress broad-band high-frequency noise or cancel narrowband (nb) interference, respectively. In both cases, the desired derivatives are zero and independent of $q$ and $k_t$, i.e. $\mathcal{D}^{\mathrm{pi}}_{k_\omega}=0$ for $0\leq k_\omega<K_\omega^{\mathrm{pi}}$ and $\mathcal{D}^{\mathrm{nb}}_{k_\omega}=0$ for $0\leq k_\omega<K_\omega^{\mathrm{nb}}$. The total number of derivative constraints applied is $K_\omega$ with $K_\omega=K_\omega^{\mathrm{dc}}+2K_\omega^{\mathrm{nb}}+K_\omega^{\mathrm{pi}}$ and for IIR MaxFlat filters, the filter order is equal to the number of constraints ($K=K_\omega$). The wide notch and the narrow transition-band formed using $K_\omega^{\mathrm{pi}}\gg 0$ is convenient for the suppression non-specific high-frequency phenomena that are out of the signal band [22]. Furthermore, given the lack of phase information at the Nyquist frequency (i.e. $F=F_s/2$, $\Omega=\pi F_s$, $f=1/2$ or $\omega=\pi$), the number of constraints thus the degrees of freedom required are halved, relative to the more general narrowband case ($\omega<\pi$).

The transfer-function $\mathcal{H}_{k_t}(z)$, of the recursive realization of the discrete-time differentiator of $k_t$th-order is expressed here as a linear combination of $K$ first-order complex basis-functions $\psi_k(z)$, of complex argument, with complex poles $p_k$

$\mathcal{H}_{k_t}(z)=\sum_{k=0}^{K-1}c_k\psi_k(z)$ where $K=K_\omega$, (14a)
$\psi_k(z)=\frac{z}{z-p_k}$ with (14b)
$\psi_k(\omega)=\psi_k(z)\big|_{z=e^{i\omega}}$ and (14c)
$\psi_k[m]=\mathcal{Z}^{-1}\{\psi_k(z)\}=p_k^m$. (14d)

The filter design process involves the selection of appropriate poles ($p_k$) for the desired bandwidth ($|p_k|<1$ for causal stability); followed by the optimal placement of zeros (by solving for $c_k$) to satisfy the complex (i.e. magnitude and phase) specified in (13) above. Using first-order basis-functions simplifies design mathematics (see the solution that follows) and reduces realization complexity (see Section 4.2). The size of the basis-set, thus the order of the IIR filter ($K$), need not be large if suitable basis-functions poles are chosen. It is *essential* for the selected basis-set to be capable of satisfying the constraints on complex frequency derivatives at dc, for unbiased estimates of temporal derivatives at steady state. This is readily achieved using a low-frequency basis set with poles near $z=1$. It is *desirable* for it to have a bandwidth that is matched to the signal and be capable of reproducing the desired complex frequency response over the passband. A wider bandwidth in the frequency domain, thus a narrower impulse response in the time domain, increases the flexibility of the estimator or its ability to adapt to changes in signal/system parameters or to accommodate modelling errors. A wide bandwidth decreases short-term bias errors, due to modelling errors or unknown process inputs and disturbances, but also increases random errors, due to measurement noise inputs. The bandwidth of FIR and IIR MaxFlat filters is usually set using the $K_\omega^{\mathrm{dc}}/K_\omega^{\mathrm{pi}}$ ratio and for narrow transition bands the orders of flatness are large. In the method proposed here, fewer constraints are needed if the basis-functions are chosen appropriately.

A non-causal Butterworth filter of $2K$th order with the desired cut-off frequency ($\Omega_c = \Omega_{\mathrm{wb}}$ radians per second) is used to determine the poles of the basis-functions for a low-frequency wideband (wb) signal. A spectral factorization of the non-causal continuous-time transfer-function

$$\mathcal{H}(s) = \frac{1}{1+\left(-s^2/\Omega_{\mathrm{wb}}^2\right)^K} \quad (15)$$

is then applied, yielding causal stable and non-causal stable parts (each of $K$th order) using roots $\mathrm{Re}(s_k) < 0$ and $\mathrm{Re}(s_k) > 0$ respectively, where $s_k$ is the $k$th pole (for $0 \le k < 2K$) in the complex $s$-domain. The causal continuous-time part is discretized using the $z_k = e^{T_s s_k}$ mapping of the impulse invariance method, where $z_k$ is the $k$th causal pole (for $0 \le k < K$) in the complex $z$-domain. Alternatively, the bilinear transformation, possibly with frequency warping, may be used.

The discrete-time causal part yields a stable and realizable filter, with a group delay that is determined by the Butterworth order and its bandwidth. This causal discrete-time Butterworth filter has a bandwidth of $\omega_{\mathrm{wb}} = \Omega_{\mathrm{wb}}/F_s$ radians per sample and $K_\omega$ derivative constraints satisfied at $\omega = 0$ (due to the maximal flatness of the Butterworth filter) and optionally at $\omega = \pi$ (due to zeros introduced by the bilinear transform, if applied) for a very effective smoother ($k_t = 0$); however, the group-delay is non-configurable and dc flatness is lost when narrowband constraints are incorporated to suppress interference. Derivatives of arbitrary order ($k_t \ge 0$) may be computed, and delays of arbitrary duration ($q$) applied, by cascading this prefilter with FIR differentiators and IIR equalizers; however, in the method presented here, the filter is configured without increasing the order of the filter using the basis-function expansion in (14). For a causal filter, the poles of the basis-functions $\psi_k(z)$, are set equal to the ($K_\omega$) poles $p_k = z_k$ of the non-causal discrete-time Butterworth filter that are inside the unit circle (i.e. $|p_k| < 1$).

The unknown linear coefficients $c_k$ in (14a) are chosen to ensure that the derivative constraints are satisfied for a $k_t$th-order differentiator with a *specified* group delay of $q$ samples, using the following system of (linear) equations (see Appendix A for computer code):

$\boldsymbol{d} = \boldsymbol{\Psi c}$ where (16)
$\boldsymbol{d}$ is a column vector of length $K$ containing the constraints on the frequency-response derivatives
$\boldsymbol{c}$ is a column vector of length $K$ with elements $c_k$ ($0 \le k < K$) and
$\boldsymbol{\Psi}$ is a square matrix containing the derivatives of the basis-functions evaluated at the constraint frequencies
($\omega = 0$, $\omega = \pm\omega_{\mathrm{nb}}$ and $\omega = \pi$).

For the system of linear equations in (16) above:
$\boldsymbol{d}$ is formed by (vertically) stacking the derivative constraints

$$\boldsymbol{d} = \begin{bmatrix} \boldsymbol{d}^{\mathrm{dc}} \\ \mathbf{0}_{2K_\omega^{\mathrm{nb}}\times 1} \\ \mathbf{0}_{K_\omega^{\mathrm{pi}}\times 1} \end{bmatrix} \text{ where}$$

$\boldsymbol{d}^{\mathrm{dc}}$ is a column vector of length $K_\omega^{\mathrm{dc}}$ containing elements $\mathcal{D}^{\mathrm{dc}}_{k_\omega,k_t}(q)$ for $0 \le k_\omega < K_\omega^{\mathrm{dc}}$ and
$\mathbf{0}_{M\times N}$ is an $M \times N$ matrix of zeros;
$\boldsymbol{\Psi}$ is formed by (vertically) concatenating the column vectors $\boldsymbol{\Psi}_k$ for $0 \le k < K$ with

$$\boldsymbol{\Psi}_k = \begin{bmatrix} \boldsymbol{\Psi}_k^{\mathrm{dc}} \\ \boldsymbol{\Psi}_k^{\mathrm{nb}} \\ \boldsymbol{\Psi}_k^{\mathrm{pi}} \end{bmatrix} \text{ where}$$

$\boldsymbol{\Psi}_k^{\mathrm{dc}}$ is a column vector of length $K_\omega^{\mathrm{dc}}$ containing elements
$\left\{\frac{d^{k_\omega}}{d\omega^{k_\omega}}\psi_k(\omega)\right\}\Big|_{\omega=0}$ for $0 \le k_\omega < K_\omega^{\mathrm{dc}}$,
$\boldsymbol{\Psi}_k^{\mathrm{nb}}$ is a column vector of length $2K_\omega^{\mathrm{nb}}$ containing elements
$\left\{\frac{d^{k_\omega}}{d\omega^{k_\omega}}\psi_k(\omega)\right\}\Big|_{\omega=-\omega_{\mathrm{nb}}}$ for $0 \le k_\omega < K_\omega^{\mathrm{nb}}$ then
$\left\{\frac{d^{k_\omega}}{d\omega^{k_\omega}}\psi_k(\omega)\right\}\Big|_{\omega=+\omega_{\mathrm{nb}}}$ for $0 \le k_\omega < K_\omega^{\mathrm{nb}}$ and
$\boldsymbol{\Psi}_k^{\mathrm{pi}}$ is a column vector of length $K_\omega^{\mathrm{pi}}$ containing elements

$\left\{\frac{d^{k_\omega}}{d\omega^{k_\omega}}\psi_k(\omega)\right\}\Big|_{\omega=\pi}$ for $0 \le k_\omega < K_\omega^{\text{pi}}$.

The elements of $\boldsymbol{\Psi}_k$ at a given constraint frequency ($\omega_d$) are evaluated recursively via the sequential application of the product rule.

$$\text{Let } \Psi_k^{(k_\omega)}(\omega_d) = \left\{\frac{d^{k_\omega}}{d\omega^{k_\omega}}\psi_k(\omega)\right\}\Big|_{\omega=\omega_d}. \tag{17}$$

Thus after $k_\omega$ consecutive derivatives of $\psi_k(\omega)$ w.r.t. $\omega$

$$\Psi_k^{(k_\omega)}(\omega_d) = i^{k_\omega}\sum_{l_\omega=0}^{k_\omega}(-1)^{l_\omega}\alpha_{k_\omega,l_\omega}\psi_k^{l_\omega+1}(\omega_d) \tag{18}$$

where

$$\alpha_{k_\omega,l_\omega} = \begin{cases} 0 & \text{for } l_\omega < 0 \\ 1 & \text{for } l_\omega = 0 \\ !\,l_\omega & \text{for } l_\omega = k_\omega \\ 0 & \text{for } l_\omega > k_\omega \end{cases}$$

with

$$\alpha_{k_\omega,l_\omega} = l_\omega\alpha_{k_\omega-1,l_\omega-1} + (l_\omega+1)\alpha_{k_\omega-1,l_\omega}$$

otherwise.

For a specified (passband) group delay $q$, the system of equations is solved for the coefficients using

$$\boldsymbol{c} = \boldsymbol{\Psi}^{-1}\boldsymbol{d} \tag{19a}$$

then the transfer function of the filter is found using (6a). For a filterbank of differentiators

$$\boldsymbol{C} = \boldsymbol{\Psi}^{-1}\boldsymbol{D} \tag{19b}$$

is instead used where $\boldsymbol{C}$ and $\boldsymbol{D}$ are $K_\omega \times K_t$ matrices formed by (horizontally) packing the respective $\boldsymbol{c}$ and $\boldsymbol{d}$ column vectors, corresponding to the $k_t$th temporal derivative for $0 \le k_t < K_t$, from left to right. This configuration is referred to here as a $K_t$th-order filterbank.

The white-noise gain ($\Sigma_{k_t}$ or $\Sigma$) of a given filter, or the white-noise cross-gain matrix ($\boldsymbol{\Sigma}$) of a filterbank of differentiators, is readily computed using

$$\Sigma = \boldsymbol{c}^\dagger\boldsymbol{S}\boldsymbol{c} \text{ or } \boldsymbol{\Sigma} = \boldsymbol{C}^\dagger\boldsymbol{S}\boldsymbol{C} \text{ where} \tag{20a}$$

$\boldsymbol{S}$ is a $K \times K$ Hermitian matrix with elements

$$S_{k_a,k_b} = \frac{1}{2\pi i}\oint_{|z|=1}\bar{\psi}_{k_a}(z)\psi_{k_b}(z)\frac{dz}{z}, \tag{20b}$$

$$S_{k_a,k_b} = \frac{1}{2\pi}\int_{-\pi}^{+\pi}\psi_{k_a}(\omega)\psi_{k_b}(\omega)\,d\omega \text{ or} \tag{20c}$$

$$S_{k_a,k_b} = \sum_{m=0}^{\infty}\psi_{k_a}[m]\psi_{k_b}[m] \text{ where} \tag{20d}$$

$\psi_k(\omega)$ and $\psi_k[m]$ are (respectively) the frequency response and impulse response of $\psi_k(z)$, $\overline{[\cdot]}$ denotes complex conjugation and $[\cdot]^\dagger$ is the Hermitian transpose operator. Evaluation of the $S$ elements in the $z$, $\omega$ and $m$ domains in (20b), (20c) and (20d) are equivalent due to Parseval's theorem. Evaluation in the $m$-domain is straightforward as the infinite summations converge rapidly for stable basis-functions with non-negligible bandwidth.

When the desired passband group-delay is *unspecified*, the optimal $q$ that minimizes the white-noise gain of a given filter in the filterbank is determined using

$$\boldsymbol{c}(q) = \boldsymbol{\Psi}^{-1}\boldsymbol{d}(q) \tag{21a}$$

$$\Sigma(q) = \boldsymbol{c}^\dagger(q)\boldsymbol{S}\boldsymbol{c}(q) \tag{21b}$$

$$\mathcal{P}(q) = \frac{d}{dq}\Sigma(q) \tag{21c}$$

where $\Sigma(q)$ and $\mathcal{P}(q)$ are polynomials in $q$. Local minima (and maxima) are determined by solving $\mathcal{P}(q) = 0$ for $q$ (see Appendix A for computer code). The optimal passband group-delay ($q_{\text{opt}}$) is then set equal to the real root with the lowest white-noise gain.

For $k_t = 0$ (i.e. a smoother) the $\Sigma(q)$ polynomial of degree $2(K-1)$ and the $\mathcal{P}(q)$ polynomial of degree $2(K-1)-1$ are found using

$$\Sigma(q) = \sum_{k_b=0}^{K_\omega^{\mathrm{dc}}-1} \sum_{k_a=0}^{K_\omega^{\mathrm{dc}}-1} J_{k_a,k_b} \bar{\delta}_{k_a} \delta_{k_b}\, q^{k_q} \text{ and} \quad (22a)$$

$$\mathcal{P}(q) = \sum_{k_b=0}^{K_\omega^{\mathrm{dc}}-1} \sum_{k_a=0}^{K_\omega^{\mathrm{dc}}-1} k_q J_{k_a,k_b} \bar{\delta}_{k_a} \delta_{k_b}\, q^{k_q-1} \quad (22b)$$

where
$k_q = k_a + k_b$
$J_{k_a,k_b}$ are the (complex) elements of the $(K \times K)$ $\boldsymbol{J}$ matrix
$\boldsymbol{J} = \boldsymbol{\Phi}^\dagger \boldsymbol{S} \boldsymbol{\Phi}$ with $\boldsymbol{\Phi} = \boldsymbol{\Psi}^{-1}$ and the factors
$\delta_k = (-i)^k$ are derived from (13) using $k_t = 0$.

Note that the optimal delay is not necessarily the same for all filters in the filterbank. Indeed, it is postulated that when $K_\omega^{\mathrm{dc}} = K_\omega = K_t$, the total uncertainty (as quantified using $|\boldsymbol{\Sigma}|$) is constant and independent of $q$.

### *4.2 Filter analysis and realization*

The filterbank of $K_t$ differentiators designed using (19b) is (at steady state) an unbiased estimator of the $K_t$ lagged derivative states of an integrating process of $K_t$th order (with $K_t$ poles at $s = 0$ and a polynomial impulse response of degree $K_t$). Convergence is readily confirmed using the final-value theorem to evaluate the lag-adjusted error at steady-state, in the absence of additive noise. In the presence of additive noise, that is white but not-necessarily Gaussian, with a variance (i.e. average power) of $\sigma_R^2$, the covariance matrix of the state estimate is $\sigma_R^2 \boldsymbol{\Sigma}$, as defined above in (20a).

When Butterworth poles are not used and the poles of an IIR MaxFlat filterbank have the same radius $|p_k|$, it may be interpreted as a recursive realization of discounted linear regression, with the form of the basis functions determined by the pole angle $\angle p_k$ and the pole multiplicity (e.g. discrete associated Laguerre polynomials for repeated real poles) [15]. The noise variance ($\sigma_R^2$) may then be estimated recursively online, using the residual of the weighted least-squares fit, where the 'memory' of the exponential weight is determined by the common pole radius.

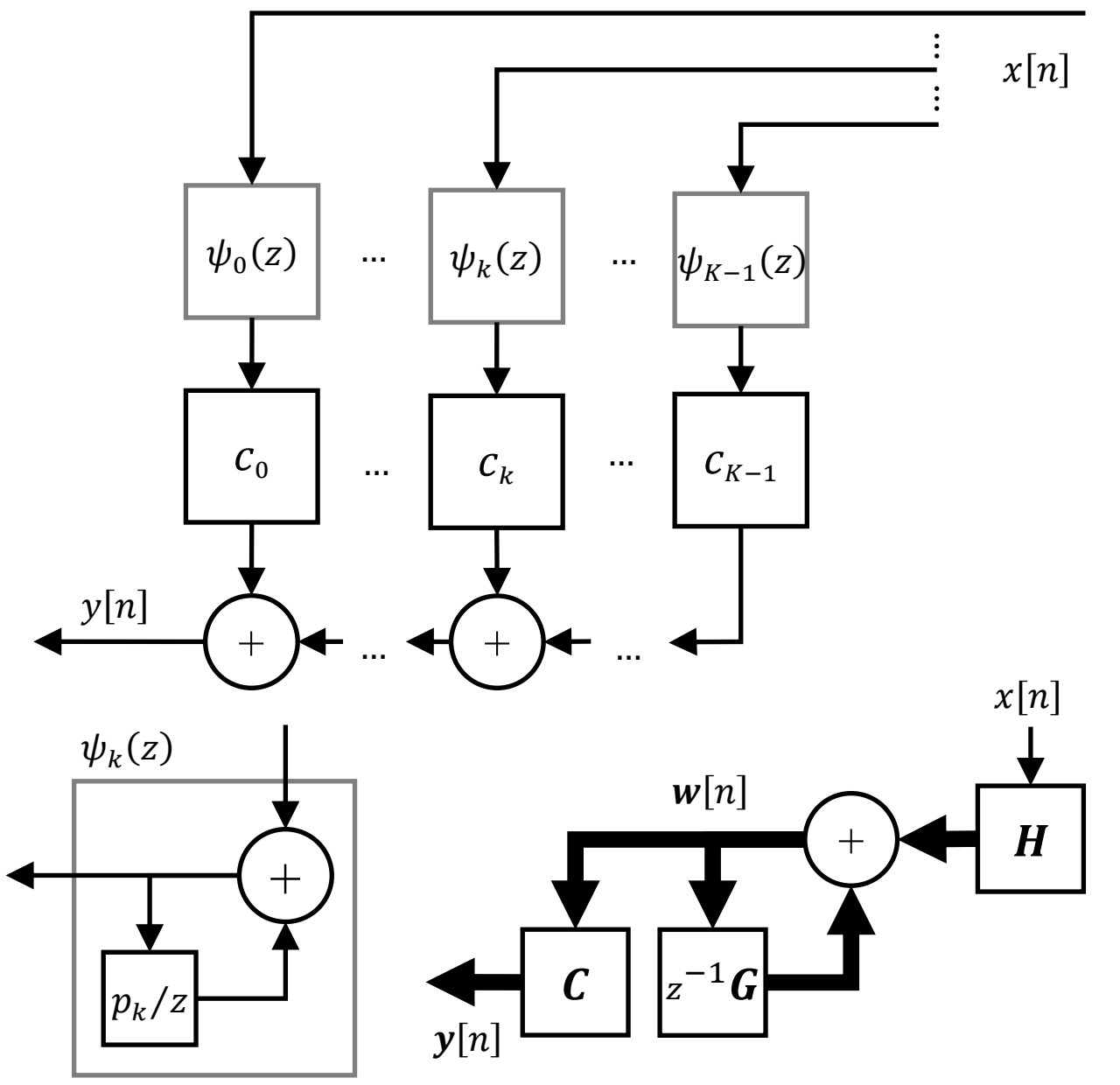

**Figure 1**. Block diagram of $k_t$th-order differentiating filter (top) composed of first-order feedback elements (lower left). Block diagram of $K_t$th-order filterbank in linear state-space form (lower right). Scalar-type and vector-type connections are represented using thin and thick arrows, respectively.

A block diagram for the realization of the transfer function of a single discrete-time filter, as defined in (6), is provided in Figure 1. All filters in a $K_t$th-order filterbank share a common set of poles, which leads to the simplified LSS representation that is also shown in Figure 1.The system poles of this filterbank are encoded in $\boldsymbol{G}$ (a $K \times K$ matrix); whereas, the unique filter zeros, and the common group delay of $q$ samples, are encoded in the rows of $\boldsymbol{C}$ (now a $K_t \times K$ matrix); $\boldsymbol{H}$ (a $K \times 1$ vector) is the input operator and $\boldsymbol{w}$ (a $K \times 1$ vector) holds the internal states of this 'synthetic', discrete-time system. The $k_t$th element of the output $\boldsymbol{y}$ (a $K_t \times 1$ vector) is the $k_t$th-order derivative of the signal component in the sampled waveform input $x$ (a scalar). Note that the discrete-time

LSS system in Figure 1, in accordance with the definition in (4), has the unit delay in the return path of the feedback loop, so that the output of the state estimator at time $T_s n$ includes all measurements up to and including that time, as is the convention in the tracking literature [1],[2]; whereas placing the delay in the forward path of the feedback loop is the convention in the control literature [26], presumably because it results in a discrete-time LSS definition $\boldsymbol{w}[n+1] = \boldsymbol{G}_{K\times K}\boldsymbol{w}[n] + \boldsymbol{H}_{K\times 1}x[n]$ and $z\boldsymbol{W}(z) = \boldsymbol{G}\boldsymbol{W}(z) + \boldsymbol{H}X(z)$ that has the same form as the corresponding continuous-time definition in (1).

The proposed filter structure and solution procedure lead directly to a diagonal canonical form (DCF) with $\boldsymbol{H}$, $\boldsymbol{G}$ and $\boldsymbol{C}$ defined as follows:

$$\boldsymbol{G}_{\mathrm{DCF}} = \begin{bmatrix} p_0 & \cdots & 0 & \cdots & 0 \\ \vdots & \ddots & \vdots & \ddots & \vdots \\ 0 & \cdots & p_k & \cdots & 0 \\ \vdots & \ddots & \vdots & \ddots & \vdots \\ 0 & \cdots & 0 & \cdots & p_{K-1} \end{bmatrix}, \boldsymbol{H}_{\mathrm{DCF}} = \begin{bmatrix} 1 \\ \vdots \\ 1 \\ \vdots \\ 1 \end{bmatrix}$$

$$\boldsymbol{C}_{\mathrm{DCF}} = \begin{bmatrix} \mathcal{C}_{0,0} & \cdots & \mathcal{C}_{k,0} & \cdots & \mathcal{C}_{K-1,0} \\ \vdots & \ddots & \vdots & \ddots & \vdots \\ \mathcal{C}_{0,k_t} & \cdots & \mathcal{C}_{k,k_t} & \cdots & \mathcal{C}_{K-1,k_t} \\ \vdots & \ddots & \vdots & \ddots & \vdots \\ \mathcal{C}_{0,K_t-1} & \cdots & \mathcal{C}_{k,K_t-1} & \cdots & \mathcal{C}_{K-1,K_t-1} \end{bmatrix}. \quad (23)$$

where $\mathcal{C}_{k,k_t}$ coefficients are the elements of the $\boldsymbol{\mathcal{C}}$ matrix in (19b) and $p_k$ are the poles of the basis functions in (14b). This low-complexity form is ideal for filter realization, although complex arithmetic is required. The following alternative (but equivalent) LSS representation is also useful:

$$\boldsymbol{G}_{\mathrm{CCF}} = \begin{bmatrix} -a[1] & -a[2] & \cdots & -a[K-2] & -a[K-1] & -a[K-0] \\ 1 & 0 & \cdots & 0 & 0 & 0 \\ 0 & 1 & \cdots & 0 & 0 & 0 \\ \vdots & \vdots & \ddots & \vdots & \vdots & \vdots \\ 0 & 0 & \cdots & 1 & 0 & 0 \\ 0 & 0 & \cdots & 0 & 1 & 0 \end{bmatrix}$$

$$\boldsymbol{C}_{\mathrm{CCF}} = \begin{bmatrix} b_0[0] & \cdots & b_0[k] & \cdots & b_0[K-1] \\ \vdots & \ddots & \vdots & \ddots & \vdots \\ b_{k_t}[0] & \cdots & b_{k_t}[k] & \cdots & b_{k_t}[K-1] \\ \vdots & \ddots & \vdots & \ddots & \vdots \\ b_{K_t-1}[0] & \cdots & b_{K_t-1}[k] & \cdots & b_{K_t-1}[K-1] \end{bmatrix}, \boldsymbol{H}_{\mathrm{CCF}} = \begin{bmatrix} 1 \\ 0 \\ 0 \\ \vdots \\ 0 \\ 0 \end{bmatrix}. \quad (24)$$

For all equivalent LSS realizations, the poles (i.e. the eigenvalues of $\boldsymbol{G}$) and the zeros are the same. In this so-called ‘controller’ canonical form (CCF), the internal states are simply a series of delay registers with no physical significance. This cascading structure results in a slight increase in complexity; however, it exposes the $b[k]$ and $a[k]$ coefficients of the linear-difference equations for each filter in the filterbank, which may be used to facilitate the generation of frequency responses and the independent realization of the filters using optimized libraries called via a standard interface, e.g. `y = filter(b,a,x)`.

By definition, $Y(z) = \mathcal{H}(z)X(z)$ thus $\mathcal{H}(z) = Y(z)/X(z)$ where $X(z)$ and $Y(z)$ are the $\mathcal{Z}$-transformed input and output sequences, i.e. $X(z) = \mathcal{Z}\{x[n]\}$ and $Y(z) = \mathcal{Z}\{y[n]\}$. $\mathcal{H}(z)$ is a rational function with ($K$th-order) numerator and denominator polynomials $B(z)$ & $A(z)$ and real coefficients $b[k]$ & $a[k]$ for $0 \le k \le K$. By convention $a[0] = 1$ and by definition $b[K] = 0$. The latter restriction follows from the chosen form of the basis functions, which are realized using a delay in the return path, i.e. $\psi_k(z) = 1/(1 - z^{-1}p_k)$, instead of the forward path, i.e. $\psi_k(z) = z^{-1}/(1 - z^{-1}p_k)$. It reduces the degrees of freedom by one (i.e. one less zero to be placed); however, it simplifies the system definitions above because an additional $\boldsymbol{D}$ term (a $K_t \times 1$ vector) is not required in (11b) for a zero-delay connection between the input and the outputs, i.e. $\boldsymbol{y}[n] = \boldsymbol{C}\boldsymbol{w}[n] + \boldsymbol{D}x[n]$.

For the $k_t$th filter, the coefficients $a[k]$ of the polynomial $A(z)$ and the coefficients $b[k]$ of the polynomial $B(z)$, are found by expanding

$$A(z) = \sum_{k=0}^{K} a[k] z^{K-k} = (z - p_0) \ldots \times (z - p_k) \ldots \times (z - p_{K-1}) \quad (25a)$$

and

$$B(z) = \sum_{k=0}^{K} b[k] z^{K-k} = \sum_{k=0}^{K-1} (z - p_0) \ldots \times (z - p_{k-1}) \times c_k z \times (z - p_{k+1}) \ldots \times (z - p_{K-1}) \, . \quad (25b)$$

The $b[k]$ coefficients of all filters and the $a[k]$ coefficients of the filterbank are used to define the $\boldsymbol{C}$ and $\boldsymbol{G}$ system matrices in (4). After dividing $B(z)$ & $A(z)$ by $z^K$, the causal transfer function $H(z)$, frequency response $H(\omega)$, and linear difference equation, of the $k_t$th filter, are derived as follows:

$$\mathcal{H}(z) = \frac{Y(z)}{X(z)} = \frac{\sum_{k=0}^{K} b[k]z^{-k}}{\sum_{k=0}^{K} a[k]z^{-k}} = \frac{\sum_{k=0}^{K-1} b[k]z^{-k}}{1+\sum_{k=1}^{K} a[k]z^{-k}} \text{ and} \quad (26a)$$

$$\mathcal{H}(\omega) = \mathcal{H}(z)|_{z=e^{i\omega}} . \quad (26b)$$

Rearranging (26a) yields

$$Y(z) = X(z)\sum_{k=0}^{K-1} b[k]z^{-k} - Y(z)\sum_{k=1}^{K} a[k]z^{-k} . \quad (27)$$

Then after taking the inverse $\mathcal{Z}$-transform of both sides, the filter may then be realized using the linear difference equation

$$y[n] = \sum_{k=0}^{K-1} b[k]x[n-k] - \sum_{k=1}^{K} a[k]y[n-k] . \quad (28)$$

The derivative state form (DSF) is also useful because the internal states of the linear state-space system defined in (4) have 'physical' significance. In this form, the first $K_t$ elements of $\boldsymbol{w}[n]$ correspond to the first $K_t$ temporal derivatives of the signal and the remaining states are used internally to apply the narrowband and Nyquist frequency nulls. It is reached by applying a coordinate transform to the internal state vector such that $\boldsymbol{C} = \boldsymbol{I}_{K\times K}$. The required transform is found using

$$\mathbb{T} = \begin{bmatrix} \boldsymbol{C}_{\text{DCF}} & \\ \boldsymbol{0}_{(K-K_t)\times K_t} & \boldsymbol{I}_{(K-K_t)\times(K-K_t)} \end{bmatrix}_{K\times K}^{-1} . \quad (29a)$$

In this coordinate system,

$$\boldsymbol{G}_{\text{DSF}} = \mathbb{T}^{-1}\boldsymbol{G}_{\text{DCF}}\mathbb{T},\ \boldsymbol{H}_{\text{DSF}} = \mathbb{T}^{-1}\boldsymbol{H}_{\text{DCF}} \text{ and} \quad (29b)$$

$$\widetilde{\boldsymbol{C}}_{\text{DSF}} = \boldsymbol{C}_{\text{DCF}}\mathbb{T} = \boldsymbol{I}_{K\times K} . \quad (29c)$$

As only the first $K_t$ elements are of $\boldsymbol{w}$ are of interest, only those rows of $\widetilde{\boldsymbol{C}}_{\text{DSF}}$ are retained, i.e. $\boldsymbol{C}_{\text{DSF}} = \boldsymbol{I}_{K_t\times K}$. In this form, the filter is simply initialized using $\boldsymbol{w}[0] = [x[0] \quad \boldsymbol{0}_{1\times(K-1)}]^{\text{T}}$. For the other forms, the initial state must be determined either analytically (via the final value theorem) or numerically (via a loop until convergence). These methods determine the steady-state values for a step input that is held for infinite time, where the magnitude of the step function is equal to the first sample that enters the filter. The internal states at infinite time, determined offline for a unit step, are then scaled by the initial sample when the filter is applied online to a data sequence.

## 5 Discrete-time machines in a continuous-time world

Many natural processes on earth and elsewhere, involving unbound objects obeying Newton's laws of motion, are well modelled as integrating systems, possibly with gradual loss or damping (e.g. due to dissipatory drag or friction forces) and low-frequency oscillation (e.g. due to perturbatory centripetal forces), i.e. they have a pair of dominant poles near the origin of the complex $s$-plane. The motion of bound objects, at celestial, human, or subatomic scales, may also exhibit resonant modes (e.g. due to gravitational, elastic, or electrostatic forces), i.e. they have a conjugate pair of dominant poles near the imaginary axis and far from the real axis of the complex $s$-plane.

Over a timescale that is sufficiently brief, all such (continuous-time LTI) systems are approximately integrating, with $K$ poles at $s \approx 0$ and $K$ (internal) derivative states, generating an output *signal* that is a $K$th-degree polynomial or a $K$th-order Taylor-series expansion over a short time interval, in response to an impulse input. If the observation interval ($T_M$, i.e. the timescale) is uniformly sampled (at a rate of $F_s$) and used to infer or estimate the internal state vector of the system at a specified time, inside or outside the interval, a minimum of $M \geq K + 1$ measurements are required. However, many more samples may be needed to reduce the effects of additive sensor-noise or *interference* to an acceptable level.

As the frequency of sampling ($F_s$ in samples per second) thus number of samples increase ($M = \lfloor T_M/T_S \rfloor$), the relative bandwidth ($\omega_c$ in radians per sample) for a signal process with finite bandwidth ($\Omega_c$ in radians per second) contracts around dc ($\omega_c = \Omega_c/F_s \rightarrow 0$) and the resonant modes of (internal or external) interference processes at higher frequencies (e.g. centred at $F_c$) are properly represented (i.e. not aliased) according to Nyquist's sampling theorem (i.e. $F_c < F_s/2$) with improved resolution and separation of signal and interference bands. In the absence of further information (i.e. models or measurements) the sum of all remaining unresolved phenomena is conveniently modelled as additive uncorrelated (i.e. white) noise, of unknown power and uniform power spectral density, that is not necessarily Gaussian.

Unfortunately, the volume of data acquired in these oversampled digital systems may be problematic when processing is done online using feeble computers in embedded devices that are unable to utilize the scale-efficiency of the fast Fourier transform (FFT) for the low-complexity realization of linear-phase FIR filters. In such systems, the recursive structure of IIR filters, albeit with their non-linear phase response, makes them an attractive alternative.

When (signal and interference) process models are well known, as represented by their rational continuous-time transfer functions $H(s)$, they are readily incorporated into recursive state estimators, with rational discrete-time transfer functions $H(z)$, for instance in a steady-state Kalman filter, if noise is Gaussian with known variance; or in an open-loop Luenberger observer (with no control command input) designed by pole placement [22]. The recursive realization of these IIR estimators, using only delay-multiply-add components, follows directly from their $z$-plane representation and their steady-state error properties are embodied within their frequency responses $H(\omega)$, which are determined by evaluating $\mathcal{H}(z)$ around the unit circle where $|z| = 1$. When models are uncertain, for instance when number and locations of poles and zeros in the complex $s$-plane are unknown and only the centre frequency and bandwidth of a process are approximately known, an alternative approach is suggested here, that focuses on the direct synthesis of the required frequency response for the estimator $H(\omega)$, via a MaxFlat digital filter.

The complex response of a digital filter for the evaluation of the $k_t$th-order derivative of a wideband signal with respect to time is readily specified in the frequency domain. For approximately polynomial signals and sinusoidal interferers, with a duration approaching infinity and a bandwidth approaching zero, the ideal response need only be defined near the dc limit, using a $K_\omega^{\mathrm{dc}}$-th order Taylor-series expansion around $\omega = 0$ and at the frequencies of the (high-power and high-frequency) narrowband interferers, e.g. at $\omega = \omega_{\mathrm{nb}}$ (and/or $\omega = \pi$) where the first $K_\omega^{\mathrm{nb}}$ (and/or $K_\omega^{\mathrm{pi}}$) derivatives are set to zero. The assumed signal and interference bandwidth around these critical frequencies, increases with the order of the local expansion. At all other frequencies, the magnitude response is ideally negligible, to attenuate white noise.

## 6 Detecting pulsed signals

A function (e.g. a transmitted or received signal) cannot be concentrated in both time and frequency [41],[42]. This follows from the mathematical definition of the Fourier transform and it influences the way we interact with our environment via our biological senses and digital sensors. Transmitted signals are therefore adapted in biological and digital systems alike (e.g. whales, bats, sonar and radar) to balance the resolution of time versus frequency according to need and circumstance. It is therefore essential that filters for processing such signals (e.g. detecting and classifying) are jointly parameterizable in time and frequency domains. Indeed, quantum theories and the concept of wave-particle duality, indicate that a constant product of frequency bandwidth and time duration is not simply a sensory phenomenon but also a fundamental property of the natural world.

For the detection of transient pulses, LSS designs are ideal because they are designed around models that capture both transient and steady-state characteristics of signal and interference processes, using a discrete-time transfer-function that is defined over the entire complex $z$-plane. However, if such details are unknown then an approximate frequency-domain model may be more appropriate. A MaxFlat design method may be used to define the process around a circular locus in the $z$-plane only (i.e. the unit circle). Derivative constraints explicitly define the response at a few critical points only (e.g. $\omega = 0$ and $\omega = \pi$); with the response at interpolating frequencies determined by the pole positions and the group delay.

### *6.1 The Teager-Kaiser operator*

A Hilbert transformer and the lesser-known Teager-Kaiser (TK) operator are similar in many respects. Both techniques operate on real waveforms and aim to reproduce the envelope of a sinusoidally modulated signal (e.g. a pulse). As such, they may be used for the detection and classification of transient signals, in biomedical sensors and instruments, for instance. Discrete-time realizations of the Hilbert transform produce a complex output from a real input (i.e. an 'in-phase' term) so that standard (RF) techniques may be applied to the complex (i.e. analytic) signal in applications where an imaginary part (i.e. a quarter-cycle phase-shifted or 'quadrature' term) cannot be provided (e.g. by analogue front-end hardware). The TK operator aims to estimate an 'energy' quantity that is derived from the sum of kinetic- and potential-energy terms of a harmonic oscillator, i.e. its Hamiltonian. Discrete-time realizations of the TK operator require estimates of signal derivatives (w.r.t time) [37],[38],[39],[40]. Digital realizations of both operators require the gain of the filter to be minimized at frequencies expected to be occupied by noise or interference. Gain should also be low where the error of the estimator is known to be large due to the sampling and truncation required for a discrete-time implementation of an ideal continuous-time convolution, e.g. at dc for a Hilbert transformer and near pi for a TK operator. The filter presented in Section 4 is configured and analysed in this context; it is used to smooth a signal and its derivatives for a digital TK operator, in a hypothetical scenario using simulated data.

The continuous-time TK operator uses signal derivatives to produce the energy quantity as follows:

$$E(t) = x^{(1)}(t)x^{(1)}(t) - x^{(0)}(t)x^{(2)}(t) \qquad (30)$$

The three-point formulae are usually used to estimate temporal derivatives above, i.e. using the convolution kernels

$h_0[m] = [0.0 \quad 1.0 \quad 0.0]/T_s^0$
$h_1[m] = [0.5 \quad 0.0 \quad -0.5]/T_s^1$
$h_2[m] = [1.0 \quad -2.0 \quad 1.0]/T_s^2$ in

$y[n]=\sum_{m=0}^{M-1}b[m]x[n-m]$ where
$y[n]=x^{(k_t)}[n]$, $b[m]=h_{k_t}[m]$ and $M=3$. (31)

The following discrete-time TK operator is then obtained:

$$E[n]=(x[n-1]x[n-1]-x[n-2]x[n])/T_s^2\ . \quad (32)$$

This causal realization has a group delay of one sample (with $q=1$). If a one-sample advance is applied, then the following more commonly used non-causal realization is reached (with $q=0$):

$$E[n]=(x[n]x[n]-x[n-1]x[n+1])/T_s^2. \quad (33)$$

As these low-order three-point derivative filters amplify high-frequency noise, a low-pass prefilter is recommended for low-frequency band-limited signals; for instance, an IIR Butterworth pre-filter is used in [38]. Its output $y_0[n]$ is then substituted for $x$ in the above equations for a *two-stage configuration*. It is suggested here that if reasonable models of the signal and interference processes are available, an (FIR or IIR) Wiener filter could also be used to remove interference before the TK operator is applied. The $k_t=0$ filter from a filterbank of (IIR) Butterworth MaxFlat filters could also be used for this purpose (see Section 4). Alternatively, the FIR differentiation stage may be omitted, if the $k_t=0\ldots2$ elements of the IIR filterbank output $\boldsymbol{y}[n]$, i.e. $y_0[n]$, $y_1[n]$ & $y_2[n]$, are used to estimate the temporal derivatives in (30) directly, yielding:

$$E[n]=y_1[n]y_1[n]-y_0[n]y_2[n]\ . \quad (34)$$

The internal derivative states of an (IIR) steady-state Kalman filter, or the outputs of an (FIR) Savitzky-Golay filterbank could also be used in (34). Such (IIR and FIR) smoothing/differentiating filterbanks, without the three-point FIR differentiators, are referred to here as *one-stage configurations*. These one-stage and two-stage filtering alternatives are explored in this section.

### *6.2 Simulation scenarios*

Monte-Carlo (MC) simulations were performed to investigate the behaviour of various derivative filters for the discrete-time TK operator. The TK operator is used to detect a pulsed signal, e.g. a heartbeat in an electrocardiogram, a gun muzzle report in an acoustic geolocation system, or a fault in an electricity distribution network, in the presence of interference. The waveform produced by the transducer is a sum of signal and interference waveforms that were generated by second-order processes ($K=2$) with poles at $\sigma_c\pm\Omega_c i$. The $\sigma_c$ parameter ($\sigma_c<0$) is derived from the coherence duration $\tau_c$ (in seconds) using $\sigma_c=-1/\tau_c$. The $\Omega_c$ parameter is derived from the wave period $\lambda_c$ (in seconds) using $\Omega_c=2\pi/\lambda_c$. The waveform generating process is defined as follows:

$$\mathcal{H}(s)=\frac{b_0}{s^2-2\sigma_c s+\sigma_c^2+\Omega_c^2}$$
$$h(t)=\frac{b_0}{\Omega_c}\exp(\sigma_c t)\sin(\Omega_c t)$$
$$\boldsymbol{A}=\begin{bmatrix}0 & 1\\ -(\sigma_c^2+\Omega_c^2) & 2\sigma_c\end{bmatrix}, \boldsymbol{B}=\begin{bmatrix}0\\1\end{bmatrix}, \boldsymbol{C}=\begin{bmatrix}b_0 & 0\end{bmatrix}$$
$$b_0=\sqrt{-4\sigma_c(\sigma_c^2+\Omega_c^2)}\ . \quad (35)$$

The process is normalized (using $b_0$) so that the power of the impulse response is equal to unity, i.e.

$$\frac{1}{2\pi}\int_{-\infty}^{\infty}P(\Omega)d\Omega=1. \quad (36)$$

For an impulse input, the generating process outputs a pulsed waveform with a mean envelope duration of $\tau_c$ and a modulation frequency of $\Omega_c$.

The sampled signal and interference waveforms were generated by driving the respective processes by a piecewise-constant input formed from a contiguous sequence of rectangular pulses, each with an amplitude that is held over the sampling period, starting at $t=T_s n$ and ending at $t=T_s(n+1)$ for $n=n_0\ldots n_1$. As shown in (3) the output at the sampling times is therefore computed using (4) with

$$\boldsymbol{G}=\begin{bmatrix}\exp(\sigma_c T_s)\left\{\cos(\Omega_c T_s)-\frac{\sigma_c}{\Omega_c}\sin(\Omega_c T_s)\right\} & \frac{1}{\Omega_c}\exp(\sigma_c T_s)\sin(\Omega_c T_s)\\ \frac{-(\sigma_c^2+\Omega_c^2)}{\Omega_c}\exp(\sigma_c T_s)\sin(\Omega_c T_s) & \exp(\sigma_c T_s)\left\{\cos(\Omega_c T_s)+\frac{\sigma_c}{\Omega_c}\sin(\Omega_c T_s)\right\}\end{bmatrix}$$

$$\boldsymbol{H}=\begin{bmatrix}\frac{1}{(\sigma_c^2+\Omega_c^2)}-\frac{1}{(\sigma_c^2+\Omega_c^2)}\exp(\sigma_c T_s)\left\{\cos(\Omega_c T_s)-\frac{\sigma_c}{\Omega_c}\sin(\Omega_c T_s)\right\}\\ \frac{1}{\Omega_c}\exp(\sigma_c T_s)\sin(\Omega_c T_s)\end{bmatrix}. \quad (37)$$

A *deterministic* waveform is generated using a single rectangular input pulse (i.e. with $n_0 = n_1$) and an amplitude of $A_c = \sqrt{P_c/T_s}$. A *non-deterministic* waveform is generated using a sequence of rectangular input pulses (i.e. with $n_0 < n_1$) and an amplitude that is randomly drawn from a Normal distribution with a variance of $P_c/T_s$ and a mean of zero. The dimensionless $P_{\text{sig}}/P_{\text{int}}$ quantity may be interpreted as a signal-to-noise ratio; $P_{\text{sig}} = 1$ was used in all scenarios while $P_{\text{int}}$ was varied between zero and one. For both signal and interference processes, $\tau_c$ was constant and assumed to be known precisely for all MC instantiations. It was set using $\tau_c = \alpha_\tau T_s/f_c$ , with $f_{\text{sig}} = 0.05$ and $f_{\text{int}} = 0.07$, where $f_c$ (i.e. $f_{\text{sig}}$ or $f_{\text{int}}$) is a normalized frequency in cycles per sample ($0.0 \le f_c \le 0.5$, with $f_{\text{sig}} < f_{\text{int}}$). The difficulty of the detection problem increases as the separation between these frequencies decreases. Various values were investigated; however, simple low-pass filters without interference models are sufficient if $f_{\text{sig}}$ is low and $f_{\text{int}}$ is high. The factor of $\alpha_\tau$ is an arbitrary multiplier that determines the bandwidth of the process, the effective duration of the deterministic waveform and the phase coherence of the non-deterministic waveform. Filters with a long impulse response that are highly frequency selective are better for narrowband waveforms generated by a process with a large $\alpha_\tau$ parameter, that have poles closer to the imaginary axis. All results presented in this sections used $\alpha_\tau = 4$. Processes with *known* and *unknown* $\lambda_c$ were considered using $\lambda_c = 1/F_s\tilde{f}_c$ , where $\tilde{f}_c = f_c$ in the known scenario and $\tilde{f}_c$ was randomly drawn from a uniform distribution over the $(0, f_c)$ interval in the unknown scenario. Only the known scenario was considered for the interference process; however, both known and unknown scenarios were considered for the signal process.

The pulse detection problem considered in this section used $F_s = 1$ kHz and a one second batch of data was generated for each MC instantiation ($N = 1000$ samples). Receiver operating characteristic (ROC) curves were generated using 5000 MC instantiations of a signal-plus-interference instance and an interference-only instance. Interference was generated using $n_0 = 0$ and $n_1 = N - 1$. Deterministic signals were generated using $n_0 = n_1 = 400$, $P_{\text{sig}} = 1.0$ & $P_{\text{int}} = 0.1$. Non-deterministic signals were generated using $n_0 = 400$ & $n_1 = 450$ and in this scenario $P_{\text{sig}} = 1.0$ & $P_{\text{int}} = 1.0$ were used to compensate for the increased signal power. Various detection thresholds were applied to the energy quantity produced by the TK operator. For a given threshold: a true detection is declared if the threshold is exceeded over the $n = (400{,}500)$ interval in the signal-plus-interference instance; a false detection is declared if the threshold is exceeded over the $n = (200{,}800)$ interval in the interference-only instance. A large margin is used at the beginning and end of the batch in the latter instance, to avoid start-up transients associated with filter initialization and interference generation.

The ROC curves for the scenario containing a deterministic signal of unknown frequency and non-deterministic interference of known frequency and an input power of $P_{\text{int}} = 0.1$ are shown in Figure 2. This is referred to as the baseline scenario; modified scenarios are considered in Section 6.4. Only the top-left quadrant is shown for all ROC plots. The ROC AUC is presented in Table I for each filter (an ideal detector has unity AUC). Important parameters derived from the realized frequency responses of the $k_t = 0$ filter in each filterbank are also provided in Table I. For two-stage filters (i.e. IIR WF and IIR BW NC), these parameters are for the response of the FIR derivative stage and the IIR smoother stage combined. The frequency responses of all filters and the rationale behind their design (for the baseline scenario) are discussed in the subsection that follows. Two random MC instantiations and the energy computed via the TK operator for the various filters in Figure 2 are shown in Figure 3 and Figure 4 for context. In Figure 3 and Figure 4, $P_{\text{int}}$ has been reduced by a factor of 10 so that the signal is not completely obscured by the interference (upper subplot) and the energy output of FIR NUL NC Detector is not shown so that it does obscure the energy outputs of the other detectors (lower subplot).

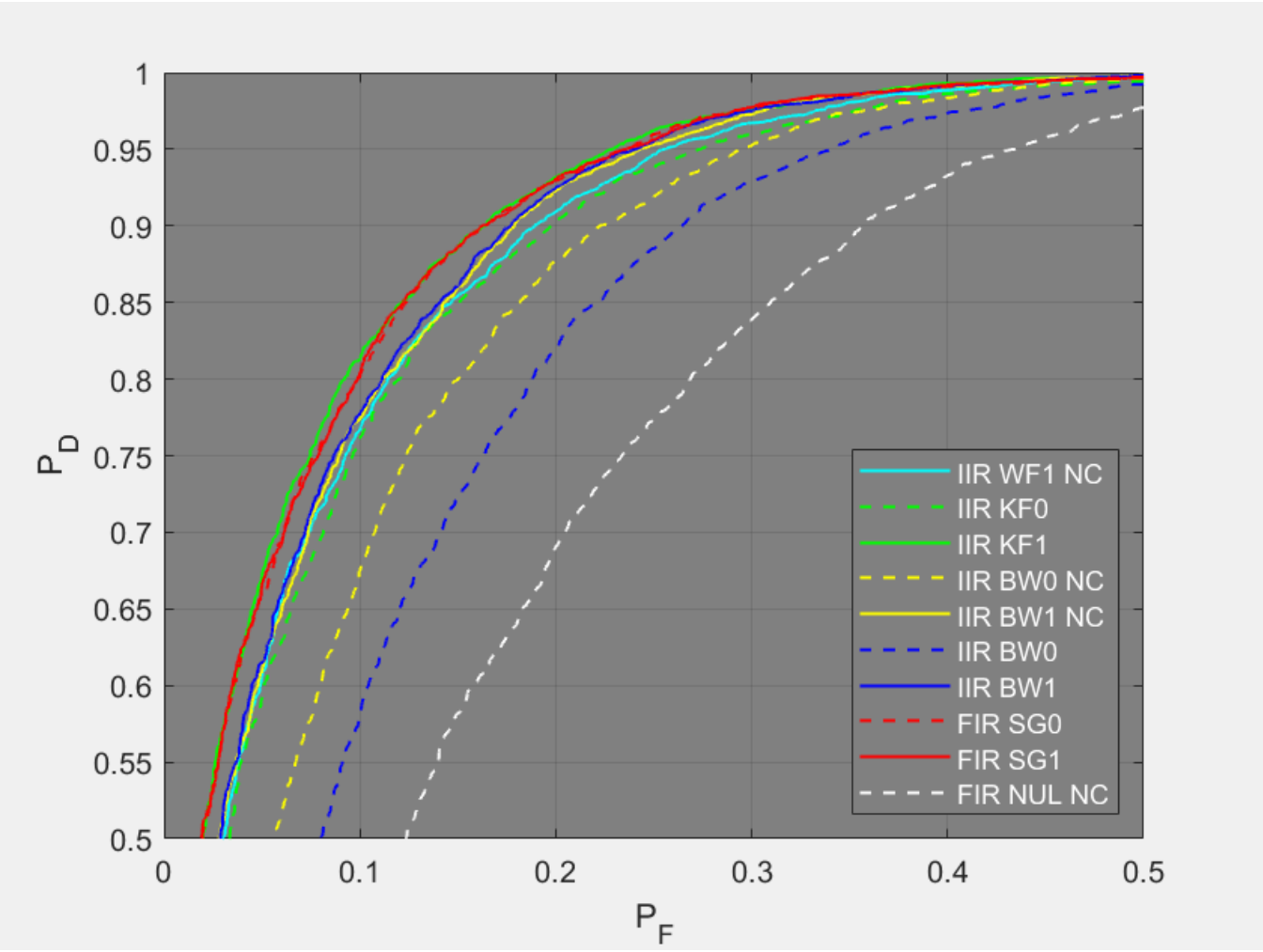

**Figure 2**. ROC of pulse detectors (with default tuning) in the baseline scenario with a deterministic signal of unknown frequency and non-deterministic interference of known frequency.

TABLE I. FILTER COMPARISON

| Filter | $q$ | $\Sigma_0$ | $\lvert\mathcal{H}_0(\omega_{\mathrm{wb}})\rvert$ | $\lvert\mathcal{H}_0(\omega_{\mathrm{nb}})\rvert$ | AUC |
|---|---|---|---|---|---|
| **IIR WF1 NC** | 00.00 | 0.084 | 9.34E-02 | 3.78E-06 | 0.933 |
| **IIR KF0** | 15.22 | 0.044 | 7.16E-02 | 2.61E-02 | 0.929 |
| **IIR KF1** | 17.87 | 0.037 | 2.83E-02 | 1.81E-03 | 0.945 |
| **IIR BW0 NC** | 00.00 | 0.089 | 4.83E-01 | 5.61E-02 | 0.911 |
| **IIR BW1 NC** | 00.00 | 0.075 | 2.43E-01 | 1.00E-14 | 0.935 |
| **IIR BW0** | 13.30 | 0.100 | 6.89E-01 | 1.20E-01 | 0.888 |
| **IIR BW1** | 12.39 | 0.066 | 1.67E-01 | 3.88E-13 | 0.937 |
| **FIR SG0** | 29.50 | 0.050 | 1.35E-02 | 3.15E-03 | 0.944 |
| **FIR SG1** | 29.50 | 0.050 | 1.98E-02 | 9.00E-15 | 0.944 |
| **FIR NUL NC** | 00.00 | 1.000 | 1.00E+00 | 1.00E+00 | 0.840 |

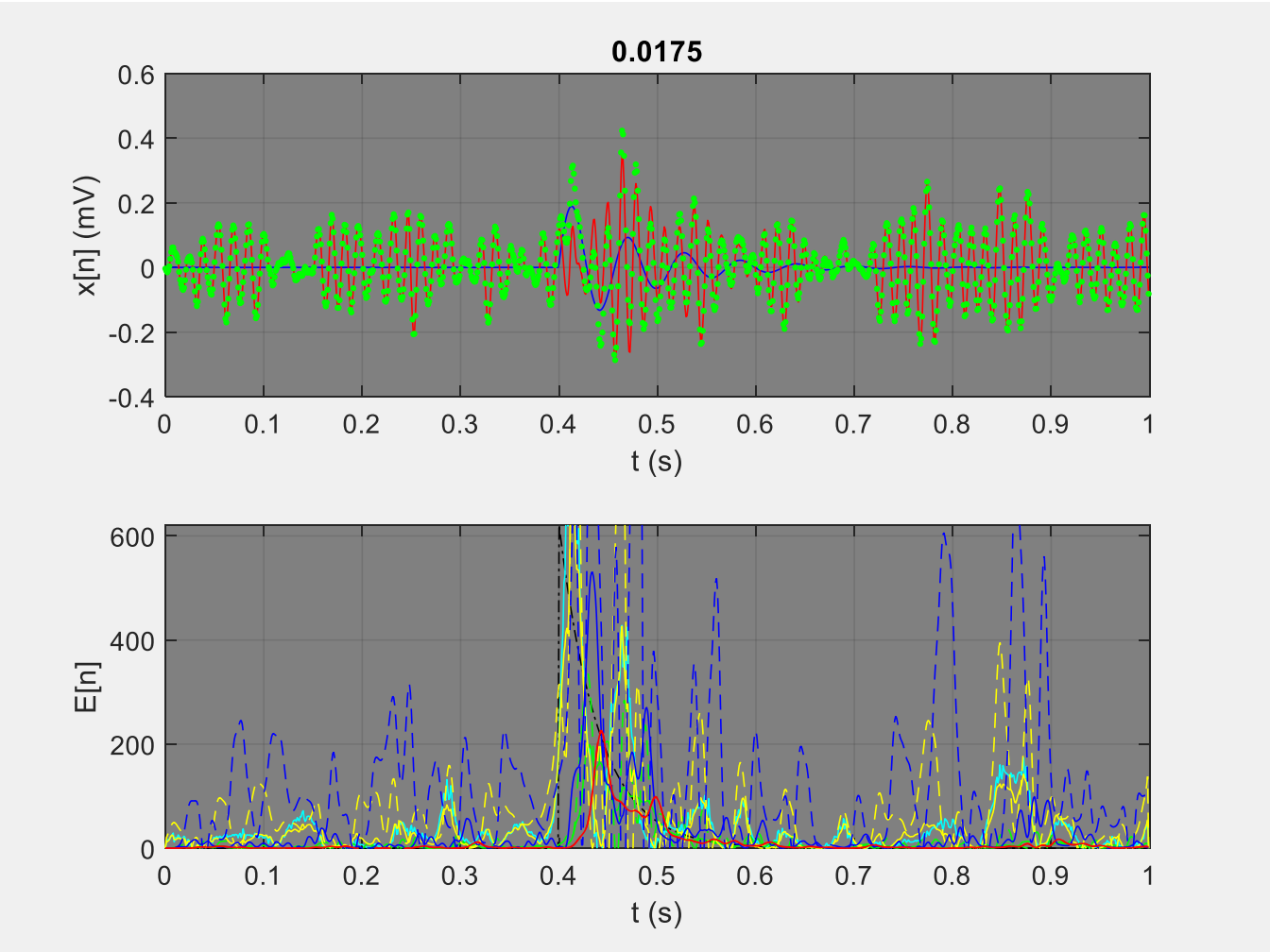

**Figure 3**. MC instantiation of a signal-plus-interference waveform (top); low-frequency signal ($f_{\mathrm{sig}} = 0.0175$) in blue, interference in red, measurements in green. Output of TK operator with various filters (bottom), see Figure 2 for legend, ideal analytic output for signal also shown (black dash-dot line).

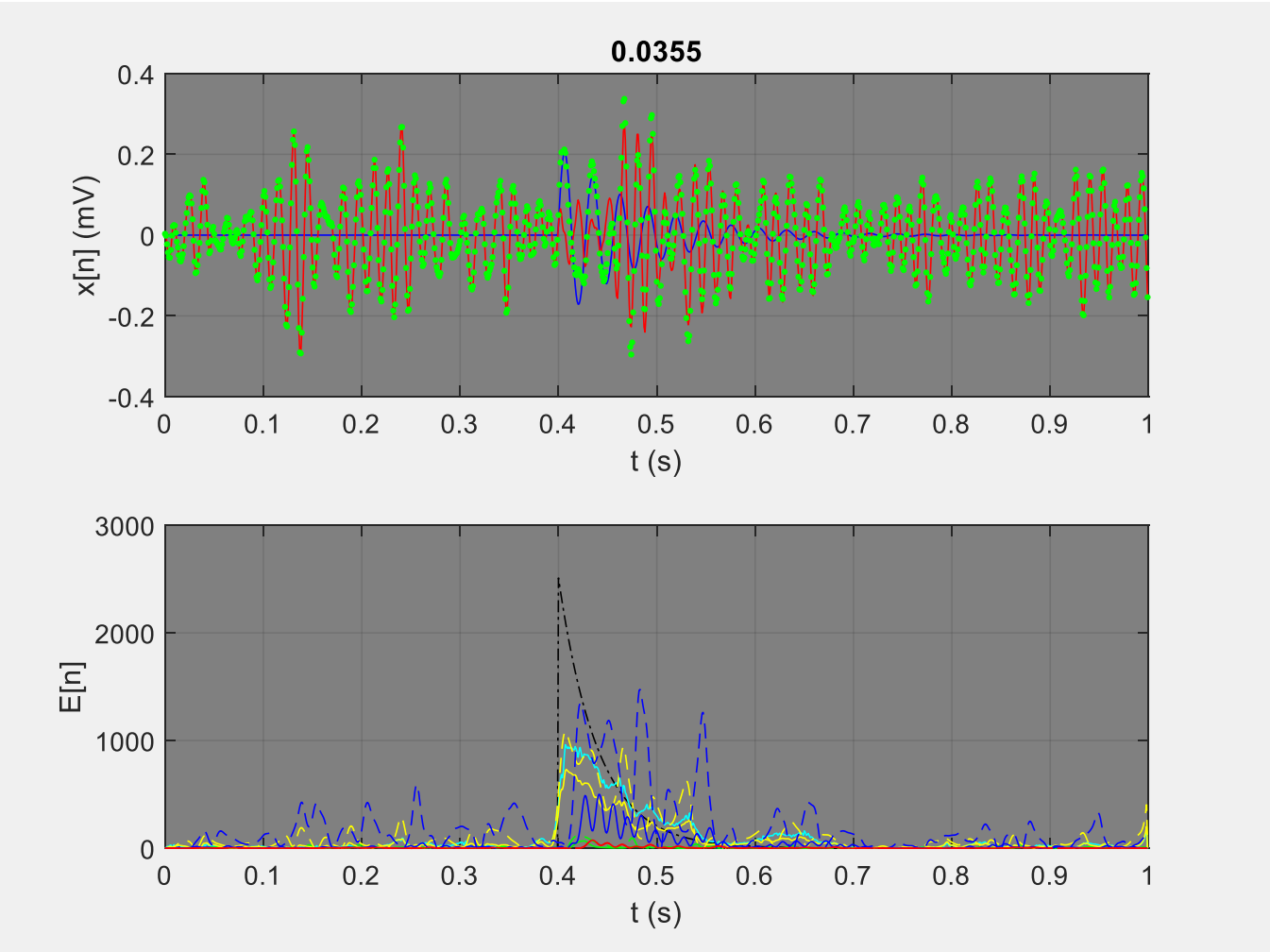

**Figure 4**. MC instantiation with a higher frequency signal ($f_{\mathrm{sig}} = 0.0355$). See Figure 3 for description.

### 6.3 Filter design

The Wiener Filter (WF) with a signal and interference model is an obvious choice for this type of problem [2],[3]. A non-causal realization with an IIR was considered here (designated **IIR WF1 NC**). The non-causal transfer-function of the optimal continuous-time WF is

$$\mathcal{H}(s) = \frac{\mathcal{H}_{\mathrm{sig}}(s)\mathcal{H}_{\mathrm{sig}}(s)}{\mathcal{H}_{\mathrm{sig}}(s)\mathcal{H}_{\mathrm{sig}}(s)+\mathcal{H}_{\mathrm{int}}(s)\mathcal{H}_{\mathrm{int}}(s)} . \quad (38)$$

The bilinear transform is then used to map this non-causal continuous-time transfer-function from the $s$-domain to the $z$-domain and the resulting non-causal discrete-time transfer-function is factored into a sum of forward and backward terms $\mathcal{H}(z) = \mathcal{H}_{\mathrm{fwd}}(z) + \mathcal{H}_{\mathrm{bwd}}(z)$, with poles inside and outside the unit circle, respectively (see Appendix B for details). As $K_{\mathrm{sig}} = 2$ and $K_{\mathrm{int}} = 2$ for the causal process, $K = 4$ for the forward and backward terms. Spectral factorization, for a *product* of forward and background terms, is generally preferred in the literature because it is a much simpler procedure that does not require the tedious algebraic manipulations of polynomials in partial-fraction expansions; however, the *sum* of forward and background terms is used for all non-casual realizations in this paper, because the forward and backward recursions are de-coupled which: firstly, allows them to be done in parallel; and secondly, there is no interaction via the initial state during the start-up transient. The magnitude response of both WF configurations has a relatively large gain at high frequencies which amplifies white noise. High frequency attenuation is greatly improved by adding a small regularization term (the white-noise power) to the denominator of (38); although, this was not necessary in the simulations considered here.

Two methods of tuning the WF were considered: one for the known signal scenario, the other for the unknown signal scenario. In the former scenario, perfectly matched signal and interference models could be used e.g. $\tau_{\mathrm{sig}} = 0.08$ s, $\lambda_{\mathrm{sig}} = 0.02$ s (50 Hz); $\tau_{\mathrm{int}} = 0.0571$ s, $\lambda_{\mathrm{int}} = 0.0143$ s (70 Hz). The magnitude and impulse responses of these assumed process models are shown in Figure 5. The tuning for the known signal scenario results in a magnitude peak at the signal frequency and a notch at the interference frequency (see Figure 6). In the unknown signal scenario, the centre frequency of the signal is set equal to the expected value of the signal frequency, using $\lambda_{\mathrm{sig}} = 0.04$ s (25 Hz, i.e. the midpoint of the uniform distribution over 0 to 50 Hz) and the bandwidth of the signal process is also widened by moving its poles away from the imaginary axis, using $\tau_{\mathrm{sig}} = 0.04$ s. These adjustments were made to better handle the unknown signal scenario (see Figure 2) and they result in a flattening of the frequency response so that it resembles a standard low-pass filter with a notch at the interferer frequency (see Figure 7). The non-causal WF is designed by considering a time-symmetric non-causal processes with four poles at $\pm\sigma_c \pm \Omega_c i$. The processes were normalized so that (36) is satisfied.

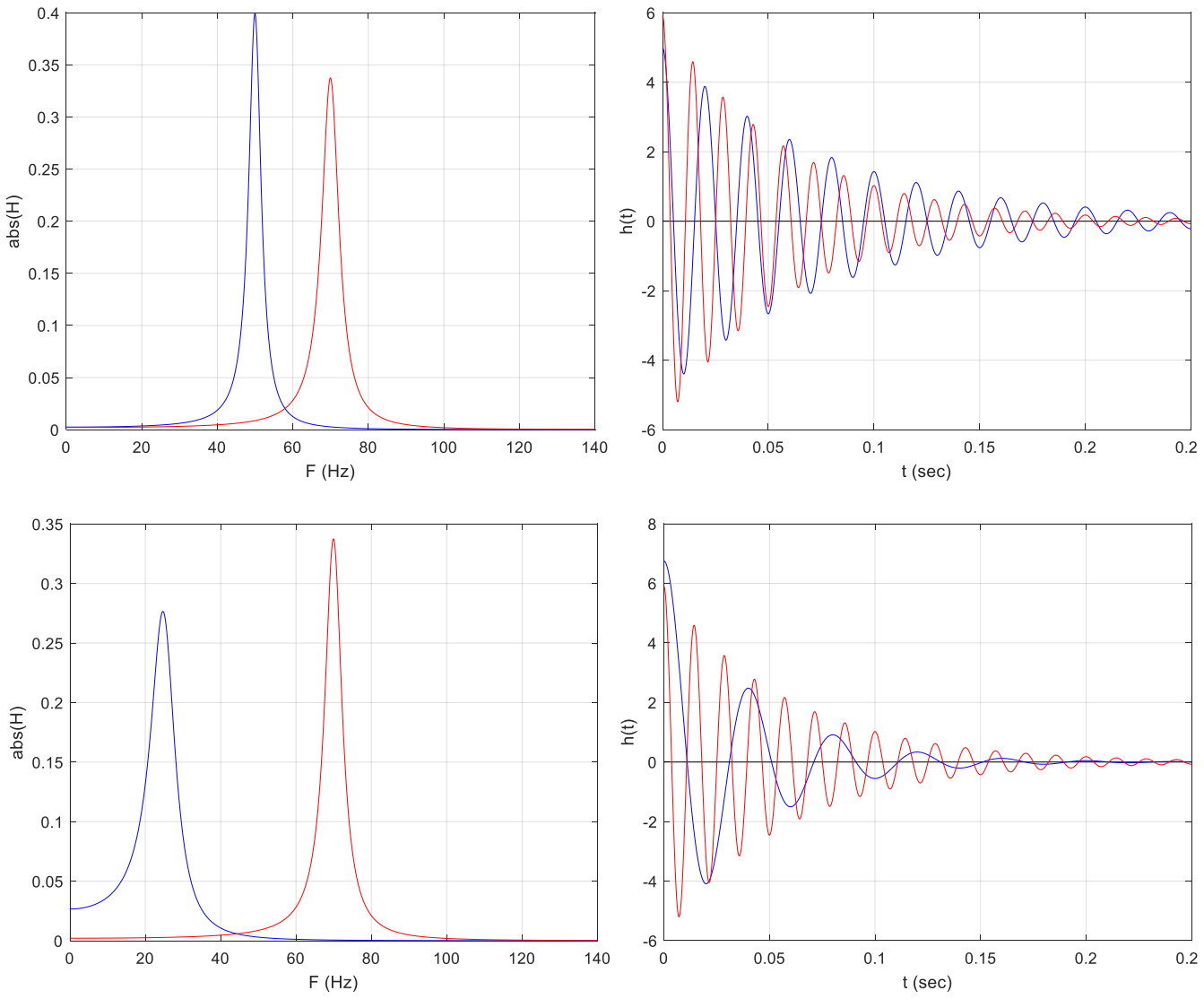

**Figure 5**. Magnitude response (left) and casual (Dirac delta) impulse response (right) of the process models used in the Wiener filter for known (top) and unknown (bottom) signal processes. Signal process in blue. Interference process in red.

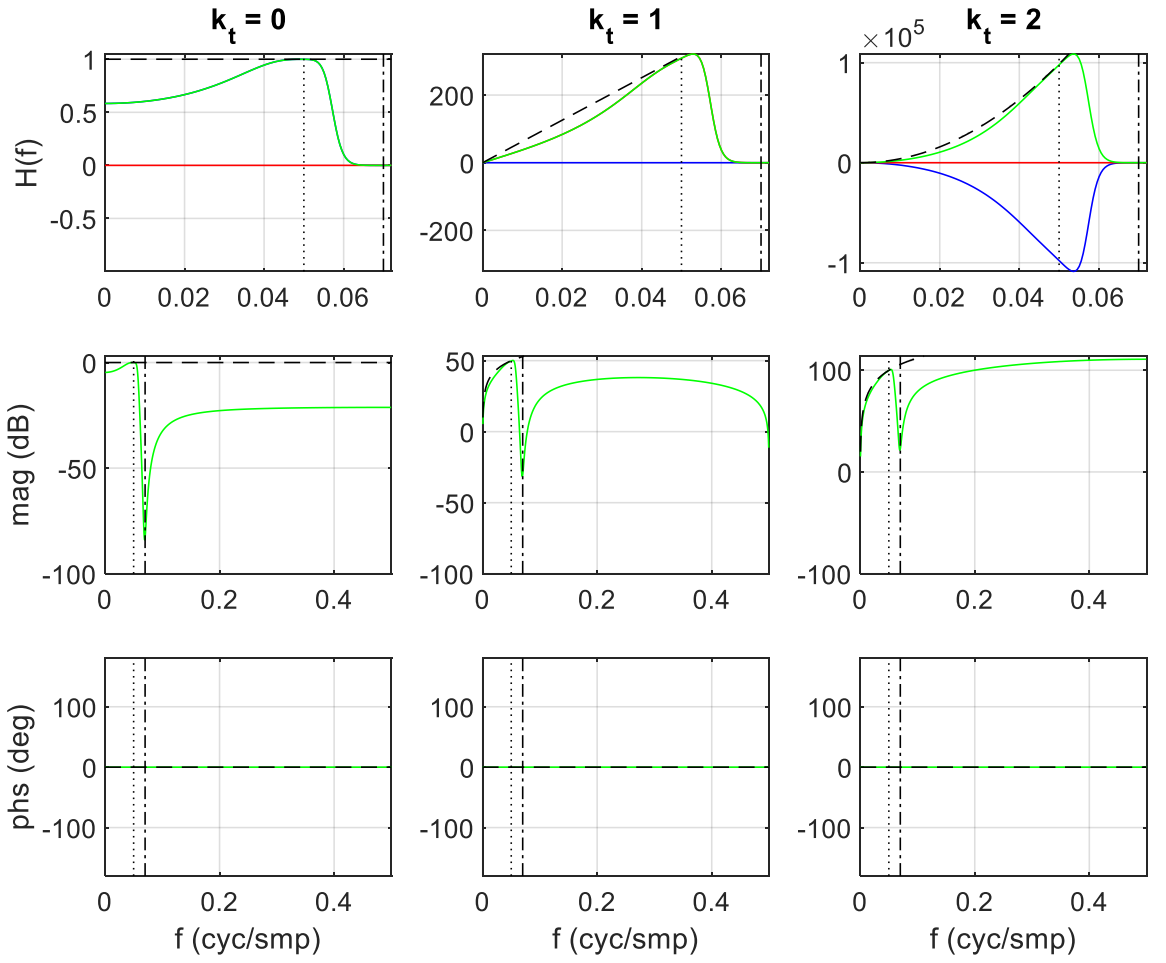

**Figure 6**. Frequency response of two-stage IIR WF1 NC filterbank for $k_t = 0$ to $k_t = 2$ (left to right) tuned for known signal at $f_{\mathrm{sig}}$. Real part (blue) imaginary part (red) and absolute value of complex response in the near-dc region shown in top row. Full-band magnitude and phase responses in the middle and bottom rows, respectively. Vertical black lines at $f_{\mathrm{sig}}$ and $f_{\mathrm{int}}$ (dotted and dash dotted, respectively). Dashed black lines show magnitude and phase response of an all-pass differentiator of $k_t$th order with a group delay of $q$ that is matched to the filterbank.

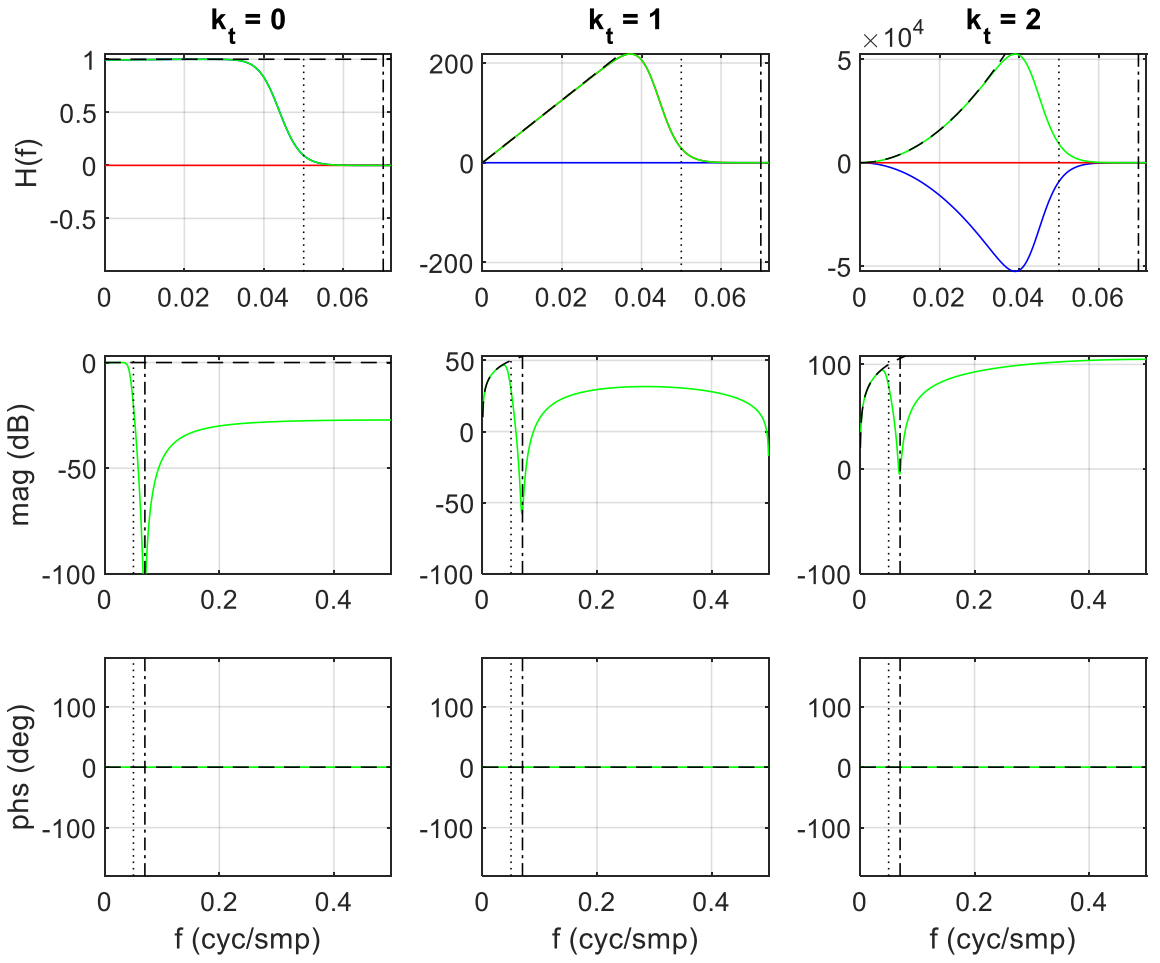

**Figure 7**. Frequency response of two-stage IIR WF1 NC filterbank tuned for unknown signal with random centre frequency on the $(0, f_{\mathrm{sig}})$ interval. See Figure 6 for description and legend.

When the parameters of the process models used in the WF are adjusted arbitrarily to achieve the desired frequency response, it is tempting to abandon the process models and to adopt a simpler or more flexible filter design procedure that yields the desired frequency response directly. The Kalman filter is certainly flexible and it has more than enough degrees of freedom to ensure that a filter with a satisfactory response is eventually reached. It is presented here as an intermediary solution. It may be designed exactly using prior signal and interference process models like the Wiener filter (along with additional information regarding second-order statistical moments if available), or it may be designed approximately using a low-order Taylor-series expansion of the essentially unknown signal and interference processes like the proposed MaxFlat design procedure. Thus, the internal states of a steady-state Kalman Filter (KF) with an IIR were also used to estimate the derivatives for the TK operator. Two variants were considered: without (KF0) and with (KF1) an interference process model. These filters are designated **IIR KF0** and **IIR KF1**, respectively. Both variants employ a third-order integrator ($K = K_{\mathrm{sig}} = 3$, i.e. constant 'acceleration') to model the unknown signal process, using

$$\mathcal{H}_{\mathrm{sig}}(s) = \frac{1}{s^3}$$
$$\boldsymbol{A}_{\mathrm{sig}} = \begin{bmatrix} 0 & 1 & 0 \\ 0 & 0 & 1 \\ 0 & 0 & 0 \end{bmatrix}, \boldsymbol{B}_{\mathrm{sig}} = \begin{bmatrix} 0 \\ 0 \\ 1 \end{bmatrix}, \boldsymbol{C}_{\mathrm{sig}} = [1 \quad 0 \quad 0]$$
$$\boldsymbol{G}_{\mathrm{sig}} = \begin{bmatrix} 1 & T_s & T_s^2/2 \\ 0 & 1 & T_s \\ 0 & 0 & 1 \end{bmatrix} \text{ and } \boldsymbol{H}_{\mathrm{sig}} = \begin{bmatrix} T_s^3/6 \\ T_s^2/2 \\ 1 \end{bmatrix}. \qquad (39)$$

The steady-state gain vector of the filter is computed using a measurement-noise variance $\boldsymbol{R} = \sigma_R^2$ and process-noise covariance matrix $\boldsymbol{Q}_{\mathrm{sig}} = \sigma_{\mathrm{sig}}^2 \boldsymbol{H}_{\mathrm{sig}} \boldsymbol{H}_{\mathrm{sig}}^{\mathrm{T}}$. The latter definition assumes that the process noise input ($x_{\mathrm{sig}}$) is held over the sampling interval. The integrating model ensures that the frequency response has the required derivatives at dc so that it responds to signals at the lower end of the frequency range; however, the KF must be manually tuned by trial and error in this application so it has sufficient gain for signals at the higher end of the frequency range and negligible gain at the centre frequency of the interferer. The KF signal bandwidth at steady state is proportional to the $\sigma_{\mathrm{sig}}^2/\sigma_R^2$ ratio. The interference model used in the KF1 variant is matched to the second-order generating model used in (35) and in this configuration, the severity (width and depth) of the interference notch (on a dB scale) increases with the $\sigma_{\mathrm{int}}^2/\sigma_R^2$ ratio however a perfect null is only possible if $\tau_{\mathrm{int}} = \infty$ is used. Thus $\sigma_R^2$ was fixed at unity for both KF variants. For the KF0 variant, $\sigma_{\mathrm{sig}}^2$ was varied until a reasonable signal bandwidth was attained (see Figure 8). For the KF1 variant, $\sigma_{\mathrm{int}}^2$ (with $\boldsymbol{Q}_{\mathrm{int}} = \sigma_{\mathrm{int}}^2 \boldsymbol{H}_{\mathrm{int}} \boldsymbol{H}_{\mathrm{int}}^{\mathrm{T}}$) was then varied for adequate interference suppression. Values of $\sigma_{\mathrm{sig}}^2 = 5.0 \times 10^{12}$ and $\sigma_{\mathrm{int}}^2 = 5.0 \times 10^{10}$ were found to give good results and were used for all results reported here (see Figure 9).

For both KF variants described above, determining the steady-state gain analytically (using the method described in [2]) and numerically (in a discrete-time iteration until convergence) produced nearly identical results, most of the time. However, the numerical method was generally more likely to produce a satisfactory gain vector, with convergence achieved after approximately 118 iterations for the KF1 variant. In an alternative KF1 tuning with $K_{\mathrm{sig}} = 6$, for a flatter passband and a narrower transition band, with process noise parameters of $\sigma_{\mathrm{sig}}^2 = 5.0 \times 10^{32}$ and $\sigma_{\mathrm{int}}^2 = 5.0 \times 10^{16}$ gave a satisfactory frequency response and an improved AUC of 0.953. For this tuning, the analytic steady-state gain computation did not yield a stable filter; however, the numeric computation converged after 792 iterations. The process-noise parameters for this alternative KF1 tuning are very large and much greater than the values used in the baseline KF1 tuning. Difficulties associated with Riccati equation solution and the fact that these process-noise parameters have no simple physical interpretation conspire to make the Kalman filter very difficult to use in this application. As the model orders were increased, it became increasingly difficult to find combinations of process-noise parameters that promoted Riccati equation convergence.

The optimal fixed-lag for the Kalman filters (i.e. the passband group-delay, $q_{\mathrm{opt}}$) was determined by adapting the procedure presented for the MaxFlat filter in (20) & (21), using

$$\boldsymbol{\Sigma} = \boldsymbol{G}\boldsymbol{S}\boldsymbol{G}^{\mathrm{T}} \text{ with} \tag{40}$$

$$S_{k_a,k_b} = \frac{1}{2\pi i}\oint_{|z|=1} \bar{\mathcal{H}}_{k_a}(z)\mathcal{H}_{k_b}(z)\frac{dz}{z} \text{ where}$$

$\mathcal{H}_{k_t}(z)$ is the $k_t$the element of of $\boldsymbol{\mathcal{H}}(z)$ in (4g) with

$\boldsymbol{G} = \boldsymbol{G}_{\mathrm{KF}}$ and $\boldsymbol{H} = \boldsymbol{H}_{\mathrm{KF}}$

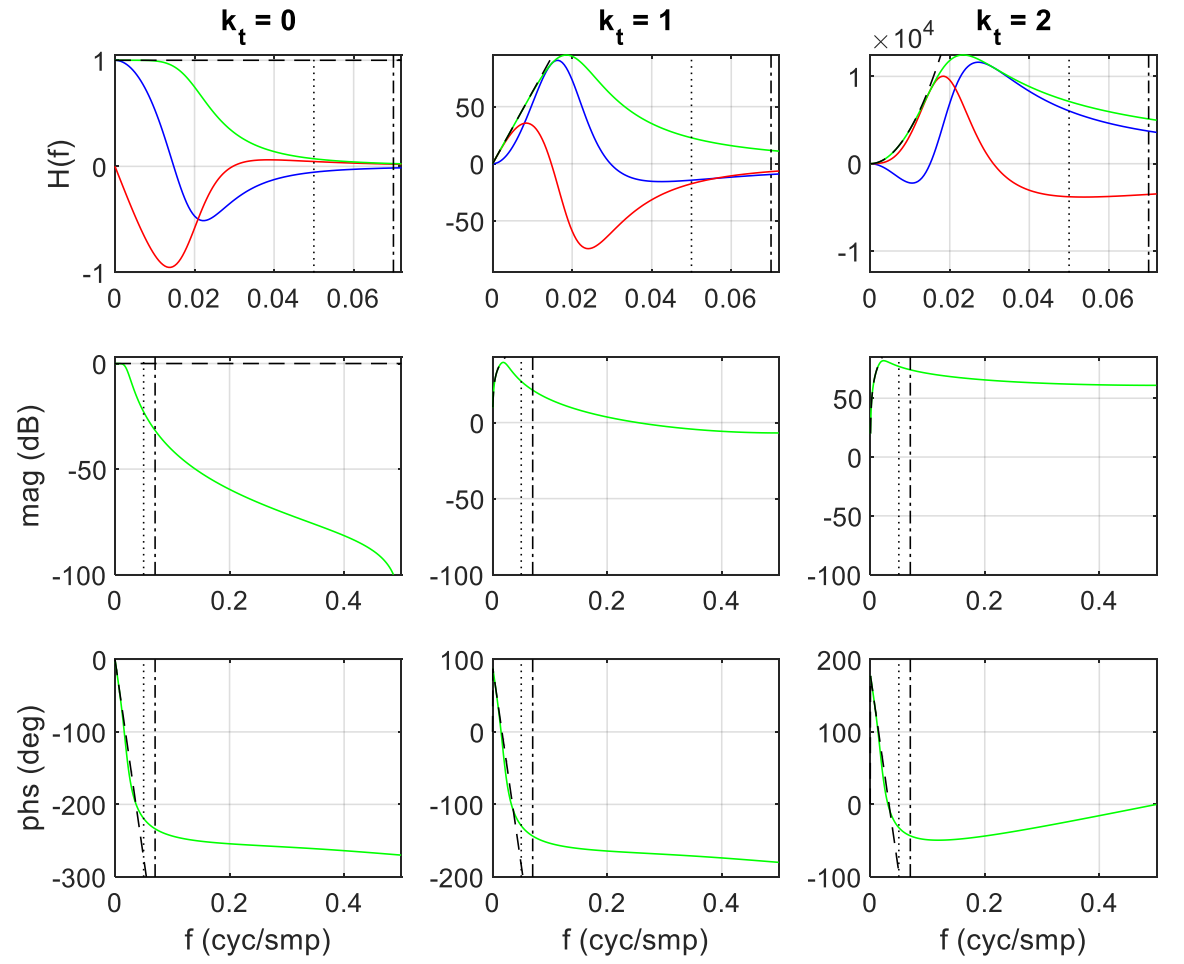

**Figure 8**. Frequency response of one-stage IIR KF0 filterbank for unknown signal. See Figure 6 for description and legend.

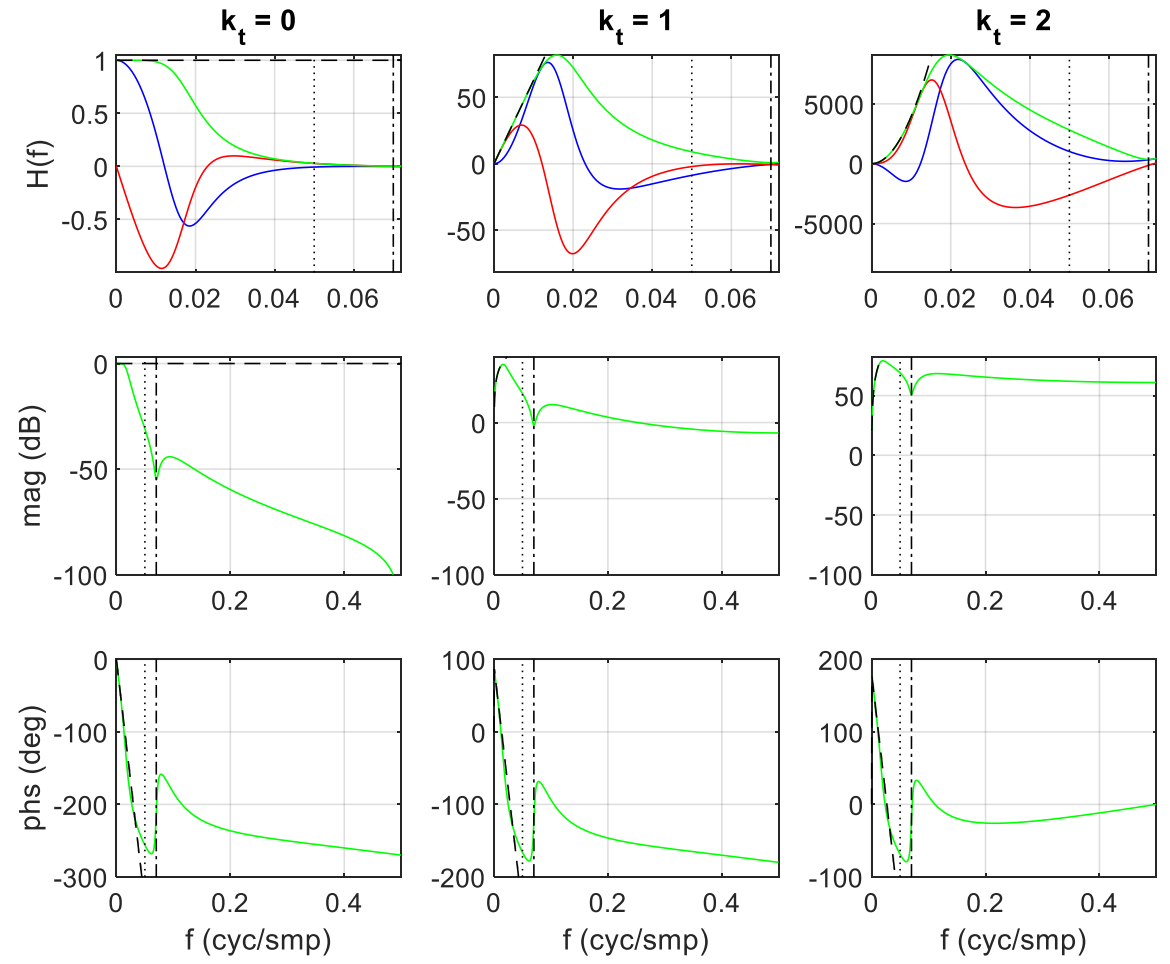

**Figure 9**. Frequency response of one-stage IIR KF1 filterbank for unknown signal. See Figure 6 for description and legend.

The Kalman filter offers the best of both worlds: it allows prior knowledge of process models to be exploited if available or it may be tuned heuristically so that is has the desired frequency response. If the process models are largely unknown and the desired frequency response is known (either intuitively or empirically) then a non-causal IIR filter is an attractive alternative. For instance, a non-causal discrete-time Butterworth (BW) filter satisfies very demanding requirements at a very low computational cost. This IIR filter (designated **IIR BW0 NC**) is maximally flat in the low-frequency signal band, has low gain at the passband edge at the interferer frequency and negligible gain elsewhere (see Figure 10). When the bilinear transform is used to discretize the continuous-time prototype of order $2K$, maximal flatness at dc is preserved ($K_\omega^{\mathrm{dc}} = 2K$) and $2K$ zeros are placed at $z = -1$ for excellent high-frequency attenuation. The ROC and AUC of a Butterworth filter with $K = 4$ in the forward and backward directions and a cut-off frequency of $\Omega_{\mathrm{wb}} = \Omega_{\mathrm{sig}}$ are slightly worse than the ROC and AUC of the Wiener filter (see Figure 2 and Table I).

If the frequency response is modified by placing two zeros on the unit circle where $\angle z = \pm\omega_{\mathrm{int}}$ (see Figure 11), its ROC and AUC are significantly improved and slightly better than the Wiener filter. This is done using (16) with $K_\omega^{\mathrm{dc}} = 4$ and $K_\omega^{\mathrm{nb}} = 2$; however, *all* Butterworth poles (inside and outside the unit circle) are used to form the $2K$ basis functions in (14a). The resulting non-realizable non-causal discrete-time transfer function is factored into forward and backward parts each with $K = 4$ (see Appendix B for details). The resulting realizable non-causal filter is designated **IIR BW1 NC**.

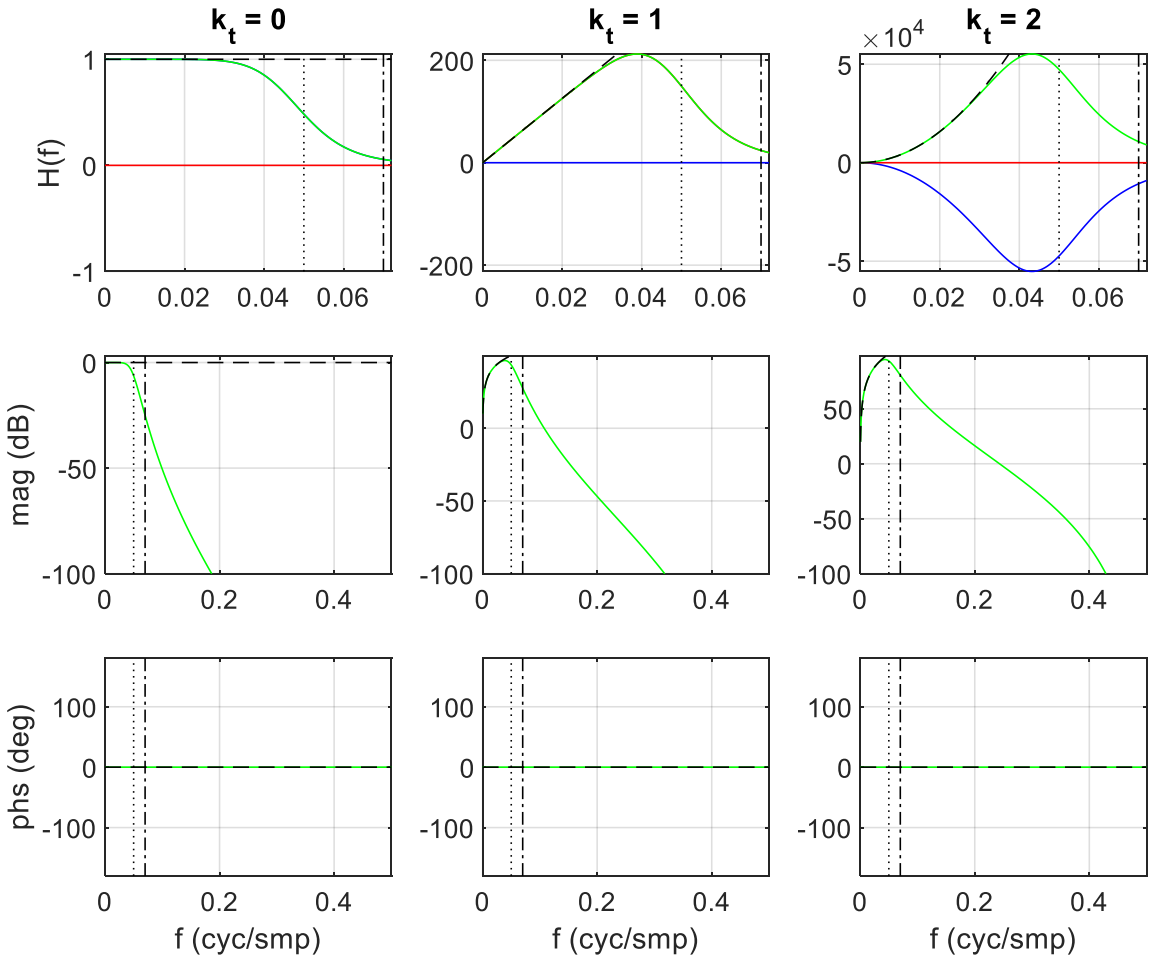

**Figure 10**. Frequency response of two-stage IIR BW0 NC filterbank for unknown signal. See Figure 6 for description and legend.

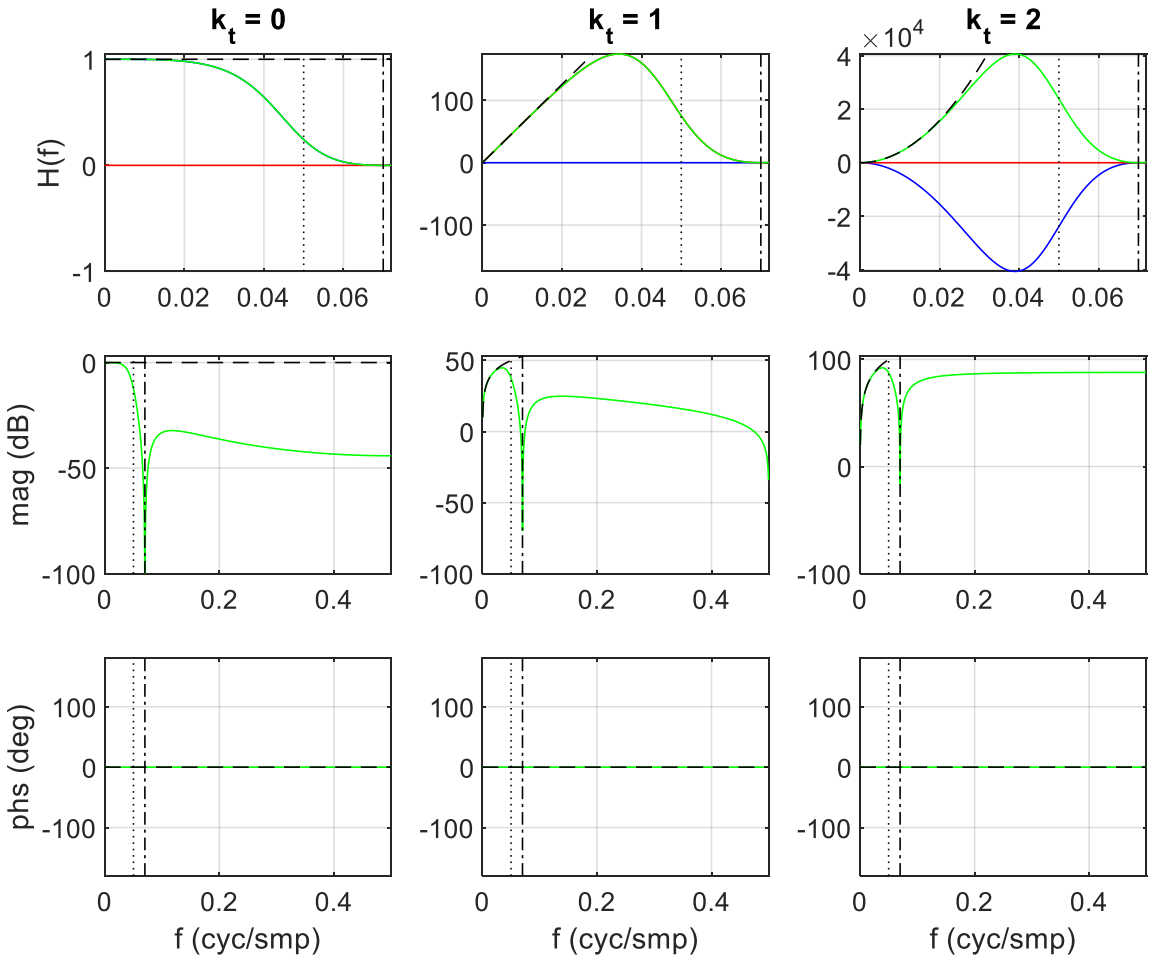

**Figure 11**. Frequency response of two-stage IIR BW1 NC filterbank for unknown signal. See Figure 6 for description and legend.

Unfortunately, non-causal filtering is an impractical luxury in embedded devices for online data-processing in real time. The non-causal Wiener and Butterworth filters are however useful reference points for more practical causal filters. A simple alternative was therefore synthesized by designing a non-casual continuous-time Butterworth prototype ($2K = 12$, $\Omega_{\mathrm{wb}} = \Omega_{\mathrm{sig}}$), performing a spectral factorization with casual stable and causal unstable parts, then discretizing the causal stable part using the bilinear transform, for a causal IIR filter (designated **IIR BW0**). As done in the Wiener filter case, this maximally flat ($K = K_{\omega}^{\mathrm{dc}} = 6$) low-pass Butterworth filter was cascaded with the three-point FIR filters for smoothed derivatives in the TK operator. This filter has a relatively low AUC because it has appreciable gain at the centre frequency of the interferer (see Figure 12 and Table I). The design procedure described Section 4 was therefore used to place a null at $\omega_{\mathrm{nb}} = \omega_{\mathrm{int}}$ with a wide notch using $K_{\omega}^{\mathrm{nb}} = 3$ (see Figure 13). The internal states of this filter were used to directly compute the derivatives for the TK operator (without the FIR stage) thus $K_{\omega}^{\mathrm{dc}} = 3$, for a causal IIR filter with Butterworth poles $K = K_{\omega}^{\mathrm{dc}} + 2K_{\omega}^{\mathrm{nb}} = 9$ (designated **IIR BW1**). The detection performance of this filter (as indicated by the ROC in Figure 2 and the AUC in Table I) is approximately the same as the corresponding non-causal filters.

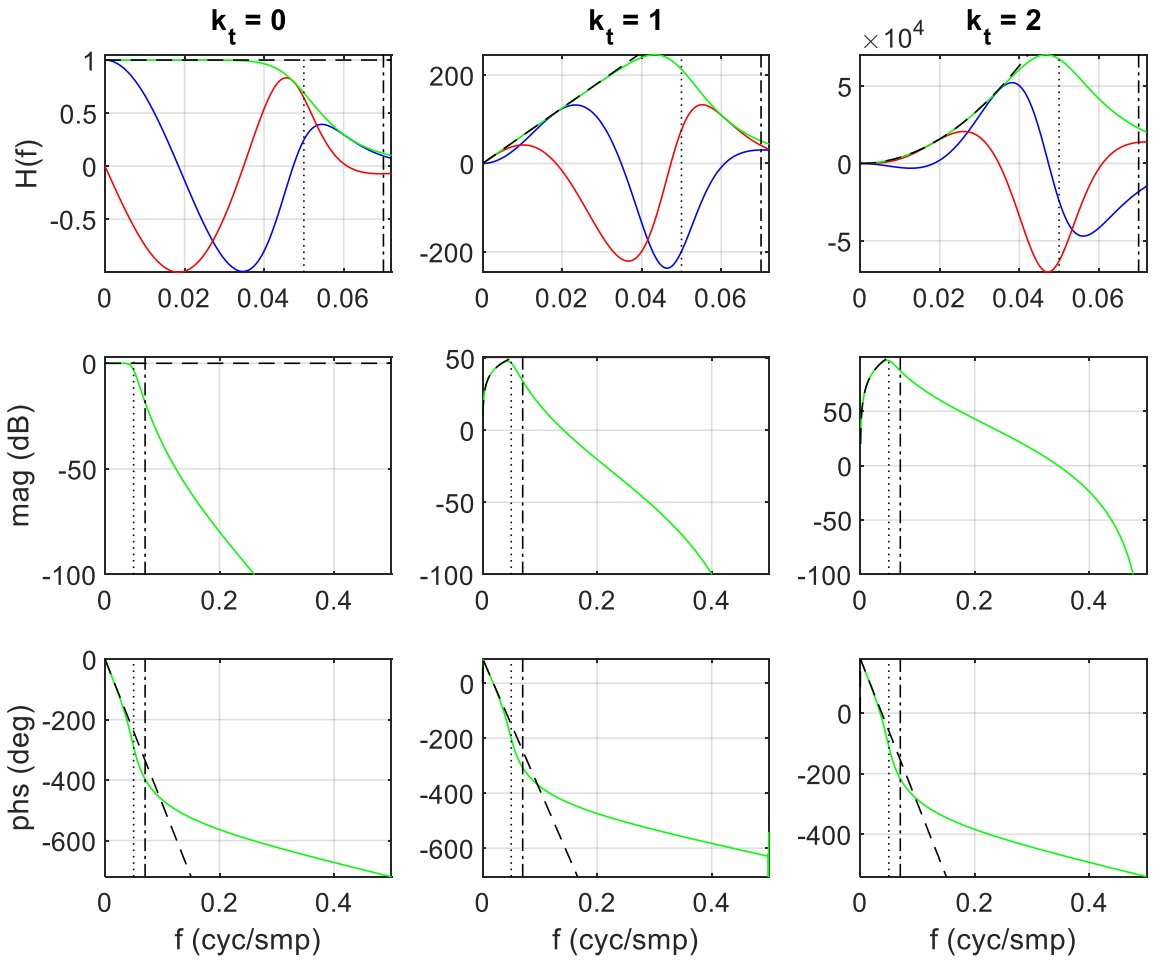

**Figure 12**. Frequency response of one-stage IIR BW0 filterbank for unknown signal. See Figure 6 for description and legend.

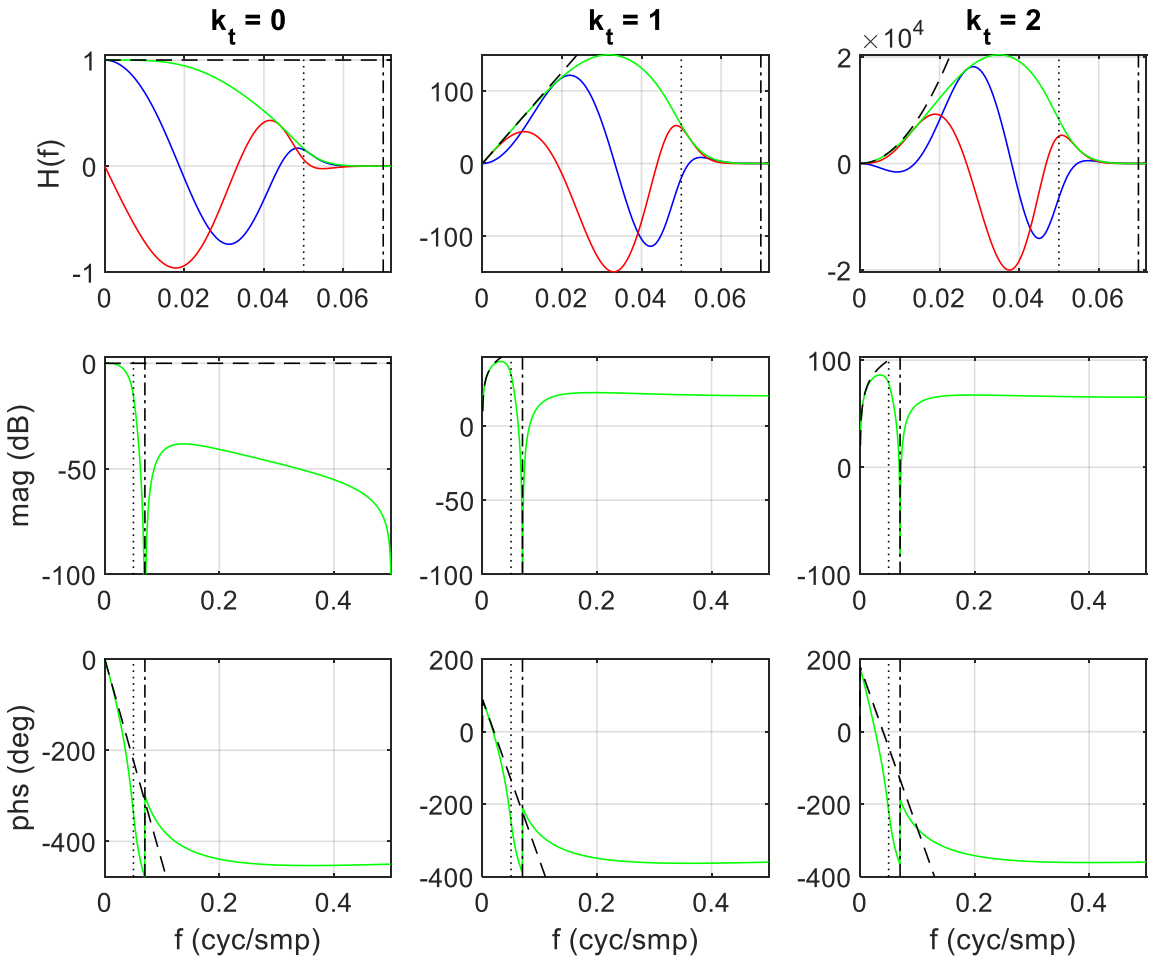

**Figure 13**. Frequency response of one-stage IIR BW1 filterbank for unknown signal. See Figure 6 for description and legend.

FIR filters have the same perfect symmetry (or anti-symmetry) in the time domain, thus perfect phase linearity in the frequency domain, as non-causal IIR filters. However, the finite time window truncates the tails of their impulse responses (multiplication by a rectangular window) which yields sidelobes in their frequency responses (convolution with the Dirichlet kernel). Although, if the window length ($M$) thus filter order ($M - 1$) are sufficiently large, and the extra computational cost can be tolerated, these artefacts have no impact on performance. A bank of 'band-limited' Savitzky-Golay (SG) smoothers/differentiators ($K_t = 3$) was designed in the frequency domain (designated **FIR SG0**), each with $K_{\omega}^{\mathrm{dc}} = 3$, $M = 3\lfloor \tau_{\mathrm{sig}}/T_{\mathrm{s}} \rfloor = 60$, $q = (M - 1)/2$ and a cut-off frequency at $\omega_{\mathrm{wb}} = \omega_{\mathrm{sig}}$ [16],[23]. In an additional design (designated **FIR SG1**), the interferer was cancelled using narrowband derivative constraints ($\omega_{\text{nb}} = \omega_{\text{int}}$ and $K_{\omega}^{\text{nb}} = 3$) [16],[23]. Using $2K_{\omega}^{\text{nb}} = 6$ degrees of freedom in this filter to satisfy narrowband derivative constraints elevates the sidelobes in the stopband slightly. The detection performance of these filters is almost identical because the gain at the interferer frequency is already negligible without explicitly placing a null there (see Figure 14 and Figure 15). Their detection performance is almost as good as the Kalman filter in the baseline scenario.

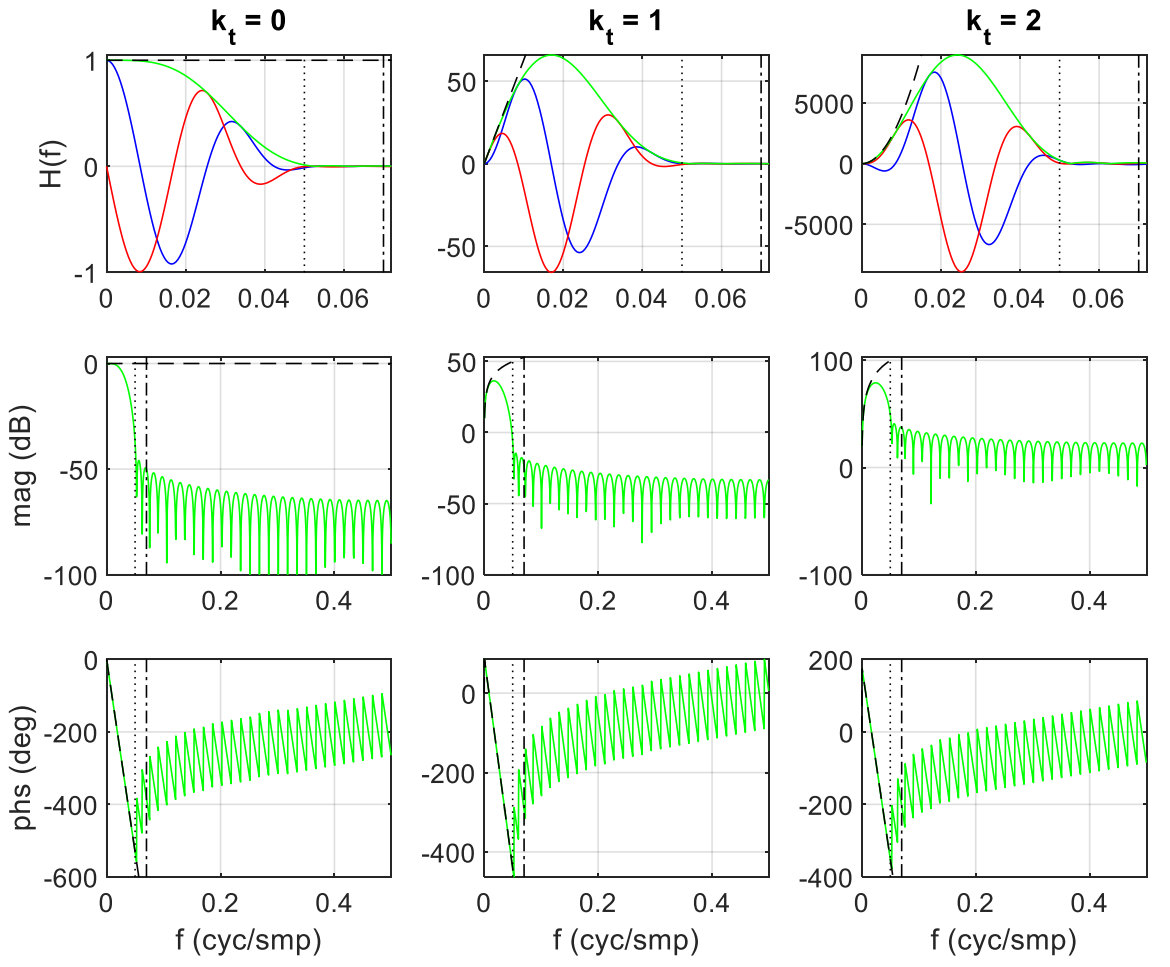

**Figure 14**. Frequency response of one-stage FIR SG0 filterbank for unknown signal. See Figure 6 for description and legend.

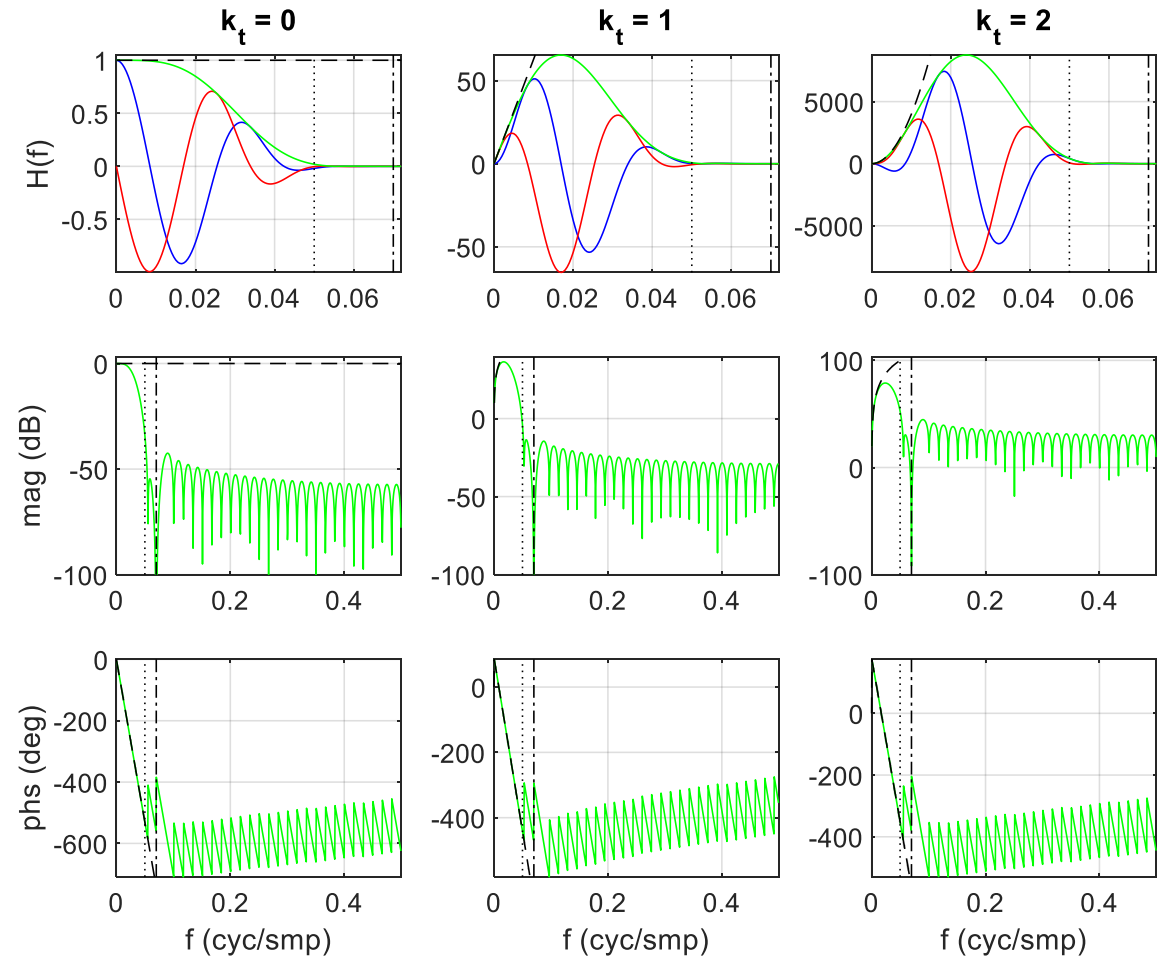

**Figure 15**. Frequency response of one-stage FIR SG1 filterbank for unknown signal. See Figure 6 for description and legend.

The basic TK operator, as specified in (33), was used as a reference implementation. It uses three-point derivatives (FIR), it does not use a pre-filter (NUL), and it is non causal (NC) due to the one-sample advance that is applied (not essential). This filterbank is designated **FIR NUL NC** and its frequency response is shown in Figure 16. The frequency responses of the filters in the two-stage filterbanks are derived by convolving the frequency response of the IIR low-pass smoother with corresponding frequency responses of these low-order FIR differentiators.

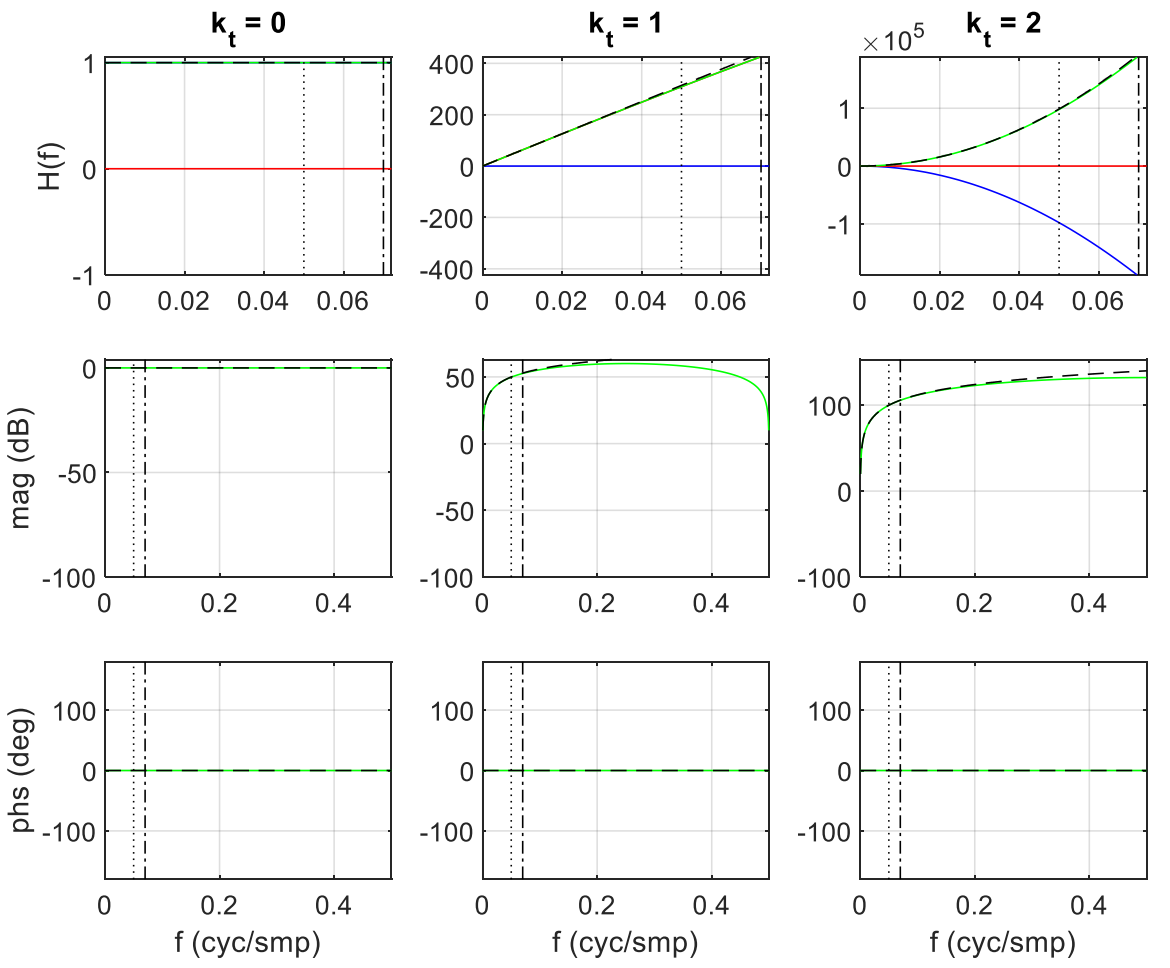

**Figure 16**. Frequency response of one-stage FIR NULL NC filterbank (with no low-pass pre-filter) for unknown signal. See Figure 6 for description and legend.

### *6.4 Discussion*

The Wiener filter and Kalman filter may both be tuned using the available degrees of freedom afforded by the process models and noise statistics so that they have the response of a low-pass filter for good detection performance in uncertain environments. Although, there are certainly simpler ways of designing a low-pass filter to a bandwidth specification. Unfortunately, the non-causal Butterworth filter, the causal Butterworth filter and the band-limited Savitztky-Golay filter have different gains at the specified cut-off frequency. In all cases, this critical frequency was set equal to the upper bound of the signal frequency ($\omega_{\text{wb}} = \omega_{\text{sig}}$). For continuous-time non-causal Butterworth filters $|H(\omega_{\text{wb}})| = 1/2 = 0.5$ and after spectral factorization $|H(\omega_{\text{wb}})| = \sqrt{1/2} \cong 0.7$ for the causal variant. After discretization, these magnitudes are approximate (i.e. $|H(\omega_{\text{wb}})| = 0.48$ & $0.69$, respectively, as shown in Table I). For the relatively narrowband low-pass filters considered here, using the bilinear method for the BW0 filters and the impulse invariance method to determine the poles of the BW1 filters was found to be adequate. For wideband low-pass filters the bilinear transform with frequency warping should be applied if exact agreement at $\omega_{\text{wb}}$ is required. For the band-limited Savitztky-Golay filters, the objective is to minimize high-frequency (coloured) noise power outside the low-frequency (polynomial) signal band (i.e. $|H(\omega)|^2$ for $\omega_{\mathrm{wb}} \leq \omega \leq \pi$), subject to derivative constraints at $\omega = 0$ and (optionally) $\omega = \omega_{\mathrm{int}}$, thus $\omega_{\mathrm{wb}}$ is a not a 3dB point; rather, it is the point at which the magnitude is ideally zero. The different filter bandwidths largely account for the observed differences in detection performance. When the (non-causal and causal) Butterworth filters were tuned using $\omega_{\mathrm{wb}} = 0.75\omega_{\mathrm{sig}}$, and the band-limited Savitztky-Golay filters using $\omega_{\mathrm{wb}} = 1.5\omega_{\mathrm{sig}}$, all filters had a similar bandwidth and differences in detection performance were less obvious (see Figure 17). For the closely spaced signal and interference frequencies considered here, the filter design problem requires the bandwidth to be optimally set so that there is sufficient gain at the upper limit of the signal frequency ($\omega_{\mathrm{sig}}$) and sufficient attenuation at the interferer frequency ($\omega_{\mathrm{int}}$). Placing a null at $\omega_{\mathrm{int}}$ simplifies this process somewhat because only the signal gain needs to be considered. Note that when the null is wide (e.g. using $K_{\omega}^{\mathrm{nb}} = 3$) and close to the passband edge the signal bandwidth of the Butterworth filters is also reduced. Sacrificing some gain near the passband edge to increase interference attenuation is a simple way of improving the ROC and increasing the AUC of a detector in the unknown/variable signal case. However, if the modulating frequency of the signal is known and fixed at the passband edge (i.e. at $\omega = \omega_{\mathrm{sig}}$) then the detection performance of the filters with a narrower bandwidth is degraded (e.g. the Kalman filters and the Savitzky-Golay filters); and in such scenarios, the Wiener filter with a matched signal model outperforms (see Figure 18).

The TK operator applies a conservation of (kinetic and potential) energy law to generate the signal envelope. If magnitude is distorted and phase is misaligned for the derivative estimate inputs, then the energy output is degraded. For the low-frequency ($\tilde{f}_{\mathrm{sig}} = 0.0175$) pulse in Figure 19, the signal is well within the passband and the pulse envelope is reproduced reasonably well for all filters. For the high-frequency ($\tilde{f}_{\mathrm{sig}} = 0.0355$) pulse in Figure 20, the signal is closer to the passband edge where differences in the (complex) frequency responses are more obvious (i.e. non-monomial magnitude and non-linear phase). For this signal, the fidelity of pulse envelope reproduction is severely degraded for filters with a narrower bandwidth. When the IIR BW1 filter was re-tuned for a wider and flatter passband using $K_{\omega}^{\mathrm{dc}} = 6$ & $K_{\omega}^{\mathrm{nb}} = 1$ (instead of $K_{\omega}^{\mathrm{dc}} = 3$ & $K_{\omega}^{\mathrm{nb}} = 3$) the pulse envelope was reproduced with high fidelity; however, detection performance was degraded (the AUC decreased from 0.937 to 0.908). When the pros and cons of FIR and IIR filters are considered in the literature, the phase non-linearity of casual IIR filters is usually emphasized; however, for the pulse detector considered in this section and the target tracker considered in the next section, phase errors are no more damaging than magnitude errors. For this reason, the magnitude of the complex error is arguably a better indication of filter error and its expected performance. For instance, the Savitzky-Golay pulse envelopes are rippled even though these FIR filters have perfectly linear phase like the non-causal IIR filters. If magnitude and phase errors are consistent over all derivatives output by the filterbank, then the ripples disappear and only scaling errors remain (with a time lag for causal filters). Envelope ripples do not necessarily degrade the only performance metric quantified here (the ROC AUC) because there is no penalty for pulse splitting, i.e. multiple detections on a single pulse; however, ripples will probably degrade the accuracy of envelope parameter estimates, e.g. duration and peak amplitude. Non-casual IIR filters have excellent frequency responses however they are not feasible in online processing. FIR filters with long impulse responses also have very good responses however they may be too slow for real-time online processing in embedded systems with feeble computers.

The results in this section indicate that all filters have sufficient degrees of freedom which may be adjusted for the desired frequency response and reasonable detection performance. The time and effort required to reach that response is the main distinguishing consideration. The Kalman filter was the most difficult to tune for this application. For the (causal and non-causal) Butterworth filters with an interference cancelling notch, good responses were obtained directly from the scenario parameters, without requiring any additional manual tuning, unlike the Kalman filter. The optimal placement of the Kalman filter poles allows very good solutions to be found; however, those solutions may be very difficult to find. The ease with which the (causal) Butterworth MaxFlat filters are tuned suggests that online reconfiguration may be feasible. Furthermore, if the bandwidth parameter remains constant (for fixed pole positions) only the zero positions need to be adjusted to cancel an interferer at a new (known) frequency, which means that the filters do not need to be re-initialized and the internal states reset. This flexibility follows from the following interpretation of a filter's transfer function: the poles determine the history that is analyzed whereas the zeros determine the way in which that history is interpreted. Adaptive configurations will be considered in future work.

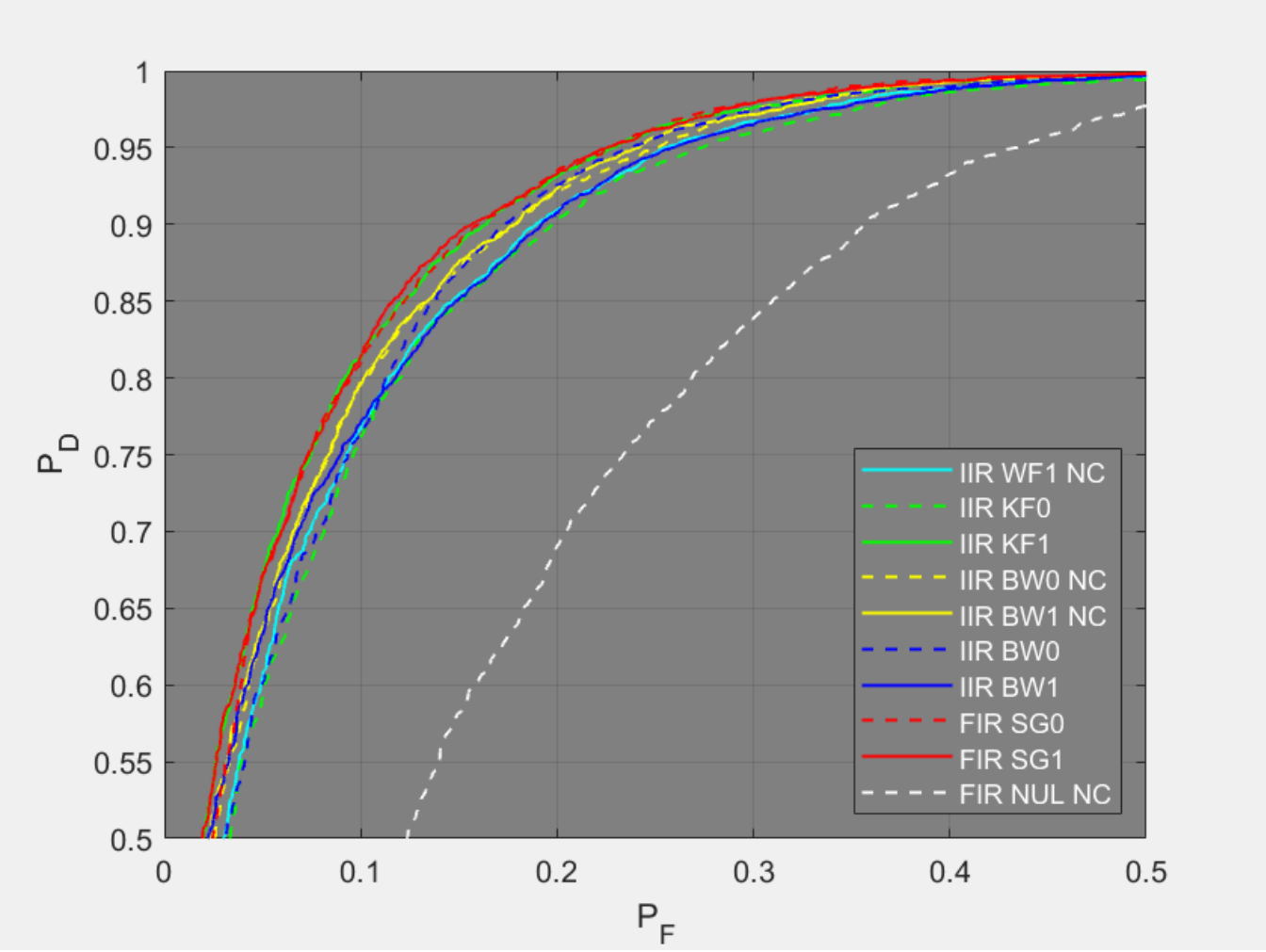

**Figure 17**. ROC of pulse detectors (with modified tuning) in the baseline scenario with a deterministic signal of unknown frequency and non-deterministic interference of known frequency. Filters that do not explicitly incorporate a noise model have been manually adjusted for improved performance in this scenario.

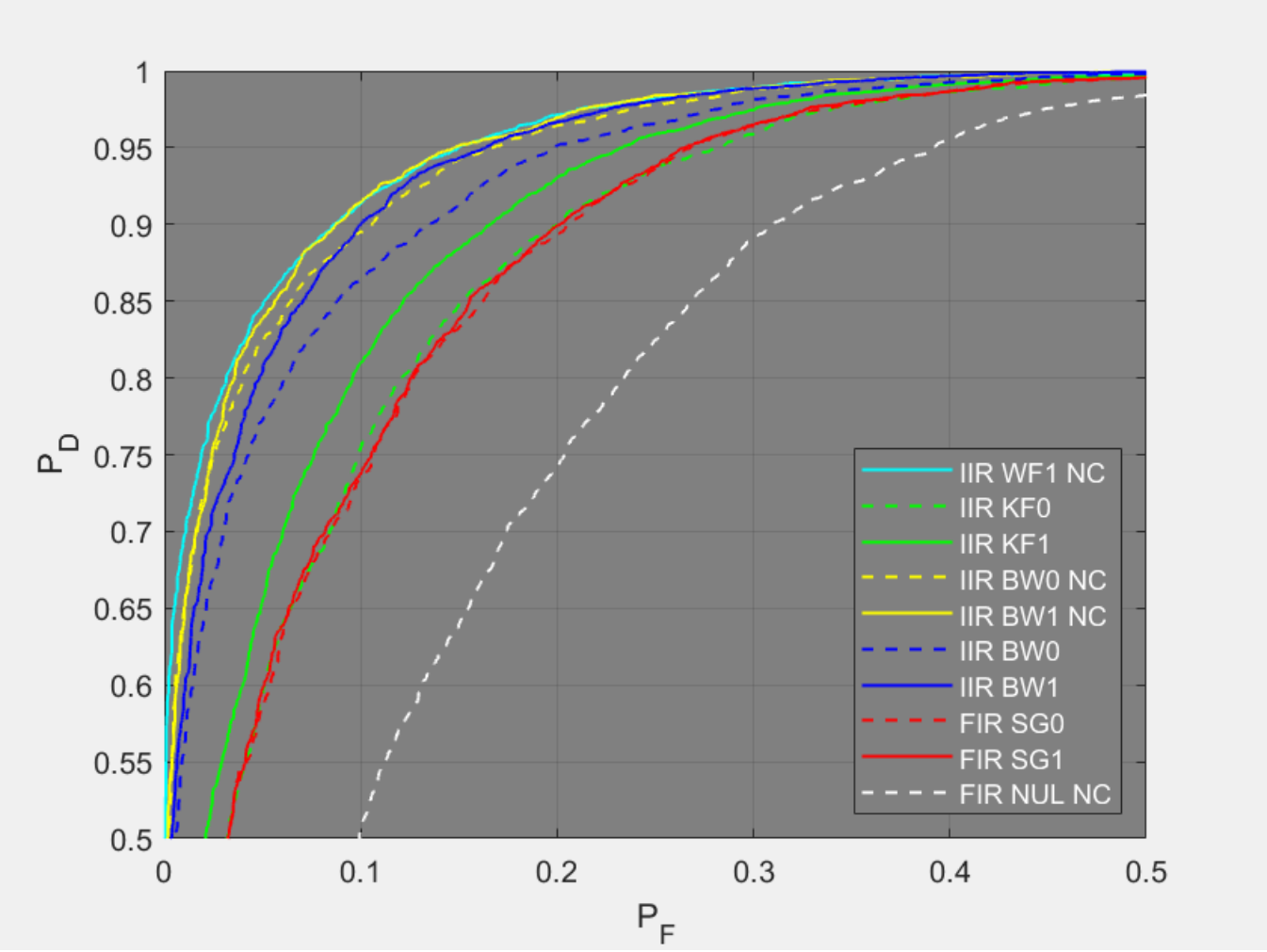

**Figure 18**. ROC of pulse detectors in a modified scenario with a non-deterministic signal of known frequency ($f_{\text{sig}}$) and non-deterministic interference of known frequency ($f_{\text{int}}$). Except for the IIR WF1 NC filter, which is tuned for the known signal (see Figure 6) all filters used the default tuning. The filters that benefited from a narrower bandwidth in the unknown signal scenario of Figure 2 (i.e. the FIR SG filters and the IIR KF filters) do not perform well in this scenario. As expected, the IIR WF1 NC pre-filter yields the best detector in this scenario.

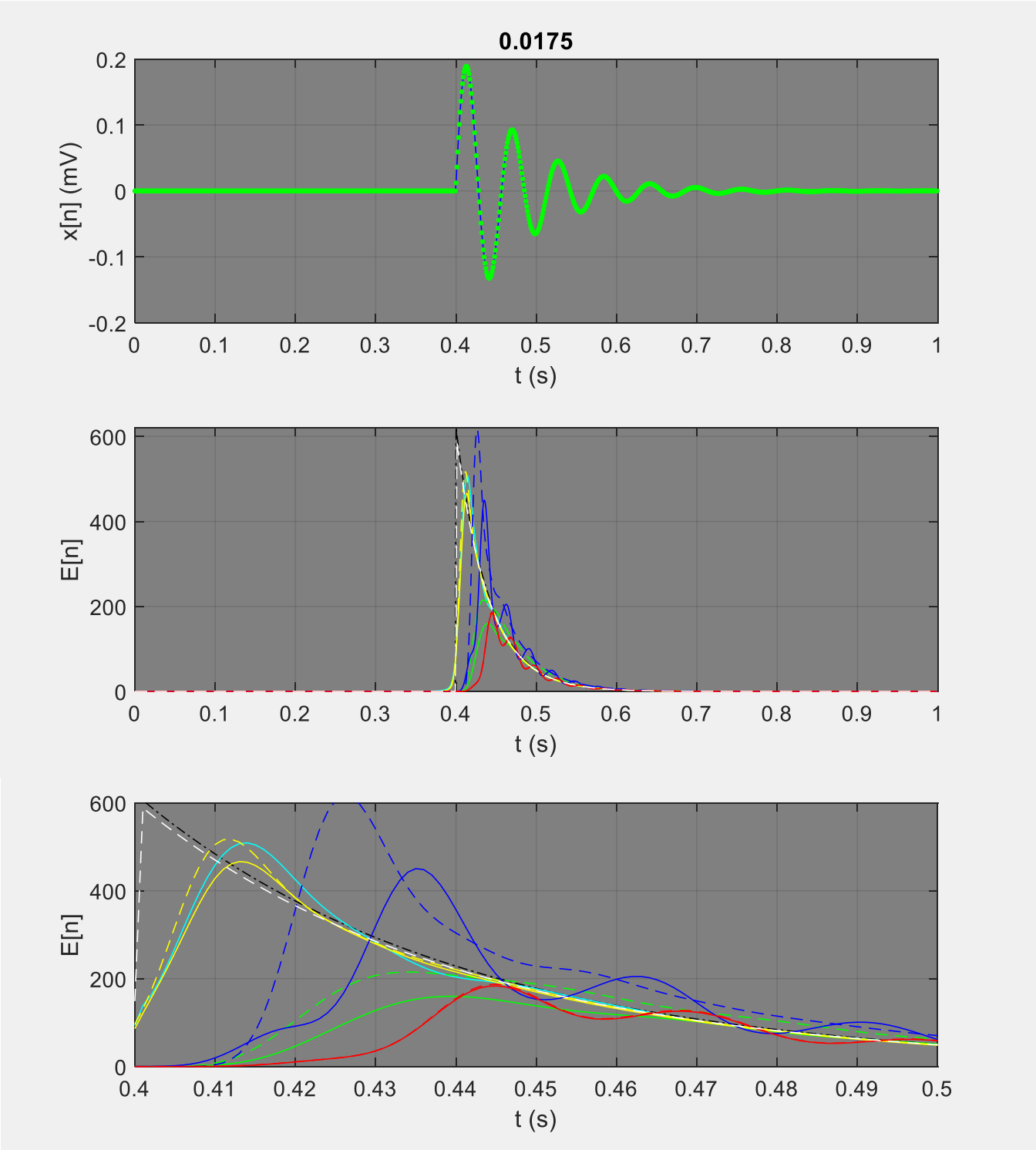

**Figure 19**. Deterministic signal ($f_{\text{sig}} = 0.0175$) with no interference (top), energy envelope for TK operator with various filterbanks (middle), detail of energy envelope (bottom). Envelope ripple is caused by phase and magnitude errors near the passband edge.

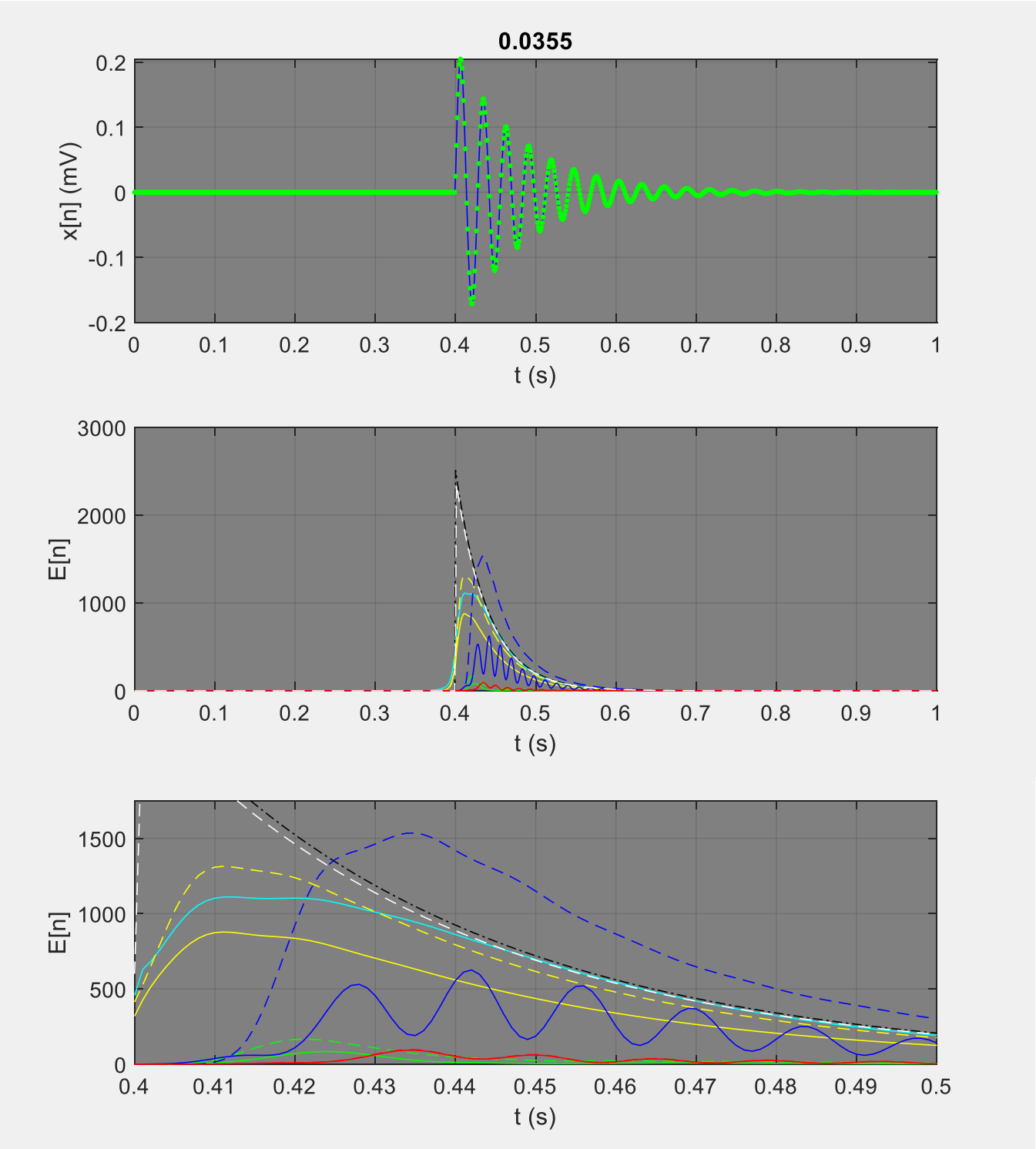

**Figure 20**. Deterministic signal ($f_{\text{sig}} = 0.0355$) with no interference (top), energy envelope for TK operator with various filterbanks (middle), detail of energy envelope (bottom). The narrower bandwidth of the KF and SG filters increases the loss for this signal of higher frequency.

## 7 Tracking manoeuvring targets

For the tracking of manoeuvring targets, LSS design is generally preferred for recursive (IIR) filters at steady state. Models of signal (i.e. target) and interference processes expose the physical states to be estimated and constrain the response of the estimator, via the appropriate placement of filter zeros, while the filter poles define the bandwidth, transient response and rate of convergence. In a Kalman filter the gain vector, thus pole positions at steady state, are set via the noise parameters [2]. In a Luenberger filter, the poles are placed arbitrarily, via the gain vector [22]. In kinematic (i.e. native or derivative state) coordinates, the internal states of the discrete-time steady-state (Kalman or Luenberger) filter are estimates of the continuous-time process states. In canonical observer or canonical controller coordinates, or any of the other direct forms with minimal complexity, a similarity transform is required to convert filter states into estimates of process states.

MaxFlat designs provide an alternative perspective. Constraints on derivatives of the frequency response at dc are imposed to match the order of the integrating signal (i.e. target) process and elsewhere in the stopband to suppress interference (e.g. due to jamming or propagation effects), for a solution that minimizes the white-noise gain. The passband width (i.e. bandwidth) of the solution defines what frequencies may also contribute to the (integrating) target process model, i.e. the extent to which other similar types of target motion are tolerated (e.g. low-frequency weaves), and the time required for the tracker to adjust to abrupt changes in the state vector (i.e. the transient response). The stopband width defines the noise-rejection properties of the tracker.

The outcome of an aerial dogfight is largely determined by the relative manoeuvrability of the aircraft involved. If one can turn at a greater rate than the other, i.e. greater speed on a smaller circle, then it has a distinct advantage. Tracking filters that estimate the kinematic states of a target, must therefore have adequate bandwidth to maintain tracks on targets executing these extreme (but uncommon) manoeuvres while also having adequate white-noise attenuation to produce smooth estimates of position, velocity, and possibly acceleration, states during less extreme (but more common) flight regimes [22],[43],[44]. Moreover, for a given steady-state error-variance of a non-manoeuvring target's state, it is also desirable to know the turn rate at which track loss is likely, due to unacceptably large steady-state bias errors. The filter presented in Section 4 is configured and analysed in this context. It is used to estimate the derivative states of a manoeuvring target, in a hypothetical scenario using randomly generated simulated data.

### *7.1 Tracker analysis and the frequency response*

The script typeface with an under-bar is used in this section to clearly distinguish the spatial coordinates ($\underline{\mathscr{x}}$ & $\underline{\mathscr{y}}$) from the system inputs/outputs ($x$ & $y$) used in prior sections. These Cartesian coordinates specify the noisy measurements, smoothed estimates $\underline{\hat{\mathscr{x}}}$ & $\underline{\hat{\mathscr{y}}}$, and true location $\underline{\mathscr{x}}$ & $\underline{\mathscr{y}}$ of a target's position, for example in a two-dimensional imaging sensor on an earth observation satellite. Two filterbanks operate independently and in parallel on each coordinate. Only the position estimate ($k_t = 0$) is considered here, although high-order derivatives, for example velocity ($k_t = 1$) and acceleration ($k_t = 2$) may also be of interest, e.g. for target classification purposes. A lag of $q$ samples is uniformly applied to all derivative estimates. Thus the $n$th output $y_0[n]$, of one smoother is $\underline{\hat{\mathscr{x}}}(t_q)$ and the output of the other smoother is $\underline{\hat{\mathscr{y}}}(t_q)$, where $t_q = (n - q)T_s$ is the lag-adjusted time. The expected value of the squared Euclidian error-norm at steady-state is

$$\sigma_y^2 = E\langle\|\boldsymbol{\varepsilon}\|_2^2\rangle = E\langle\varepsilon_{\underline{\mathscr{x}}}^2\rangle + E\,\langle\varepsilon_{\underline{\mathscr{y}}}^2\rangle \text{ where} \tag{41}$$

$E\langle\varepsilon_{\underline{\mathscr{x}}}^2\rangle$ and $E\,\langle\varepsilon_{\underline{\mathscr{y}}}^2\rangle$

are the expected squared errors in the two spatial coordinates $\left(\underline{\mathscr{x}}, \underline{\mathscr{y}}\right)$ with

$$\varepsilon_{\underline{\mathscr{x}}}[n] = \underline{\hat{\mathscr{x}}}(t_q) - \underline{\mathscr{x}}(t_q) \text{ and}$$
$$\varepsilon_{\underline{\mathscr{y}}}[n] = \underline{\hat{\mathscr{y}}}(t_q) - \underline{\mathscr{y}}(t_q). \tag{42}$$

When the input contains only white noise, or when the filter is perfectly matched to the signal and interference processes, for instance using derivative constraints in a MaxFlat filter, (41) is simply evaluated from the frequency response using (20) yielding

$$\sigma_y^2 = 2\Sigma_0 \cdot \sigma_x^2 \text{ where} \tag{43}$$

$\Sigma_0$ is the white-noise gain of the $k_t = 0$ filter.

When an undamped ($\tau_c = \infty$) sinusoidal signal or interferer $A_c e^{i\Omega_c}$ is applied in *one* dimension, the magnitude at steady state is simply $A_c|H(\omega)|\big|_{\omega=\Omega_c/F_s}$ and the phase shift is $\angle H(\omega)|_{\omega=\Omega_c/F_s}$. This follows directly from the definition of the frequency response. To analyze the tracker response for a target on a circular orbit (orb) around $\left(\underline{\mathscr{x}}_0, \underline{\mathscr{y}}_0\right)$ with a radius of $r_{\text{orb}}$ and an angular velocity of $\Omega_{\text{orb}}$ (radians per second) in *two* dimensions, it is convenient to use complex notation: $v = \underline{\mathscr{x}} + i\underline{\mathscr{y}}$. The radial ($\varepsilon_r$) and angular ($\varepsilon_\theta$) tracking errors may now simply be defined as

$$\varepsilon_r[n] = \left|\hat{v}(t_q) - v_0\right| - \left|v(t_q) - v_0\right| \text{ and}$$
$$\varepsilon_\theta[n] = \angle\{\hat{v}(t_q) - v_0\} - \angle\{v(t_q) - v_0\} \text{ where}$$
$$v_0 = \underline{x}_0 + i\underline{y}_0 \text{ (i.e. the centre of the circular orbit)}$$
$$v(t) = \underline{x}(t) + i\underline{y}(t) \text{ and}$$
$$\hat{v}(t) = \hat{\underline{x}}(t) + i\hat{\underline{y}}(t). \quad (44)$$

The radial tracking error ($\varepsilon_r$) is a magnitude shift and the angular tracking error ($\varepsilon_\theta$) is a phase shift, between the true signal $v$ and the estimated signal $\hat{v}$. The discrete-time transfer function of the smoother $\mathcal{H}(z)$ describes the time evolution and dynamics (i.e. over transient and steady-state regimes) between the system input $v(t)$ and the system output $\hat{v}(t)$; whereas the frequency response of the smoother $\mathcal{H}(\omega)$ represents the steady-state properties of the discrete-time system. The angular frequency is a rate of phase change (radians per unit of time). Equivalently, a target's angular velocity (radians per unit of time) on a circular orbit is also a rate of phase change or an angular frequency. As $\mathcal{H}(\omega)$ describes the magnitude scaling and phase shift at the output for a sinusoidal input $e^{i\omega}$, the steady-state tracking errors, in the absence of measurement noise (i.e. when $\sigma_R^2 = 0$), are simply determined by evaluating the frequency response of the smoother at the angular frequency ($\omega$) that is matched to the angular velocity ($\Omega_{\text{orb}}$) of the orbit using

$$\varepsilon_r[\infty] = \left\{|\mathcal{H}(\omega)| - \left|e^{-iq\omega}\right|\right\}r_{\text{orb}} \text{ and}$$
$$\varepsilon_\theta[\infty] = \angle\mathcal{H}(\omega) - \angle e^{-iq\omega} \text{ where}$$
$$\omega = \Omega_{\text{orb}}/F_s\,. \quad (45)$$

As discussed above (and in [22],[43],[44]) the frequency response may be used to determine the following three important steady-state tracking metrics: **1)** a track converges on the target if the order of a polynomial target trajectory is less than or equal to the flatness order of the frequency response at dc ($\omega = 0$); **2)** the steady-state bias for a target with a constant rate ($\Omega_c$) and radius of turn is determined by the phase and magnitude of the frequency response at that rate of turn ($\omega = \Omega_{\text{orb}}/F_s$); **3)** the error variance at steady state, for a non-divergent and unbiased track, is equal to the integral of the squared magnitude of the frequency response (over $\omega = 0 \dots \pi$). The proposed MaxFlat design procedure allows the frequency response to be shaped so that these requirements are satisfied. Complications arising from ambiguous measurement-to-track assignments (i.e. data association) are not considered here.

### *7.2 Analysis of MaxFlat tracking filters*

The tracking metrics described in the previous subsection are used in this subsection to analyse various tunings of the Butterworth Maxflat filterbanks discussed in Section 4. Then in the subsection that follows, the behaviour of the filters is illustrated in MC target-tracking simulations.

Four filter tunings were considered: $K_\omega^{\text{dc}} = 3$ & $K_\omega^{\text{nb}} = 0$ (**Tracker A**), $K_\omega^{\text{dc}} = 3$ & $K_\omega^{\text{nb}} = 1$ (**Tracker B**), $K_\omega^{\text{dc}} = 3$ & $K_\omega^{\text{nb}} = 3$ (**Tracker C**) and $K_\omega^{\text{dc}} = 6$ & $K_\omega^{\text{nb}} = 1$ (**Tracker D**). All were configured for a sampling rate $F_s = 100$ Hz, a bandwidth of $F_{\text{wb}} = 0.05F_s$ Hz, and an interferer at $F_{\text{nb}} = 0.05F_s$ Hz. The normalized frequencies (i.e. $f_{\text{wb}}$ & $f_{\text{nb}}$) are the same as those used in the pulse detector of Section 6. The lower sampling rate only affects the response of the $k_t > 0$ filters, thus the $k_t = 0$ frequency response of Tracker C and the IIR BW1 detector of the previous section is the same. The four tracking filters are analyzed in Figure 21 to Figure 24. In these figures, the columns correspond to the $k_t = 0$, $k_t = 1$, and $k_t = 2$ filters that output estimates of the position, velocity, and acceleration of the target in one of the Cartesian dimensions, respectively. The first row shows the magnitude of the complex error, where the error is the difference between the realized response of the filter and the desired response of the corresponding 'ideal' all-pass differentiator. Outside the passband of the filter, large (magnitude and phase) errors are inconsequential if the gain of the filter is negligible. A palette of 'autumnal' colours is used for filters designed with different group delays ($q$ in samples). The optimal passband group-delay ($q_{\text{opt}}$), that minimizes the white-noise gain using (21) is shown in green. The second and third rows show the magnitude and phase of the (complex) filter frequency-responses, respectively. The magnitude response of ideal $k_t$th-order all-pass differentiators (dashed white line) and the phase response of an all-pass delay (dashed autumnal and green lines) are also shown. Only the low-frequency range is shown to reveal the passband and transition-band response. These plots show that using the optimal passband group-delay ($q = q_{\text{opt}}$) promotes magnitude flatness and phase linearity in the passband, in addition to minimizing the white-noise gain. They also show that when $K_\omega^{\text{dc}} = K_t$ the $k_t = K_t - 1$ response is independent of $q$ because $\mathcal{P}(q)$ in (21c) is a constant and has no roots.

The bottom panel is an illustration of the 2-D orbit response for the filter with the optimal passband group-delay applied. A target is on circular trajectory ($r_{\text{orb}} = 10$ pixels or 'pix') and various (constant) angular velocities are considered ($f_{\text{orb}} = \Omega_{\text{orb}}/2\pi F_s$). The target trajectory is depicted using a white circle. A palette of 'prismatic' colours is used for tracks on targets at the various rates of turn (all with the same radius). The frequencies corresponding to these turn rates are shown in the magnitude-response and phase-response plots, using dotted vertical lines. As discussed in the previous subsection, the value of the complex frequency response at these frequency points determines the angular and radial tracking error at steady state. The target originated at (10,0) and completed 10 revolutions to ensure that steady state was reached. Solid radial lines depict the final position of the target, delayed by $T_\mathrm{s}q_\mathrm{opt}$ seconds to compensate for the delay applied by the filter. Dashed radial lines depict the final estimate of the target's position that is output by the independent $k_t = 0$ filters in both Cartesian dimensions. The radial and angular errors computed from these radial vectors agrees with those computed using (44) to at least four decimal places.

The yellow track is informative as it corresponds to a rate of turn at the passband edge of the filters. For $K_\omega^\mathrm{dc} = 3$, the radius of this track decreases as $K_\omega^\mathrm{nb}$ is increased from 0 to 1 then to 3 (i.e. Trackers A, B & C), due to the signal bandwidth contraction caused by the interferer notch dilation. The high turn-rate tracks (orange and red) contract to the origin when $K_\omega^\mathrm{nb} = 3$ is used (in Tracker C). The extra bandwidth afforded using $K_\omega^\mathrm{dc} = 6$ (in Tracker D) restores the radius of the yellow band-edge track whereas the increased near-dc flatness reduces the radial and angular errors of the blue low-frequency tracks. The large delay ($q_\mathrm{opt} = 19.4$) applied in Tracker D 'pushes' the filter zeros well outside the unit circle for a so-called non-minimum-phase response. Loosely speaking, such filters may exhibit curious start-up transients as they are effectively 'idle' while they 'wait' to accumulate enough samples for an appropriate output to be determined. The initial track whorl in Figure 24 is an illustration of such behavior.

Most aspects of digital filter design are a trade-off and the optimal balance between conflicting performance metrics is ultimately determined by the application, requirements and other (economic or physical) constraints. For example, is it better to have: a wide or narrow bandwidth (i.e. a long or a short impulse response), a wide or narrow transition band, a long or short passband group-delay, etc.? These issues complicate any detailed quantitative analysis of filter quality or performance in any hypothetical simulated scenario. However, the material in Section 7.1 and the response plots in Figure 21 to Figure 24 clearly show the balance that has been reached and the performance that can be expected for a given design. The proposed MaxFlat design procedure provides a simple way of configuring filters to meet these objectives directly.

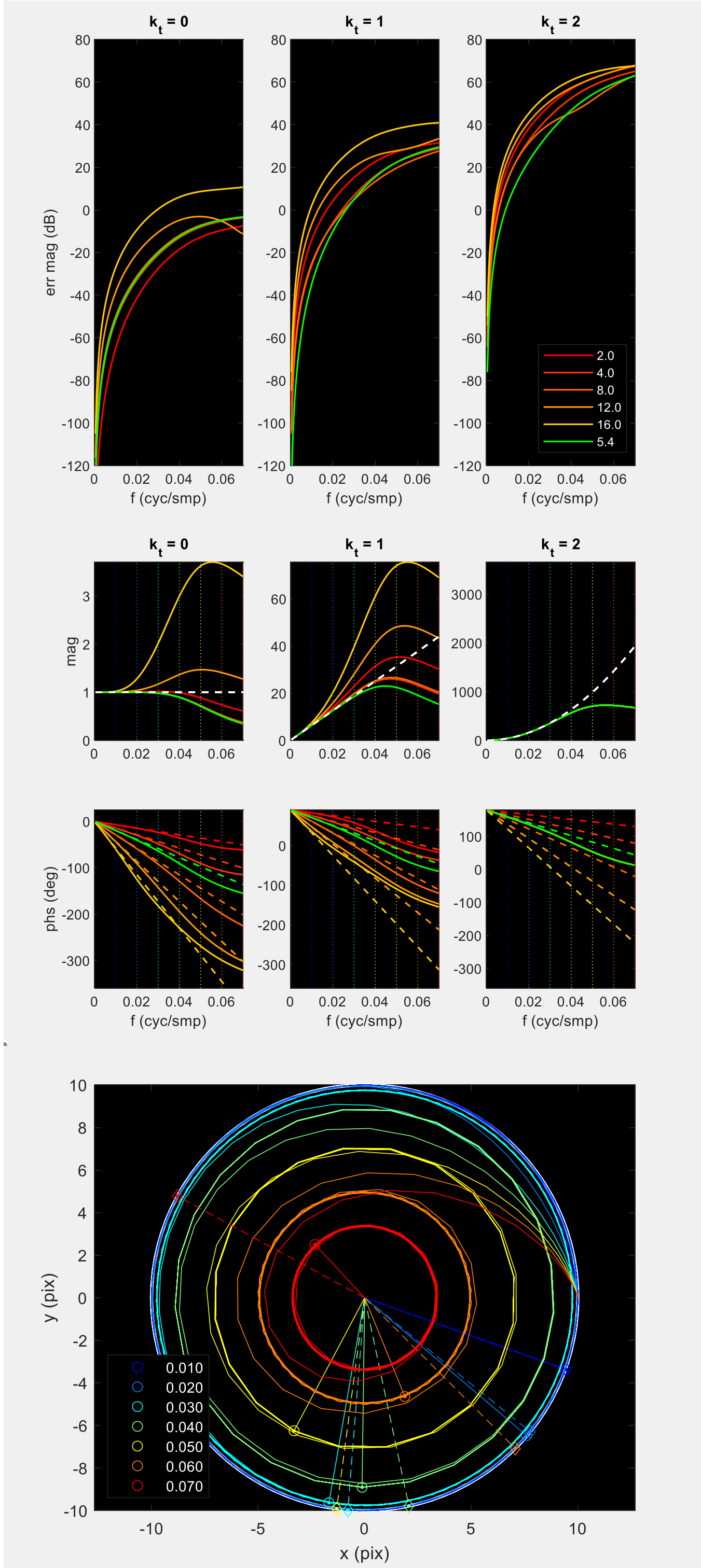

**Figure 21**. Response of Tracker A. See text for description.

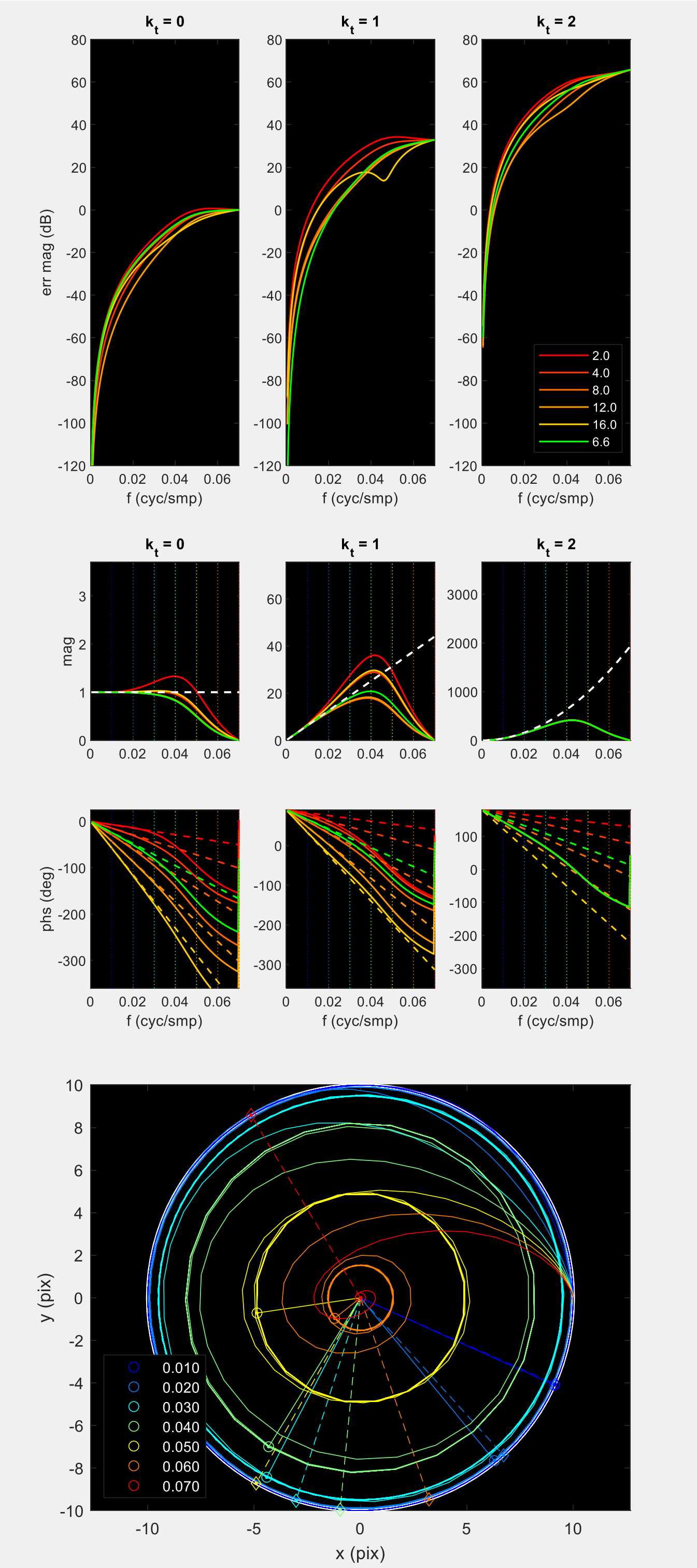

**Figure 22**. Response of Tracker B. See text for description.

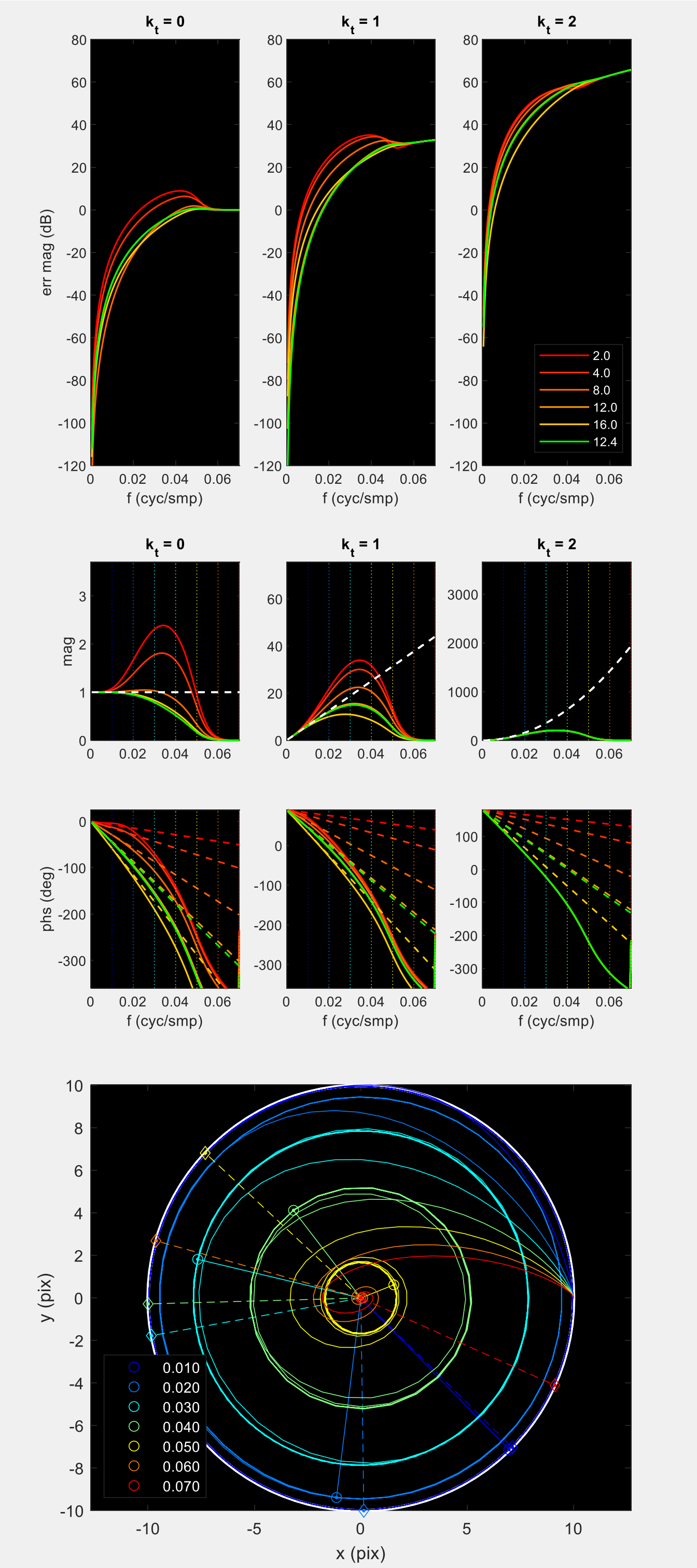

**Figure 23**. Response of Tracker C. See text for description.

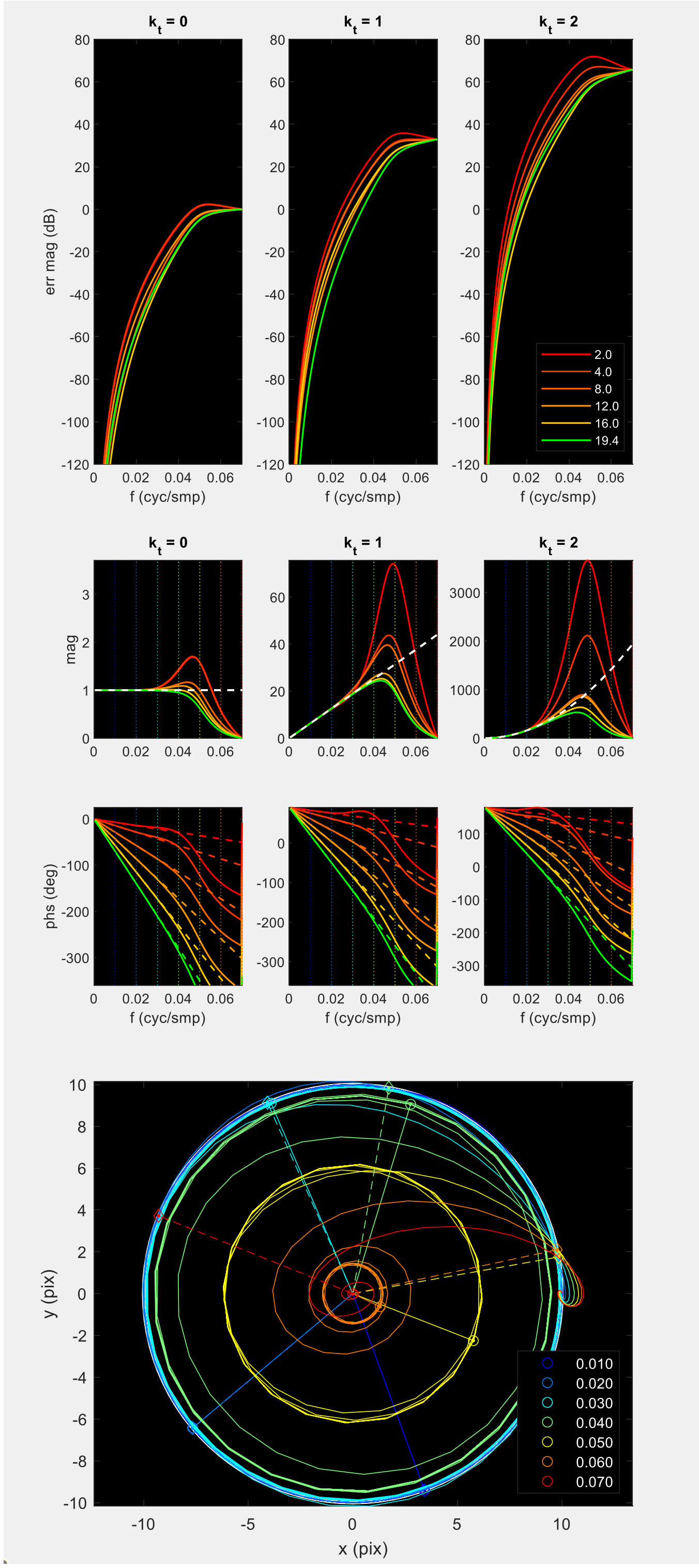

**Figure 24**. Response of Tracker D. See text for description.

### *7.3 Monte Carlo Simulations*

Trackers A-D with a passband group-delay of $q_{\mathrm{opt}}$ were used to process randomly generated sequences of measurements that are imagined to be from a high-altitude platform with a wide-area electro-optic imaging sensor observing a manoeuvring aircraft below. The true trajectory of the aircraft (i.e. the ‘target’) is randomly generated by the signal process. The measurements (in Cartesian image coordinates) are perturbed by uncompensated platform jitter/vibration that is randomly generated by the interference process. Similar processes could also be used to model drift in global-positioning-system telemetry or ionospheric disturbances in skywave surveillance radar.

As in the detection application, the measurements were formed by summing then sampling waveforms independently generated by the signal and interference processes that are driven by Gaussian noise, using the second-order system defined in (35) and (37). The process parameters, $\tau = \alpha_\tau T_s/f_c$ and $\lambda = \alpha_\lambda T_s/f_c$, were set using $T_s = 0.1$ with $f_{\text{sig}} = 0.05$ and $f_{\text{int}} = 0.07$. Two scenarios were considered: one containing an aircraft on patrol executing gentle turns (**Lo-G**); the other containing a dogfighting aircraft executing extreme turns (**Hi-G)**. The Lo-G scenario used $\alpha_\lambda = 8$ for the signal process so that the target's motion is well within the passband of the filters ($f_{\text{lo}} = 1/\lambda_{\text{lo}} = 6.25 \times 10^{-3}$ cyc/smp); the Hi-G scenario used $\alpha_\lambda = 2$ so the target is closer to the passband edge ($f_{\text{hi}} = 1/\lambda_{\text{hi}} = 2.50 \times 10^{-2}$ cyc/smp). In both scenarios: $\alpha_\lambda = 1$ for the interference process; $\alpha_\tau = 8$ for both signal and interference processes; and the average power of the random white-noise input was $P_{\text{sig}} = 1.0 \times 10^4$ & $P_{\text{int}} = 1.0 \times 10^2$. The target originated at a randomly generated position.

Two random MC instantiations of the Lo-G and Hi-G scenarios are shown in Figure 25 and Figure 26. The true target trajectory is shown in blue, the additive interference in red, the measurement samples in green, along with the outputs of Trackers A-D. In the Lo-G scenario, due to the high coherence of the interference, using $K_\omega^{\text{nb}} \geq 1$ for a narrow null (in trackers B, C & D), is sufficient to attenuate the correlated measurement noise (see Figure 25). In the Hi-G scenario, the moderate signal attenuation caused by using $K_\omega^{\text{nb}} = 1$ (in Tracker B) and the severe signal attenuation caused by using $K_\omega^{\text{nb}} = 3$ for a wider notch (in Tracker C) is evident during the long and tight turns (see Figure 26). Increasing the degree of passband flatness using $K_\omega^{\text{dc}} = 6$ (in Tracker D) instead of $K_\omega^{\text{dc}} = 3$ restores the signal response, at the expense of a much greater delay. In some applications large latencies may be unacceptable (e.g. in guidance systems); in other applications where there are already long delays due to data acquisition, pre-processing and transmission, lags of less than one second will likely go un-noticed (e.g. in strategic wide-area surveillance systems).

As in the detector application of Section 6, both tracking scenarios in this subsection are challenging because the centre frequency of the interference (i.e. narrowband coloured noise) is close to the band edge of the signal. Thus, some distortion of the signal (i.e. magnitude scaling and/or phase shifting) must be accepted if (white or coloured) noise is to be attenuated. The frequency responses in Figure 21 to Figure 24 quantify the distortion that is to be expected at steady state for a given degree of noise cancellation whereas the orbit responses are qualitative illustration of the distortion in a more practical context.

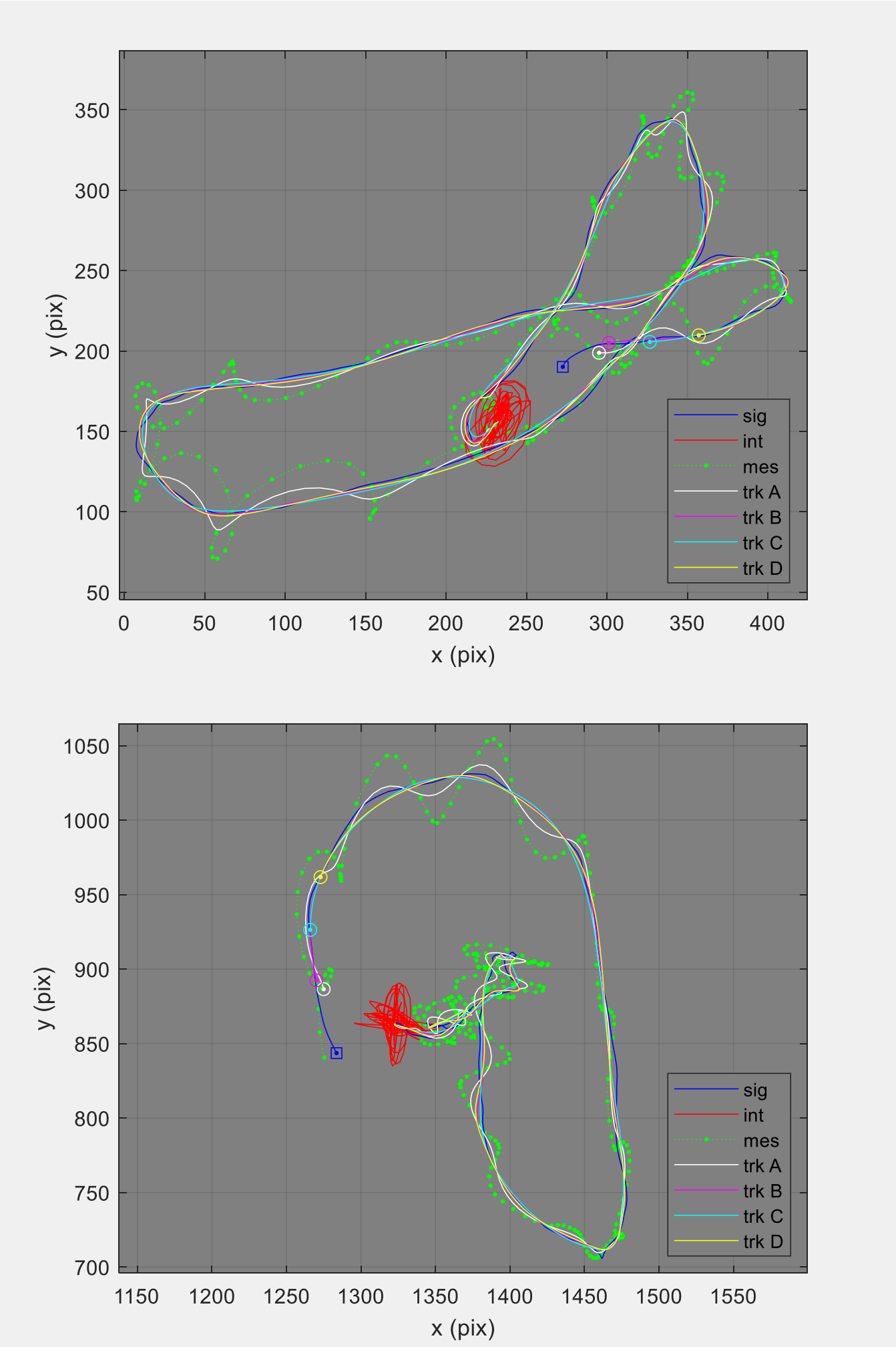

**Figure 25**. Random MC instantiations of Lo-G scenario (patrolling aircraft) with tracks.

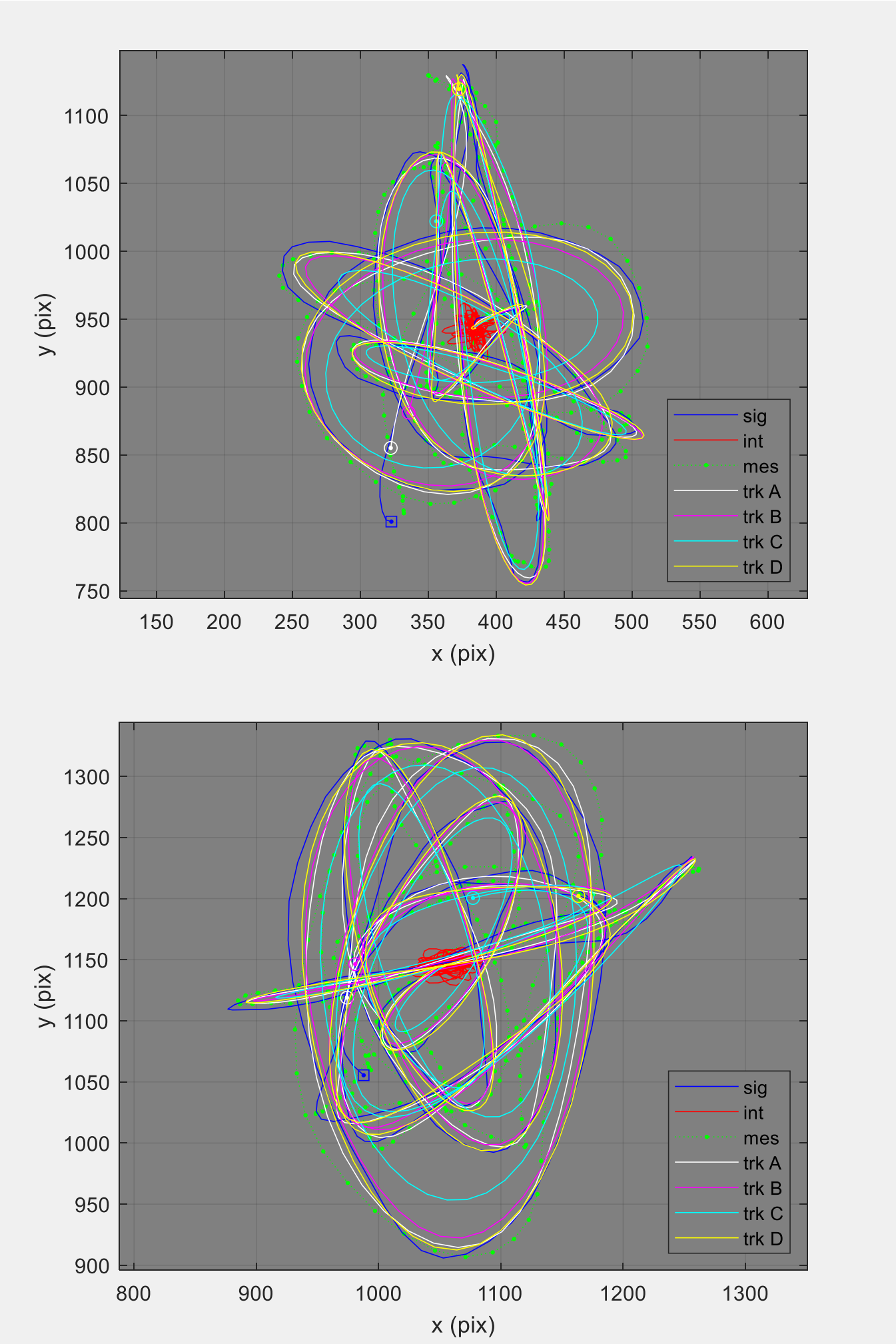

**Figure 26**. Random MC instantiations of Hi-G scenario (dogfighting aircraft) with tracks.

## 8 Conclusion

A common objective for signal processors, target trackers, and feedback controls, is the amplification of signal, the attenuation of interference and the minimization of noise. And in these systems, many signal analysis functions can be performed via Taylor-series expansions (in both time and frequency) using estimates of discrete-time derivatives, particularly when the sampling frequency is many multiples of the signal-process bandwidth (i.e. in oversampled systems). The delay (i.e. latency or lag) applied in such filtering components is a critical design parameter that determines its effectiveness in isolation and the responsiveness of the integrated system, e.g. a command-and-control network. When a long lag is applied it is possible to design filters that are highly frequency selective; however, it may also lead to unbounded divergence (i.e. instability) in high-gain closed-loop (i.e. feedback) systems, e.g. when applied in a regulator, servomechanism, or seeker system. The pervasiveness of digital filters in this age of automation, the ubiquity of fast digitizers in small computers, and the paucity of simple yet effective design procedures involving the optimization of group delay in recursive smoothers and differentiators, motivated the research presented here.

A novel design procedure for the determination of the group-delay that minimizes the white-noise variance of a discrete-time differentiator is presented. It is suitable for processing low-frequency signals in noise and high-frequency interference that are sampled at relatively high frequencies. The recursive structure of the low-order IIR filters is ideal for embedded devices that are unable to utilize the scale-efficiencies of the FFT for the realization of high-order FIR filters. The derivative order (in the time domain) is specified via constraints imposed on derivatives of the complex response at dc (in the frequency domain). A general expression for high-order dc constraints (to encourage passband phase linearity) in a filter with an arbitrary group delay is derived. Additional constraints are optionally applied to suppress narrowband interference and broad-band high-frequency noise. Banks of such filters, or the internal states of an appropriately designed IIR filter, may also be used to estimate the derivative states of an (approximately) integrating process that obeys Newton's laws of motion. The filters are configured for detecting pulsed signals and tracking manoeuvring targets. In these simulations, the signal process is only partially known; however, the noise process is assumed to be known and invariant.

Frequency-domain representations are sometimes used to design robust digital filters for feedback-control systems, where closed-loop stability is more important than the characteristics of the transient response, as an alternative to state-space controls and observers. They are routinely used in RF signal-processing systems, where a greater emphasis is placed on the steady-state response rather than the transient response. It is suggested here that they may also be used for pulse detection, as a simple alternative to the Wiener filter, when process models are only partially known, and for target tracking, as a simple alternative to the Kalman filter, when noise statistics are unknown or non-Gaussian. It is shown that when the frequency-domain properties of a good state estimator are understood, the filter coefficients of recursive detectors and trackers are readily derived via derivative constraints. The proposed MaxFlat procedure allows the bandwidth to be configured directly (using Butterworth poles), the passband group delay to be set optimally (by minimizing the white-noise gain), and interference to be cancelled (by placing notches or nulls). Due to the *a-priori* placement of the poles, the resulting causal IIR filters are guaranteed to be stable.

The once inexorable increase in the clock frequencies of new computers has stalled in recent years (calling Moore's 'law' into question) however the sampling frequencies of digitizers continues to increase unabated. This presents two serious challenges in the field of digital (sensor) signal processing. The first, and more obvious one, is how to design low-complexity digital filters to maintain real-time throughput in online systems. The second, and less obvious one, is how to model and handle the extra resolution or fine structure that the higher sampling frequencies reveal. At higher sampling rates, what may have previously appeared to be uncorrelated (i.e. white) noise (and approximately Gaussian), becomes correlated (i.e. coloured) noise or interference. Unfortunately, a common solution to both problems is to simply decimate the data and ignore the new information provided. This paper attempts to address both questions by incorporating non-trivial process models, that are only partially understood, into low-complexity state-estimators.

**Appendix A**

```
%%%%%%%%%%%%%%%%%%%%%%%%%%%%%%%%%%%%%%%%%%%%%%%%%%%%%%%%%%%%%%%%%%%%%%%%%%%%
% dtn_and_trk_1.m
%%%%%%%%%%%%%%%%%%%%%%%%%%%%%%%%%%%%%%%%%%%%%%%%%%%%%%%%%%%%%%%%%%%%%%%%%%%%
%
% A script for the design of MaxFlat filterbanks with Butterworth poles.
% To compute derivatives of a signal in noise and interference.
% For use in pulse detection and target tracking problems, for example.
%
%%%%%%%%%%%%%%%%%%%%%%%%%%%%%%%%%%%%%%%%%%%%%%%%%%%%%%%%%%%%%%%%%%%%%%%%%%%%

clear
close all

%%%%%%%%%%%%%%%%%%%%%%%%%%%%%%%%%%%%%%%%%%%%%%%%%%%%%%%%%%%%%%%%%%%%%%%%%%%%
% Filter parameters
% User adjustable

% Sampling frequency
F_s = 1000.0;     % smp/sec (Hz)

% Signal bandwidth
F_wb = 0.050*F_s; % smp/sec (Hz)

% Interference centre frequency
F_nb = 0.100*F_s; % smp/sec (Hz)

% Number of temporal derivatives to compute.
% Number of filters in filterbank.
% k_t = 0 ... K_t-1
% (k_t = 0 is a smoother)
%
K_t = 3;

% Design filter with this passband group delay (q, in samples)
% A value of Inf indicates that the optimal grp del
% should be computed and used.
% grp del in seconds = T_s*q
grp_del = Inf;

% Number of derivative constraints at w_dc, w_nb and w_pi
K_w_dc = 3;
K_w_nb = 2;
K_w_pi = 1;

%%%%%%%%%%%%%%%%%%%%%%%%%%%%%%%%%%%%%%%%%%%%%%%%%%%%%%%%%%%%%%%%%%%%%%%%%%%%
% Check filter parameters

assert(K_w_dc>=K_t)
assert(F_nb>F_wb)

%%%%%%%%%%%%%%%%%%%%%%%%%%%%%%%%%%%%%%%%%%%%%%%%%%%%%%%%%%%%%%%%%%%%%%%%%%%%
% Constants

NUP = 0;
YEP = 1;

% dc
len_dc = Inf;
w_dc = 0;

% pi
len_pi = 2;
w_pi = pi;

% When real values are expected
% check that abs val of imag part is less than this value
% before taking real part.
tol_cpx = 1.0E-3;
```

```
%%%%%%%%%%%%%%%%%%%%%%%%%%%%%%%%%%%%%%%%%%%%%%%%%%%%%%%%%%%%%%%%%%%%%%%%%%%
% Derived parameters

% Sampling period
T_s = 1/F_s;    % sec/smp

% Signal bandwidth
frq_wb = F_wb/F_s;      % cyc/smp
len_wb = 1/frq_wb;      % smp/cyc
omg_wb = 2*pi*frq_wb;   % rad/smp
w_wb = omg_wb;

% Interference centre frequency
frq_nb = F_nb/F_s;      % cyc/smp
len_nb = 1/frq_nb;      % smp/cyc
omg_nb = 2*pi*frq_nb;   % rad/smp
w_nb = omg_nb;

% omg and w are used interchangeably here
% they are both the angular frequency (rad/smp)

bnd_wid_fac = 1.0;
w_c = bnd_wid_fac*w_wb;
omg_c = w_c;        % rad/smp
OMG_c = omg_c*F_s;  % rad/sec

% Total number of constraints
K_w = K_w_dc+2*K_w_nb+K_w_pi;

% Number of IIR poles, i.e. order of filter
K = K_w;

%%%%%%%%%%%%%%%%%%%%%%%%%%%%%%%%%%%%%%%%%%%%%%%%%%%%%%%%%%%%%%%%%%%%%%%%%%%
% Define basis functions
% Use Butterworth poles


% Define the basis functions f_k(z)
% Use a linear combination of these to satisfy the derivative constraints.

% Denominator polynomial of Butterworth filter
% H(s) = B(s)/A(s)
%
a_vec = zeros(1,2*K+1);
a_vec(2*K+1) = 1;
a_vec(0+1) = (-1/OMG_c^2)^K;

% Get (causal) stable poles
%
a_rts_s = roots(a_vec);
fnd = find(real(a_rts_s)<0);
a_rts_s = a_rts_s(fnd);

% Map s-plane poles to z-plane poles
a_rts_z = exp(a_rts_s*T_s);
a_rts = a_rts_z;
%
% ... this method (impulse invariance?)
% is much simpler than bi-linear transformation.
% It is sufficient.
% Don't need bnd wid to be matched exactly.

%%%%%%%%%%%%%%%%%%%%%%%%%%%%%%%%%%%%%%%%%%%%%%%%%%%%%%%%%%%%%%%%%%%%%%%%%%%
% Determine the derivatives
% of the complex frequency response f_k(w)
% of the basis functions f_k(z)
% at the specified frequencies.
% (psi in paper)

% This is an intermediate qty
% Used to evaluate derivatives of basis functions.

alp = zeros(K,K);
for k_w = 0:K-1 % order of derivative
```

```
    for l_w = 0:K-1 %
        if l_w<0
            alp_l = 0;
        elseif l_w==0
            alp_l = 1;
        elseif l_w==k_w
            alp_l = factorial(l_w);
        elseif l_w>k_w
            alp_l = 0;
        else
            alp_l = l_w*alp(k_w-1+1,l_w-1+1)+(l_w+1)*alp(k_w-1+1,l_w+1);
        end
        alp(k_w+1,l_w+1) = alp_l;
    end
end

omg_vec = [w_dc -w_nb +w_nb w_pi];
K_w_vec = [K_w_dc K_w_nb K_w_nb K_w_pi];
F = [];
for frq_ind = 1:length(omg_vec)
    w_ = omg_vec(frq_ind);
    K_ = K_w_vec(frq_ind);
    F_ = zeros(K_,K);
    for k_f = 0:K-1 % basis function index
        p_k = a_rts(k_f+1); % pole of the kth basis function
        f_w_k = exp(i*w_)/(exp(i*w_)-p_k); % cpx frq rsp of kth basis function
        for k_w = 0:K_-1 % order of derivative
            F_sum = 0;
            for l_w = 0:k_w
                F_sum = F_sum+alp(k_w+1,l_w+1)*(-1)^l_w*f_w_k^(l_w+1);
            end
            F_(k_w+1,k_f+1) = i^k_w*F_sum;
        end
    end
    F = [F;F_];
end

F_inv = inv(F);

%%%%%%%%%%%%%%%%%%%%%%%%%%%%%%%%%%%%%%%%%%%%%%%%%%%%%%%%%%%%%%%%%%%%%%%%%%%
% Evaluate frequency domain integrals (S)
% using inf summations in sample domain
% (Thank you Marc-Antoine Parseval)
% These will be used to evaluate the white-noise gain (wng)
% even if the opt grp del is not computed

M_inf = 10000; % used for inf sum
x_vec_k = [1 zeros(1,M_inf-1)]; % impulse in
tol_inf = 1.0E-12; % check for convergence

S = zeros(K,K);

for k_row = 0:K-1
    for k_col = 0:K-1

        % Do integral as an inf sum in m domain ...

        p_k = a_rts(k_row+1);
        b_vec_k = [1 0];
        a_vec_k = [1 -p_k];
        y_vec_k = filter(b_vec_k,a_vec_k,x_vec_k);
        y_row_k = y_vec_k;

        p_k = a_rts(k_col+1);
        b_vec_k = [1 0];
        a_vec_k = [1 -p_k];
        y_vec_k = filter(b_vec_k,a_vec_k,x_vec_k);
        y_col_k = y_vec_k;

        y_vec_k = conj(y_row_k).*y_col_k;
        assert(abs(y_vec_k(end))<tol_inf);
        %
        % If this fails, then extend summation
        % by increasing length of impulse input (x_vec_k)
        % i.e. M_inf
```

```
        S_k = sum(y_vec_k);
        S(k_row+1,k_col+1) = S_k;

    end

end

%%%%%%%%%%%%%%%%%%%%%%%%%%%%%%%%%%%%%%%%%%%%%%%%%%%%%%%%%%%%%%%%%%%%%%%%%%%
% Set passband group-delay (q) of filterbank

if grp_del<Inf

    % Use specified group delay
    q_val = grp_del

else

    % Determine the optimal group delay

    J = F_inv'*S*F_inv;

    k0_max = 2*(K_w_dc-1);
    q0_vec = zeros(1,k0_max+1);

    k1_max = 2*(K_w_dc-1)-1;
    q1_vec = zeros(1,k1_max+1);

    for k_b = 0:K_w_dc-1
        del_b = (-i)^k_b;
        for k_a = 0:K_w_dc-1
            del_a = (-i)^k_a;

            fac = conj(del_a)*del_b;
            k_q = k_a+k_b;
            if k_q>=0
                q0_vec(k_q+1) = q0_vec(k_q+1)+fac*J(k_a+1,k_b+1);
            end

            fac = (k_a+k_b)*conj(del_a)*del_b;
            k_q = k_a+k_b-1;
            if k_q>=0
                q1_vec(k_q+1) = q1_vec(k_q+1)+fac*J(k_a+1,k_b+1);
            end

        end
    end

    q0_vec = fliplr(q0_vec);

    q1_vec = fliplr(q1_vec);
    q_sol_all = roots(q1_vec);

    wng_sol_all = polyval(q0_vec,q_sol_all);

    % Only consider real solutions
    q_sol_r = real(q_sol_all(find(abs(imag(q_sol_all))<tol_cpx)));
    % Use the grp del that yields the lowest wng
    wng_sol_r = polyval(q0_vec,q_sol_r);
    [wng_min,ind_min] = min(wng_sol_r);
    % If multiple real solutions have this wng
    % Then use the solution that has the lowest grp del
    tol_wng = 1.0E-6;
    fnd_sol = find(abs(wng_sol_r-wng_min)<tol_wng);
    [q_min,ind_min] = min(q_sol_r(fnd_sol));
    q_sol = q_min;
    wng_sol = wng_min;

    q_val = q_sol;

end

%%%%%%%%%%%%%%%%%%%%%%%%%%%%%%%%%%%%%%%%%%%%%%%%%%%%%%%%%%%%%%%%%%%%%%%%%%%
% Determine the desired derivatives of the cpx frq rsp
% at dc for the (specified or computed) grp del
% to pass the signal
%
```

```
D_dc = zeros(K_w_dc,K_t);
for k_w = 0:K_w_dc-1
    for k_t = 0:K_t-1
        if k_w>=k_t
            D_dc(k_w+1,k_t+1) =  ...
                (i^k_w)*(-q_val)^(k_w-k_t)*(1/T_s)^k_t* ...
                factorial(k_w)/factorial(k_w-k_t);
        end
    end
end

% And at other frq ...

% Desired derivatives of the cpx frq rsp
% at nb and pi are all zero
% to cancel interference.

D_nb = zeros(2*K_w_nb,K_t);

D_pi = zeros(1*K_w_pi,K_t);

D = [D_dc;D_nb;D_pi];

%%%%%%%%%%%%%%%%%%%%%%%%%%%%%%%%%%%%%%%%%%%%%%%%%%%%%%%%%%%%%%%%%%%%%%%%%%%
% Determine linear coefficients

% H(z) is expressed as a linear combination of f(z)
% i.e. 1st order basis functions.
% Determine those coefficients.
%
c = F_inv*D;
%
% columns of c are the coeffs of the k_t th filter in the filterbank

% Determine the wng mtx of the filters in the filterbank
wng_mtx = c'*S*c;
assert(all(all(abs(imag(wng_mtx))<tol_cpx)))
wng_mtx = real(wng_mtx);

%%%%%%%%%%%%%%%%%%%%%%%%%%%%%%%%%%%%%%%%%%%%%%%%%%%%%%%%%%%%%%%%%%%%%%%%%%%
% Determine
% a_vec (common)
% and
% b_vec (unique).
% for filters in the filterbank
% For use in x_vec = filter(b_vec,a_vec,x_vec)

% H(z) = B(z)/A(z)

% Determine A(z) polynomial for the filterbank
%
a_vec = poly(a_rts);
% ... same ...
a_vec = 1;
for k = 0:K-1
    a_vec_k = [1 -a_rts(k+1)];
    a_vec = conv(a_vec,a_vec_k);
end
assert(all(abs(imag(a_vec))<tol_cpx));
a_vec = real(a_vec);

% Determine B(z) polynomial for each filter
% H(z) is a sum of 1st-order terms.
% Multiply top and bottom of each term by A(z)
% cancel and sum all terms to get b_vec
%
b_arr = zeros(K_t,K+1);
for k_t = 0:K_t-1
    b_vec = zeros(1,K+1);
    for k = 0:K-1
        b_vec_k = [c(k+1,k_t+1) 0];
        for k_ = 0:K-1
            if k_~=k
                a_vec_k = [1 -a_rts(k_+1)];
                b_vec_k = conv(b_vec_k,a_vec_k);
            end
```

```
        end
        b_vec = b_vec+b_vec_k;
    end
    assert(all(abs(imag(b_vec))<tol_cpx));
    b_vec = real(b_vec);

    b_arr(k_t+1,:) = real(b_vec);
end

%%%%%%%%%%%%%%%%%%%%%%%%%%%%%%%%%%%%%%%%%%%%%%%%%%%%%%%%%%%%%%%%%%%%%%%%%%%%
% Define linear state-space system using the computed coefficients

% Diagonal Canonical Form
% (simplest form)
%
G_dcf = diag(a_rts);
H_dcf = ones(K,1);
C_dcf_aug = eye(K,K);
C_dcf_aug([0:K_t-1]+1,[0:K-1]+1) = conj(c)';
C_dcf = C_dcf_aug([0:K_t-1]+1,:);

% Derivative State Form (DSF)
% (kinenmatic form)
%
dcf_T_dsf = inv(C_dcf_aug);
dsf_T_dcf = C_dcf_aug;
C_dsf_aug = C_dcf_aug*dcf_T_dsf;
C_dsf = C_dsf_aug([0:K_t-1]+1,:);
G_dsf = dsf_T_dcf*G_dcf*dcf_T_dsf;
H_dsf = dsf_T_dcf*H_dcf;

% Controller canonical form (CCF)
% (this form exposes filter coeffs)
%
g_vec = -a_vec([1:K]+1);
G_ccf = [g_vec;[eye(K-1),zeros(K-1,1)]];
H_ccf = [1;zeros(K-1,1)];
C_ccf = b_arr(:,[0:K-1]+1);

% Doesn't matter which LSS coord sys is used
% response should be the same.

sys_DCF = 0;
sys_DSF = 1;
sys_CCF = 2;

sys = sys_CCF; % pick a coord sys

if sys==sys_DCF
    G = G_dcf;
    H = H_dcf;
    C = C_dcf;
elseif sys==sys_DSF
    G = G_dsf;
    H = H_dsf;
    C = C_dsf;
elseif sys==sys_CCF
    G = G_ccf;
    H = H_ccf;
    C = C_ccf;
end

% Generate imp rsp using LSS recursion
% Should be the same as using filter()

N = 100;
m_vec = [0:N-1];
x_vec = [1 zeros(1,N-1)]; % impulse
y_vec = zeros(K_t,N);

w_n = zeros(K,1);

for n = 0:N-1

    x_n = x_vec(n+1);
    w_n = G*w_n+H*x_n;
```

```
    y_n = C*w_n;
    y_vec(:,n+1) = y_n;

end

assert(all(all(abs(imag(y_vec))<tol_cpx)))
y_vec = real(y_vec);
fig_imp_rsp = {};
for k_t = 0:K_t-1

    fig_imp_rsp{k_t+1} = figure;
    figure(fig_imp_rsp{k_t+1});
    hold on
    grid on
    box on
    ylabel('h(m)')
    xlabel('m')
    stem(m_vec,y_vec(k_t+1,:),'ob')

end

%%%%%%%%%%%%%%%%%%%%%%%%%%%%%%%%%%%%%%%%%%%%%%%%%%%%%%%%%%%%%%%%%%%%%%%%%%%%
```

**Appendix B**

```
%%%%%%%%%%%%%%%%%%%%%%%%%%%%%%%%%%%%%%%%%%%%%%%%%%%%%%%%%%%%%%%%%%%%%%%%%%%
% fac_fwd_bwd_1.m
%%%%%%%%%%%%%%%%%%%%%%%%%%%%%%%%%%%%%%%%%%%%%%%%%%%%%%%%%%%%%%%%%%%%%%%%%%%
%
% A script to factor a non-causal discrete-time transfer function
% H(z) = B(z)/A(z)
% into a sum of realizable forward (fwd) and backwards (bwd) parts
% with poles inside and outside the unit circle, respectively
% i.e.
% H(z) = H_fwd(z)+H_bwd(z)
% or
% B(z)/A_z = B_fwd(z)/A_fwd(z)+B_bwd(z)/A_bwd(z)
%
%%%%%%%%%%%%%%%%%%%%%%%%%%%%%%%%%%%%%%%%%%%%%%%%%%%%%%%%%%%%%%%%%%%%%%%%%%%

clear
close all

NUP = 0;
YEP = 1;

%%%%%%%%%%%%%%%%%%%%%%%%%%%%%%%%%%%%%%%%%%%%%%%%%%%%%%%%%%%%%%%%%%%%%%%%%%%
% Randomly generate H(z)

K_ = 4;
K = 2*K_; % order of H(z)

% B(z)
b_rts = randn(K_ ,1)+i*randn(K_ ,1);
b_rts = [b_rts;conj(b_rts)];
[val,ind] = sort(abs(b_rts));
b_rts = b_rts(ind);
b_vec = poly(b_rts);

% A(z)
a_rts = 2.0*rand(K_ ,1)+0.5*rand(K_ ,1)*i;
a_rts = [a_rts;conj(a_rts)];
[val,ind] = sort(abs(a_rts));
a_rts = a_rts(ind);
a_vec = poly(a_rts);

% Normalize for unity dc gain
w_dc = 0;
h_dc = polyval(b_vec,exp(i*w_dc))/polyval(a_vec,exp(i*w_dc));
b_vec = b_vec/abs(h_dc);

%%%%%%%%%%%%%%%%%%%%%%%%%%%%%%%%%%%%%%%%%%%%%%%%%%%%%%%%%%%%%%%%%%%%%%%%%%%
% Factor H(z)

rts_rad = abs(a_rts);

% Get poles inside unit circle and make A_fwd(z)

fnd_ltu = find(rts_rad<1);
K_ltu = length(fnd_ltu);
a_rts_ltu = a_rts(fnd_ltu);
a_ltu = poly(a_rts_ltu);
a_fwd = a_ltu;
a_fwd_0 = a_fwd(0+1);

% Get poles outside unit circle and make A_bwd(z)

fnd_gtu = find(rts_rad>1);
K_gtu = length(fnd_gtu);
a_rts_gtu = a_rts(fnd_gtu);
a_gtu = poly(a_rts_gtu);
a_bwd = fliplr(a_gtu);
a_bwd_0 = a_bwd(0+1);

K_all = K_ltu+K_gtu; % = K
```

```
% Now solve a set of linear equations to get B_fwd(z) and B_bwd(z)

A = zeros(K_all+1,K_all+1);
for k_ltu = 0:K_ltu-1
    A(k_ltu+[0:K_gtu]+1,k_ltu+1) = a_gtu';
end
for k_gtu = 0:K_gtu
    A(k_gtu+[0:K_ltu]+1,K_ltu+k_gtu+1) = a_ltu';
end

b = b_vec';

c = inv(A)*b;

b_ltu = [c([0:K_ltu-1]+1)' 0];
b_fwd = b_ltu;

b_gtu = [c(K_ltu+[0:K_gtu]+1)'];
b_bwd = fliplr(b_gtu);

b_fwd = b_fwd/a_fwd_0;
a_fwd = a_fwd/a_fwd_0;

b_bwd = b_bwd/a_bwd_0;
a_bwd = a_bwd/a_bwd_0;

% Check equivalence

a_vec_chk = conv(a_ltu,a_gtu);
a_vec_err = sum(abs(a_vec-a_vec_chk)) % = 0?

b_vec_chk = conv(b_ltu,a_gtu)+conv(b_gtu,a_ltu);
b_vec_err = sum(abs(b_vec-b_vec_chk)) % = 0?

% Can now realize the filter using fwd and bwd parts
% Apply filter
% to generate the impulse response of H(z)

M = 100;
m_vec = [-M:+M];
x_vec = [zeros(1,M) 1 zeros(1,M)];

y_vec_fwd = filter(b_fwd,a_fwd,x_vec);
y_vec_bwd = fliplr(filter(b_bwd,a_bwd,fliplr(x_vec)));
y_vec = y_vec_fwd+y_vec_bwd;

h_m_vec = y_vec;

fig_imp_rsp = figure;
hold on
grid on
box on
xlabel('m (smp)')
ylabel('h[m]')
stem(m_vec,h_m_vec,'ob')

%%%%%%%%%%%%%%%%%%%%%%%%%%%%%%%%%%%%%%%%%%%%%%%%%%%%%%%%%%%%%%%%%%%%%%%%%%%%
```